
\input phyzzx.tex
%
\newbox\hdbox%
\newcount\hdrows%
\newcount\multispancount%
\newcount\ncase%
\newcount\ncols
\newcount\nrows%
\newcount\nspan%
\newcount\ntemp%
\newdimen\hdsize%
\newdimen\newhdsize%
\newdimen\parasize%
\newdimen\spreadwidth%
\newdimen\thicksize%
\newdimen\thinsize%
\newdimen\tablewidth%
\newif\ifcentertables%
\newif\ifendsize%
\newif\iffirstrow%
\newif\iftableinfo%
\newtoks\dbt%
\newtoks\hdtks%
\newtoks\savetks%
\newtoks\tableLETtokens%
\newtoks\tabletokens%
\newtoks\widthspec%
%
%
\immediate\write15{%
CP SMSG GJMSINK TEXTABLE --> TABLE MACROS V. 851121 JOB = \jobname%
}%
%
%
\tableinfotrue%
\catcode`\@=11
%
%
\def\tstrut{\vrule height3.1ex depth1.2ex width0pt}%
\def\and{\char`\&}
\def\tablerule{\noalign{\hrule height\thinsize depth0pt}}%
\thicksize=1.5pt
\thinsize=0.6pt
\def\thickrule{\noalign{\hrule height\thicksize depth0pt}}%
\def\ctr#1{\hfil\ #1\hfil}%
%
%
%
%
\tablewidth=-\maxdimen%
\spreadwidth=-\maxdimen%
\def\tabskipglue{0pt plus 1fil minus 1fil}%
%
%
\centertablestrue%
%
%
%
%
\parasize=4in%
\long\def\para#1{
   {%
      \vtop{%
         \hsize=\parasize%
         \baselineskip14pt%
         \lineskip1pt%
         \lineskiplimit1pt%
         \noindent #1%
         \vrule width0pt depth6pt%
      }%
   }%
}%
\gdef\ARGS{########}
\gdef\headerARGS{####}
\def\@mpersand{&}
{\catcode`\|=13
\gdef\letbarzero{\let|0}
\gdef\letbartab{\def|{&&}}%
\gdef\letvbbar{\let\vb|}%
}
{\catcode`\&=4
\def\ampskip{&\omit\hfil&}
\catcode`\&=13
\let&0
\xdef\letampskip{\def&{\ampskip}}%
\gdef\letnovbamp{\let\novb&\let\tab&}
}
\def\begintable{
   \begingroup%
   \catcode`\|=13\letbartab\letvbbar%
   \catcode`\&=13\letampskip\letnovbamp%
   \def\multispan##1{
      \omit \mscount##1%
      \multiply\mscount\tw@\advance\mscount\m@ne%
      \loop\ifnum\mscount>\@ne \sp@n\repeat%
   }
   \def\|{%
      &\omit\widevline&%
   }%
   \ruledtable
}
\long\def\ruledtable#1\endtable{%
%
%
%
   \offinterlineskip
   \tabskip 0pt
   \def\widevline{\vrule width\thicksize}
   \def\endrow{\@mpersand\omit\hfil\crnorm\@mpersand}%
   \def\crthick{\@mpersand\crnorm\thickrule\@mpersand}%
   \def\crthickneg##1{\@mpersand\crnorm\thickrule
          \noalign{{\skip0=##1\vskip-\skip0}}\@mpersand}%
   \def\crnorule{\@mpersand\crnorm\@mpersand}%
   \def\crnoruleneg##1{\@mpersand\crnorm
          \noalign{{\skip0=##1\vskip-\skip0}}\@mpersand}%
   \let\nr=\crnorule
   \def\endtable{\@mpersand\crnorm\thickrule}%
   \let\crnorm=\cr
%
%
   \edef\cr{\@mpersand\crnorm\tablerule\@mpersand}%
   \def\crneg##1{\@mpersand\crnorm\tablerule
          \noalign{{\skip0=##1\vskip-\skip0}}\@mpersand}%
   \let\ctneg=\crthickneg
   \let\nrneg=\crnoruleneg
   \the\tableLETtokens
%
%
   \tabletokens={&#1}
%
%
   \countROWS\tabletokens\into\nrows%
   \countCOLS\tabletokens\into\ncols%
%
%
   \advance\ncols by -1%
   \divide\ncols by 2%
   \advance\nrows by 1%
%
%
   \iftableinfo %
      \immediate\write16{[Nrows=\the\nrows, Ncols=\the\ncols]}%
   \fi%
%
%
   \ifcentertables
      \ifhmode \par\fi
      \line{
      \hss
   \else %
      \hbox{%
   \fi
      \vbox{%
         \makePREAMBLE{\the\ncols}
         \edef\next{\preamble}
         \let\preamble=\next
         \makeTABLE{\preamble}{\tabletokens}
      }
      \ifcentertables \hss}\else }\fi
   \endgroup
   \tablewidth=-\maxdimen
   \spreadwidth=-\maxdimen
}
\def\makeTABLE#1#2{
   {
   \let\ifmath0
   \let\header0
   \let\multispan0
%
%
   \ncase=0%
   \ifdim\tablewidth>-\maxdimen \ncase=1\fi%
   \ifdim\spreadwidth>-\maxdimen \ncase=2\fi%
   \relax
%
   \ifcase\ncase %
      \widthspec={}%
   \or %
      \widthspec=\expandafter{\expandafter t\expandafter o%
                 \the\tablewidth}%
   \else %
      \widthspec=\expandafter{\expandafter s\expandafter p\expandafter r%
                 \expandafter e\expandafter a\expandafter d%
                 \the\spreadwidth}%
   \fi %
   \xdef\next{
      \halign\the\widthspec{%
      #1
      \noalign{\hrule height\thicksize depth0pt}
      \the#2\endtable
%
      }
   }
   }
   \next
}
\def\makePREAMBLE#1{
   \ncols=#1
   \begingroup
   \let\ARGS=0
   \edef\xtp{\widevline\ARGS\tabskip\tabskipglue%
   &\ctr{\ARGS}\tstrut}
   \advance\ncols by -1
   \loop
      \ifnum\ncols>0 %
      \advance\ncols by -1%
      \edef\xtp{\xtp&\vrule width\thinsize\ARGS&\ctr{\ARGS}}%
   \repeat
   \xdef\preamble{\xtp&\widevline\ARGS\tabskip0pt%
   \crnorm}
   \endgroup
}
\def\countROWS#1\into#2{
   \let\countREGISTER=#2%
   \countREGISTER=0%
   \expandafter\ROWcount\the#1\endcount%
}%
\def\ROWcount{%
   \afterassignment\subROWcount\let\next= %
}%
\def\subROWcount{%
   \ifx\next\endcount %
      \let\next=\relax%
   \else%
      \ncase=0%
      \ifx\next\cr %
         \global\advance\countREGISTER by 1%
         \ncase=0%
      \fi%
      \ifx\next\endrow %
         \global\advance\countREGISTER by 1%
         \ncase=0%
      \fi%
      \ifx\next\crthick %
         \global\advance\countREGISTER by 1%
         \ncase=0%
      \fi%
      \ifx\next\crnorule %
         \global\advance\countREGISTER by 1%
         \ncase=0%
      \fi%
      \ifx\next\crthickneg %
         \global\advance\countREGISTER by 1%
         \ncase=0%
      \fi%
      \ifx\next\crnoruleneg %
         \global\advance\countREGISTER by 1%
         \ncase=0%
      \fi%
      \ifx\next\crneg %
         \global\advance\countREGISTER by 1%
         \ncase=0%
      \fi%
      \ifx\next\header %
         \ncase=1%
      \fi%
      \relax%
      \ifcase\ncase %
         \let\next\ROWcount%
      \or %
         \let\next\argROWskip%
      \else %
      \fi%
   \fi%
   \next%
}
\def\counthdROWS#1\into#2{%
\dvr{10}%
   \let\countREGISTER=#2%
   \countREGISTER=0%
\dvr{11}%
\dvr{13}%
   \expandafter\hdROWcount\the#1\endcount%
\dvr{12}%
}%
\def\hdROWcount{%
   \afterassignment\subhdROWcount\let\next= %
}%
\def\subhdROWcount{%
   \ifx\next\endcount %
      \let\next=\relax%
   \else%
      \ncase=0%
      \ifx\next\cr %
         \global\advance\countREGISTER by 1%
         \ncase=0%
      \fi%
      \ifx\next\endrow %
         \global\advance\countREGISTER by 1%
         \ncase=0%
      \fi%
      \ifx\next\crthick %
         \global\advance\countREGISTER by 1%
         \ncase=0%
      \fi%
      \ifx\next\crnorule %
         \global\advance\countREGISTER by 1%
         \ncase=0%
      \fi%
      \ifx\next\header %
         \ncase=1%
      \fi%
\relax%
      \ifcase\ncase %
         \let\next\hdROWcount%
      \or%
         \let\next\arghdROWskip%
      \else %
      \fi%
   \fi%
   \next%
}%
{\catcode`\|=13\letbartab
\gdef\countCOLS#1\into#2{%
   \let\countREGISTER=#2%
   \global\countREGISTER=0%
   \global\multispancount=0%
   \global\firstrowtrue
   \expandafter\COLcount\the#1\endcount%
   \global\advance\countREGISTER by 3%
   \global\advance\countREGISTER by -\multispancount
}%
\gdef\COLcount{%
   \afterassignment\subCOLcount\let\next= %
}%
{\catcode`\&=13%
\gdef\subCOLcount{%
   \ifx\next\endcount %
      \let\next=\relax%
   \else%
      \ncase=0%
      \iffirstrow
         \ifx\next& %
            \global\advance\countREGISTER by 2%
            \ncase=0%
         \fi%
         \ifx\next\span %
            \global\advance\countREGISTER by 1%
            \ncase=0%
         \fi%
         \ifx\next| %
            \global\advance\countREGISTER by 2%
            \ncase=0%
         \fi
         \ifx\next\|
            \global\advance\countREGISTER by 2%
            \ncase=0%
         \fi
         \ifx\next\multispan
            \ncase=1%
            \global\advance\multispancount by 1%
         \fi
         \ifx\next\header
            \ncase=2%
         \fi
         \ifx\next\cr       \global\firstrowfalse \fi
         \ifx\next\endrow   \global\firstrowfalse \fi
         \ifx\next\crthick  \global\firstrowfalse \fi
         \ifx\next\crnorule \global\firstrowfalse \fi
         \ifx\next\crnoruleneg \global\firstrowfalse \fi
         \ifx\next\crthickneg  \global\firstrowfalse \fi
         \ifx\next\crneg       \global\firstrowfalse \fi
      \fi
\relax
      \ifcase\ncase %
         \let\next\COLcount%
      \or %
         \let\next\spancount%
      \or %
         \let\next\argCOLskip%
      \else %
      \fi %
   \fi%
   \next%
}%
\gdef\argROWskip#1{%
   \let\next\ROWcount \next%
}
\gdef\arghdROWskip#1{%
   \let\next\ROWcount \next%
}
\gdef\argCOLskip#1{%
   \let\next\COLcount \next%
}
}
}
\def\spancount#1{
   \nspan=#1\multiply\nspan by 2\advance\nspan by -1%
   \global\advance \countREGISTER by \nspan
   \let\next\COLcount \next}%
\def\dvr#1{\relax}%
\def\header#1{%
\dvr{1}{\let\cr=\@mpersand%
\hdtks={#1}%
\counthdROWS\hdtks\into\hdrows%
\advance\hdrows by 1%
\ifnum\hdrows=0 \hdrows=1 \fi%
\dvr{5}\makehdPREAMBLE{\the\hdrows}%
\dvr{6}\getHDdimen{#1}%
{\parindent=0pt\hsize=\hdsize{\let\ifmath0%
\xdef\next{\valign{\headerpreamble #1\crnorm}}}\dvr{7}\next\dvr{8}%
}%
}\dvr{2}}
\def\makehdPREAMBLE#1{
\dvr{3}%
\hdrows=#1
{
\let\headerARGS=0%
\let\cr=\crnorm%
\edef\xtp{\vfil\hfil\hbox{\headerARGS}\hfil\vfil}%
\advance\hdrows by -1
\loop
\ifnum\hdrows>0%
\advance\hdrows by -1%
\edef\xtp{\xtp&\vfil\hfil\hbox{\headerARGS}\hfil\vfil}%
\repeat%
\xdef\headerpreamble{\xtp\crcr}%
}
\dvr{4}}
\def\getHDdimen#1{%
\hdsize=0pt%
\getsize#1\cr\end\cr%
}
\def\getsize#1\cr{%
\endsizefalse\savetks={#1}%
\expandafter\lookend\the\savetks\cr%
\relax \ifendsize \let\next\relax \else%
\setbox\hdbox=\hbox{#1}\newhdsize=1.0\wd\hdbox%
\ifdim\newhdsize>\hdsize \hdsize=\newhdsize \fi%
\let\next\getsize \fi%
\next%
}%
\def\lookend{\afterassignment\sublookend\let\looknext= }%
\def\sublookend{\relax%
\ifx\looknext\cr %
\let\looknext\relax \else %
   \relax
   \ifx\looknext\end \global\endsizetrue \fi%
   \let\looknext=\lookend%
    \fi \looknext%
}%
%
%
\def\tablelet#1{%
   \tableLETtokens=\expandafter{\the\tableLETtokens #1}%
}%
\catcode`\@=12

\def\plotpicture#1#2#3{\vskip#2}
\def\getpicture#1#2#3{\relax}
\singlespace
\vsize=51pc \hsize=35.5pc
\def\gev{{\rm GeV}}
\def\crr{\cr\noalign{\vskip 5pt}}
\def\us#1{\undertext{#1}}
\def\pri{^{\, \prime }}
\def\hf{\hfill}
\def\bold#1{\setbox0=\hbox{$ #1$}
     \kern-.015em\copy0\kern-\wd0    
     \kern.03em\copy0\kern-\wd0
     \kern-.015em\raise.0033em\box0 }
\def\9{\phantom 0}
\def\ifmath#1{\relax\ifmmode #1\else $#1$\fi}
\def\PRL#1&#2&#3&{\sl Phys.~Rev.~Lett.\ \bf #1\ \rm (19#2)\ #3}
\def\PRB#1&#2&#3&{\sl Phys.~Rev.\ \bf #1\ \rm (19#2)\ #3}
\def\NP#1&#2&#3&{\sl Nucl.~Phys.\ \bf #1\ \rm (19#2)\ #3}
\def\PRP#1&#2&#3&{\sl Phys.~Rep.\ \bf #1\ \rm (19#2)\ #3}
\def\PL#1&#2&#3&{\sl Phys.~Lett.\ \bf #1\ \rm (19#2)\ #3}
\def\normalrefmark#1{~[#1]}
\unlock
\def\refitem#1{\r@fitem{#1.}}
\refindent=20pt
\chapterminspace=6pc
\def\chapter#1{\par \penalty-300 \vskip30pt
   \spacecheck\chapterminspace
   \chapterreset \noindent{\ifcn@@\fourteenbf \chapterlabel.~\fi  #1}
   \nobreak\vskip\headskip \penalty 30000
   {\pr@tect\wlog{\string\chapter\space \chapterlabel}}
       \ifcontentson%
          \c@ntentscheck%
          \CONTENTS{C};{#1}%
       \fi%
}
\def\section#1{\par \ifnum\lastpenalty=30000\else
   \penalty-200 \vskip12pt plus 2pt minus 0.5pt
   \spacecheck\sectionminspace\fi
   \gl@bal\advance\sectionnumber by 1
   {\pr@tect
   \xdef\sectlabel{\ifcn@@ \chapterlabel.\fi
       \the\sectionstyle{\the\sectionnumber}}%
   \wlog{\string\section\space \sectlabel}}%
   \noindent {\bf \sectlabel.~~#1}\par
   \nobreak  \penalty 30000
       \ifcontentson%
          \c@ntentscheck%
          \CONTENTS{S};{#1}%
       \fi%
   }
   \def\chapterinfo#1{%
      \line{%
         \ifcn@@%
            \hbox to \itemsize{\hfil\chapterlabel .~~}%
         \fi%
         \noexpand{#1}\leaderfill\the\pagenumber%
      }%
   }

   \def\sectioninfo#1{%
      \line{%
         \ifcn@@%
            \hbox to 2\itemsize{\hfil\sectlabel ~~}%
          \else%
            \hbox to \itemsize{\hfil\quad}%
          \fi%
          \ \noexpand{#1}%
          \leaderfill \the\pagenumber%
      }%
   }
\lock

\font\mathbf = cmbsy10 scaled \magstep1
\let\ds=\displaystyle
\def\sss{\scriptscriptstyle}
\def\frac#1#2{{#1\over #2}}
\def\smallfrac#1#2{{\textstyle{#1\over #2}}}
\def\nicefrac#1/#2{\leavevmode\kern.1em\raise.5ex\hbox{\the\scriptfont0
         #1}\kern-.1em/\kern-.15em\lower.25ex\hbox{\the\scriptfont0 #2}}
\def\hs#1{\ifmath{_{\raise1.5pt\hbox{$\!\scriptstyle #1$}}}}
\def\vph{\vphantom(}
\def\ls#1{\ifmath{_{\lower1.5pt\hbox{$\scriptstyle #1$}}}}
\def\lsup#1{^{\lower 6pt\hbox{$\scriptstyle#1$}}}
\def\ptildeV#1{{\buildrel{\sss{(\sim)}}\over V}^{\lower10pt%
               \hbox{$\scriptstyle#1$}} }
\def\lrarrow#1#2{{\buildrel{\sss{\leftrightarrow}}\over #1}%
      \lower5pt\hbox{$\scriptstyle#2$} }
\def\Re{{\rm Re}}
\def\Tr{{\rm Tr}}
\def\Str{{\rm Str}}
\def\wh{\widehat}
\def\wt{\widetilde}
\def\mpl{M\ls{NP}}
\def\vevs{VEVs}
\def\eqalignalign#1{\vcenter{\openup1\jot
  \ialign{\strut\hfil
  $\displaystyle{##}$&$\displaystyle{{}##}$\quad
 &$\displaystyle{##}$&$\displaystyle{{}##}$\hfil\crcr
       #1\crcr}}}
\def\crrr{\cr\noalign{\vskip8pt}}
\def\epem{e^+e^-}
\def\calm{{\cal M}}

\def\calv{{\cal V}}

\def\sw{s_W}
\def\cw{c_W}
\def\sb  {s_{\beta}}
\def\cb  {c_{\beta}}

\def\tanb{\tan\beta}

\def\sww{s_W^2}
\def\hl{h^0}
\def\hh{H^0}
\def\ha{A^0}
\def\mhl{m_{\hl}}
\def\mhh{m_{\hh}}
\def\mha{m_{\ha}}
\def\hpm{H^{\pm}}
\def\mhpm{m_{\hpm}}
\def\mhsm{m_{\phi^0}}

\def\mb{m_b}
\def\mt{m_t}
\def\mw{m_W}
\def\mzz{m_Z^2}
\def\mz{m_Z}
\def\mweak{M\ls{{\rm weak}}}
\def\msusy{M\ls{{\rm SUSY}}}
\def\msusyy{M\ls{{\rm SUSY}}^2}
\def\half{\ifmath{{\textstyle{1 \over 2}}}}
\def\threehalf{\ifmath{{\textstyle{3 \over 2}}}}
\def\fivehalf{\ifmath{{\textstyle{5 \over 2}}}}
\def\ninehalf{\ifmath{{\textstyle{9 \over 2}}}}
\def\third{\ifmath{{\textstyle{1 \over 3}}}}

\def\sixteenthirds{\ifmath{{\textstyle{16\over 3}}}}
\def\twentythirds{\ifmath{{\textstyle{20 \over 3}}}}
\def\fortythirds{\ifmath{{\textstyle{40 \over 3}}}}
\def\sixtyfourthirds{\ifmath{{\textstyle{64 \over 3}}}}
\def\third{\ifmath{{\textstyle{1 \over 3}}}}
\def\fourth{\ifmath{{\textstyle{1\over 4}}}}
\def\ninefourth{\ifmath{{\textstyle{9 \over 4}}}}
\def\fifteenfourth{\ifmath{{\textstyle{15\over 4}}}}
\def\threefifths{\ifmath{{\textstyle{3 \over 5}}}}
\def\eighth{\ifmath{{\textstyle{1\over 8}}}}
\def\threeighth{\ifmath{{\textstyle{3 \over 8}}}}
\def\sevenninths{\ifmath{{\textstyle{7\over 9}}}}
\def\thirteenninths{\ifmath{{\textstyle{13\over 9}}}}
\def\fivetwelfth{\ifmath{{\textstyle{5 \over 12}}}}
\def\seventeentwelfth{\ifmath{{\textstyle{17 \over12}}}}

\centerline{\fourteenbf  INTRODUCTORY LOW-ENERGY SUPERSYMMETRY}
\vskip18pt
\centerline{\caps Howard E. Haber}
\vskip2pt
\centerline{\it Santa Cruz Institute for Particle Physics}
\centerline{\it University of California, Santa Cruz, CA 95064}
\vskip24pt
\vbox{ \narrower
\centerline{Abstract}
\vskip6pt
Low-energy supersymmetry is a theoretical extension of the Standard
Model of particle physics in which supersymmetry is invoked to
explain the origin of the electroweak scale.  In this approach,
the energy scale of supersymmetry breaking can be no larger than
about 1 TeV.  In these lectures, a pedagogical account of
softly broken supersymmetric gauge theories is presented.  The
minimal supersymmetric extension of the Standard Model (MSSM) is
defined and constraints on its parameters are explored.  The
implications of supersymmetric tree-level interactions and its
one-loop corrections are discussed.   Low-energy supersymmetric
model alternatives to the MSSM are briefly mentioned.
}
\vskip24pt
\vbox{%
\itemsize=10pt
\def\leaderfill{\leaders\hbox to0.9em{\hss.\hss}\hskip10pt plus1fill}%

\centerline{Table of Contents}
\vskip6pt
\line{{Introduction}              \leaderfill2}
\line{\hbox to \itemsize {\hfil {1}.}\quad
  {How to Build a Low-Energy Supersymmetric Model}
                                            \leaderfill5}
\line{\hbox to 4\itemsize {\hfil {1}.1}\quad
  {How to Create a Supersymmetric Extension of the Standard Model}
                                           \leaderfill5}
\line{\hbox to 4\itemsize {\hfil {1}.2}\quad
  {The Low-Energy Supersymmetric Spectrum} \leaderfill8}
\line{\hbox to 4\itemsize {\hfil {1}.3}\quad
  {The Supersymmetric Lagrangian}           \leaderfill{10}}
\line{\hbox to 4\itemsize {\hfil {1}.4}\quad
  {Supersymmetry Breaking}                  \leaderfill{16}}
\line{\hbox to 4\itemsize {\hfil {1}.5}\quad
  {The Definition of the MSSM}              \leaderfill{24}}
\line{\hbox to \itemsize {\hfil {2}.}\quad
  {What Do We Know About the Supersymmetric Parameters?}
                                           \leaderfill{33}}
\line{\hbox to 4\itemsize {\hfil {2}.1}\quad
 {The MSSM Parameter Space}                 \leaderfill{33}}
\line{\hbox to 4\itemsize {\hfil {2}.2}\quad
  {Theoretical Constraints and Biases}      \leaderfill{34}}
\line{\hbox to 4\itemsize {\hfil {2}.3}\quad
{Experimental Searches for Supersymmetry}   \leaderfill{49}}
\line{\hbox to 4\itemsize {\hfil {2}.4}\quad
{Alternative Low-Energy Supersymmetry Phenomenologies}
                                           \leaderfill{52}}
\line{\hbox to \itemsize {\hfil {3}.}\quad
  {Radiative Corrections to Tree-level Low-Energy Supersymmetry}
                                           \leaderfill{55}}
\line{\hbox to 4\itemsize {\hfil {3}.1}\quad
  {MSSM Contributions to Precision Electroweak Measurements}
                                           \leaderfill{55}}
\line{\hbox to 4\itemsize {\hfil {3}.2}\quad
  {MSSM Radiative Corrections to Processes Involving $b$ Quarks}
                                           \leaderfill{70}}
\line{\hbox to 4\itemsize {\hfil {3}.3}\quad
  {Natural Relations of the MSSM Higgs Sector}
                                           \leaderfill{72}}
\line{\hbox to 4\itemsize {\hfil {3}.4}\quad
  {Radiative Corrections to MSSM Higgs Masses}
                                            \leaderfill{76}}
\line{\hbox to 4\itemsize {\hfil {3}.5}\quad
  {Implications of the Radiatively Corrected MSSM Higgs Sector}
                                            \leaderfill{84}}
\line{{Concluding Remarks}  \leaderfill{88}}
 }
\vfill \endpage
\vskip30pt
\leftline{\fourteenbf Introduction}

\REF\suss{E. Gildener, {\sl Phys. Rev.} {\bf B14} (1976) 1667;
S. Weinberg, {\sl Phys. Lett.} {\bf 82B} (1979) 387;
L. Susskind, {\sl Phys. Rep.} {\bf 104} (1984) 181.}
\REF\thooft{S. Weinberg, {\sl Phys. Rev.} {\bf D13} (1976) 974;
{\bf D19} (1979) 1277; L. Susskind, {\sl Phys. Rev.} {\bf D20} (1979)
2619; G. 't Hooft, in {\it Recent Developments in Gauge
Theories,} Proceedings of the NATO Advanced Summer Institute,
Cargese, 1979, edited by G. 't~Hooft \etal\ (Plenum, New York,
1980) p.~135.}
\REF\gutproton{P. Langacker, {\sl Phys. Rep.} {\bf 72} (1981) 185;
G.G. Ross, \it Grand Unified Theories \rm (Addison-Wesley Publishing
Company, Reading, MA, 1984);
M. Srednicki, in {\it From the Planck Scale to the Weak Scale:
Toward a Theory of the Universe}, Proceedings of the 1986 Theoretical
Advanced Study Institute, University of California, Santa Cruz,
edited by H.E. Haber (World Scientific, Singapore, 1987) p.~1.}
The Standard Model of particle physics provides an extremely successful
description of all particle physics phenomena accessible to present
day accelerators.  No experimentally confirmed
deviations from the Standard Model have yet been
found.  It is clear that the Standard Model is a accurate ``effective
low-energy theory'' at energy scales up to 100 GeV.
However, theorists strongly believe that the success of the Standard
Model will not persist to higher energy scales.  This belief arises
{}from attempts to embed the Standard Model in a more fundamental
theory.  We know that the Standard Model cannot be the ultimate
theory, valid to arbitrarily high energy scales.  Even in the absence
of grand unification of strong and electroweak forces at a very high
energy scale, it is clear that the Standard Model must be
modified  to incorporate the effects of gravity at the Planck
scale ($\mpl\simeq 10^{19}$~GeV).  In this context, it is a mystery
why the ratio $\mw/\mpl\simeq 10^{-17}$ is so small.  This is called
the hierarchy problem\refmark\suss.  Moreover,
in the Standard Model, the scale of
the electroweak interactions derives from an elementary scalar field
which acquires a vacuum expectation value of $v=2\mw/g=246$~GeV.
However, if one couples a theory of scalar particles to new physics at
some arbitrarily high scale $\Lambda$, radiative corrections to the
scalar squared-mass are of ${\cal O}(\Lambda^2)$, due to the quadratic
divergence in the scalar self-energy (which indicates quadratic
sensitivity to the largest energy scale in the theory).  Thus, the
``natural'' mass for any scalar particle is $\Lambda$ (which is
presumably equal to $\mpl$).  Of course,
in order to have a successful electroweak theory, the Higgs mass must
be of order the electroweak scale.  The fact that the Higgs mass
cannot be equal to its natural value of $\mpl$ is called the
``naturalness'' problem\refmark\thooft.

The SU(5) grand unified model provides a nice example
of these points\refmark\gutproton.
In this model, $\Lambda\simeq 10^{15}$~GeV.
The SU(5) is broken down to SU(3)$\times$U(1)$\ls{EM}$ by vacuum
expectation values (\vevs) of a ${\bf 24}$-plet and a
${\bf 5}$-plet of Higgs scalars.  The respective \vevs\ are of order
$\Lambda$ and $v=246$~GeV respectively.  How is the hierarchy of
$v/\Lambda\simeq 10^{-13}$ arranged?  The dimensionful
parameters of the Higgs potential must be of order $\Lambda$
(raised to the appropriate power according to the dimension of the
corresponding term in the potential).  One can arrange the necessary
hierarchy only by carefully fine-tuning the squared-mass
parameters of the Higgs
potential to one part in $10^{26}$.  More generally, if the
Standard Model is to embedded in a more fundamental theory  that
includes gravity, the radiative corrections to the Higgs squared mass
will be dominated by physics at the Planck scale resulting in a
squared-mass shift of order $\mpl^2$.  A Higgs mass of order the
electroweak scale can be arranged only if the bare Higgs squared mass
is of ${\cal O}(\mpl^2)$ in such a way that when this is added to the
squared mass shift arising from the radiative corrections, the result
is a mass that is 17 orders of magnitude smaller than the Planck
scale!  This is the ``fine-tuning'' problem, which is related to
the hierarchy and naturalness problems.

\REF\technipapers{E. Farhi and L. Susskind, {\sl Phys. Rep.} {\bf 74}
(1981) 277; R.K. Kaul, {\sl Rev. Mod. Phys.} {\bf 55} (1983) 449.}
\REF\preons{I.A. D'Souza and C.S. Kalman,
{\it Preons} (World Scientific, Singapore, 1992).}
\REF\quaddiv{%
M. Veltman, {\sl Acta Phys. Pol.} {\bf B12} (1981) 437;
R.K. Kaul and P. Majumdar, {\sl Nucl. Phys.} {\bf B199} (1982) 36;
T. Inami, H. Nishino and S. Watamura, {\sl Phys. Lett.} {\bf 117B}
(1982) 197;
N.G. Deshpande, R.J. Johnson and E. Ma, {\sl Phys. Lett.} {\bf 130B}
(1983) 61; {\sl Phys. Rev.} {\bf D29} (1984) 2851.}
Theorists have been hard at work for more than a decade in an attempt
to circumvent the problems raised above.   The proposed solutions
involve removing the quadratic divergences from the theory that are
the root cause of the naturalness and fine-tuning problems.  Two
classes of solutions have been proposed.  In one class, the elementary
scalars are removed altogether.  One then must add new fundamental
fermions and new fundamental forces.  For example, in technicolor
models, new fermions $F$ are introduced such that
$\VEV{F\overline F}\neq0$ due to technicolor forces,
which results in the breaking of the electroweak
interactions\refmark\technipapers.
Other models of this class are composite models, where
some (or all) of the
particles that we presently regard as fundamental are
bound states of more fundamental fermions\refmark\preons.
In this class of models,
the physics that is responsible for electroweak symmetry breaking is
strong and its implementation requires non-perturbative techniques.
I believe that it is fair to say that no compelling realistic model of
this type has ever been constructed.  I will say no more about this
approach, since it is not the subject of these lectures.  The second
class of models are those where new particles are introduced to the
Standard Model in such a way that all quadratic divergences exactly
cancel.  Since we retain the Higgs scalars as elementary, the
cancellation of quadratic divergences can only be the result of a new
symmetry\refmark\quaddiv.
This symmetry is called supersymmetry which relates
fermions to bosons.  Because fermion self-energies have no quadratic
divergences, it is possible in a theory with a symmetry that relates
fermions to bosons to guarantee that
no quadratic divergences arise in scalar self-energies.

\REF\softsusy{L. Girardello and M.T. Grisaru, {\sl Nucl. Phys.}
{\bf B194} (1984) 419.}
In these lectures, I will explore the possibility that supersymmetry
is an approximate symmetry of the ``low-energy'' effective theory at
the electroweak scale.  That is, I shall invoke supersymmetry as a
solution to the naturalness problem.
However, supersymmetry cannot be an exact symmetry of
nature.  In a theory with exact supersymmetry, fermions and their
bosonic superpartners must be degenerate in mass.  The Standard Model
spectrum clearly does not satisfy this requirement.  Thus, if
supersymmetry is realized in nature, it must be broken.  It turns out
that supersymmetry-breaking can be implemented without introducing new
quadratic divergences to spoil the naturalness of the theory.  Such
breaking terms are called {\it soft}-supersymmetry-breaking
terms\refmark\softsusy.
These will be discussed in more detail in section 1.4.  Thus, the
approach of ``low-energy'' supersymmetry is one in which supersymmetry
is invoked to explain the origin of the electroweak scale.  The
soft-supersymmetry breaking mass parameters required for a
phenomenologically acceptable model cannot be larger than ${\cal
O}(1$~TeV).  The simplest model of this type is called the Minimal
Supersymmetric Standard Model (MSSM), which is the main subject of
these lectures.

In Lecture 1, I discuss theoretical aspects of supersymmetric model
building---how to construct a supersymmetric field theory including
the required soft-supersym\-metry-breaking terms.
The MSSM will then be
defined and elucidated.  In Lecture 2, I discuss the present knowledge
of the supersymmetric parameters.  Both theoretical and
phenomenological constraints are discussed.  Many of the
theoretical constraints discussed in the literature derive from
specific assumptions about the fundamental theory at high energy.  For
the most part, I will try to remain agnostic as to the details of the
underlying fundamental theory,
although I will try to alert the reader when some implicit
property of the high-energy theory is being used.  Finally, in Lecture
3, I examine radiative corrections to the tree-level results of
low-energy supersymmetry.  Any successful
model that supersedes the
Standard Model must be consistent with the precision electroweak data
{}from LEP and other experiments.  In addition, supersymmetric theories
possess many ``natural'' relations.  These are tree-level
relations among parameters that would each be infinitely renormalized
in a theory with less symmetry.  Radiative corrections to natural
relations are finite and calculable and can lead to important changes
in the tree-level phenomenology.

\REF\fayet{P. Fayet, {\sl Nucl. Phys.} {\bf B90} (1975) 104;
{\sl Phys. Lett.} {\bf 69B} (1977) 489; G.R. Farrar and P. Fayet,
{\sl Phys. Lett.} {\bf 76B} (1978) 575.}
\REF\wittensusy{%
E.~Witten, \sl Nucl.~Phys. \bf B188 \rm (1981) 513.}
\REF\susysol{%
S.~Dimopoulos and H.~Georgi, \sl Nucl.~Phys. \bf B193 \rm (1981) 150;
N.~Sakai, \sl Z.~Phys. \bf C11 \rm (1981) 153; L. Iba\~nez and G. Ross,
{\sl Phys. Lett.} {\bf 105B} (1981) 439; R.K. Kaul,
{\sl Phys. Lett.} {\bf 109B} (1982) 19.}
\REF\fish{M. Dine, W. Fischler,
and M. Srednicki, {\sl Nucl. Phys.} {\bf B189} (1981) 575;
S. Dimopoulos and S. Raby, {\sl Nucl. Phys.} {\bf B192} (1981) 353.}
\REF\oraifrev{L. O'Raifeartaigh, {\sl Communications
of the Dublin Inst.~for Advanced Studies}, Series A (Theor. Phys.)
No.~22 (1975).}
\REF\salam{A. Salam and J. Strathdee, {\sl Phys. Rev.} {\bf  D11}
(1975) 1521; {\sl Fortschr.~Phys.} {\bf 26} (1978) 57.}
\REF\fayfer{P. Fayet and S. Ferrara, {\sl Phys. Rep.} {\bf 32}
(1977) 249.}
\REF\nilles{H.P.  Nilles, {\sl Phys. Rep.} {\bf 110} (1984) 1.}
\REF\haberkane{H.E. Haber and G.L. Kane, {\sl Phys. Rep.} {\bf 117}
(1985) 75.}
\REF\sohnius{M.F. Sohnius, {\sl Phys. Rep.}, {\bf 128} (1985) 39.}
\REF\nanop{A.B. Lahanas and D.V. Nanopoulos, {\sl Phys. Rep.}
{\bf 145} (1987) 1.}
\REF\jacob{M. Jacob, editor, \it Supersymmetry and Supergravity, \sl
A Reprint Volume in  Physics Reports \rm (North-Holland Publishing
Company, Amsterdam, 1986).}
\REF\nath{P. Nath, R. Arnowitt and A.H. Chamseddine, {\it Applied
$N=1$ Supergravity} (World Scientific, Singapore, 1984).}
\REF\indiahall{L.J. Hall,
in {\it Supersymmetry and Supergravity, Nonperturbative QCD},
Proceedings of the Winter School in Theoretical
Physics, \rm Mahabaleshwan, India, 5-19 January 1984
(Springer Verlag, New York, 1984) p.~197; S.D. Joglekar, {\it op. cit.}
p.~1.}
\REF\sussp{J. Ellis, in {\it Supergravity and Superstrings}, Proceedings
of the 28th Scottish Universities Summer School in Physics, Edinburgh,
1985, edited by A.T. Davies and D.G. Sutherland (SUSSP Publications,
Edinburgh, 1986) p.~399; M.T. Grisaru, {\it op. cit.} p.~209.}
\REF\xtata{X. Tata, in {\it The Standard Model and Beyond}, Proceedings
of the 9th Symposium on Theoretical Physics, Mt. Sorak, Korea,
20-25 August, 1990 (World Scientific, Singapore, 1991) p.~304.}
\REF\wess{J. Wess and J. Bagger,
\it Introduction to Supersymmetry \rm
(Princeton University Press, Princeton, NJ, 1991).}
\REF\brazil{P.P. Srivastava,
\it Supersymmetry and Superfields, \rm (Adam-Hilger
Publishing, Bristol, England, 1986).}
\REF\kirsten{%
H.J.W. M\"uller-Kirsten and A. Wiedemann, \it Supersymmetry, \rm
Lecture Notes in Physics, vol.~7 (World Scientific, Singapore, 1987).}
\REF\freund{P.G.O. Freund, {\it Introduction to Supersymmetry}
(Cambridge University Press, Cambridge, England, 1986).}
\REF\pwest{P.C. West, \it Introduction to Supersymmetry and Supergravity
\rm (World Scientific, Singapore, 1990).}
\REF\mohapatra{%
R.N. Mohapatra, \it Unification and Supersymmetry, \rm
(Springer-Verlag, New York, 1991).}
\REF\misra{S.P. Misra, \it Introduction to Supersymmetry and Supergravity
\rm (Wiley Eastern Publishing Company, New Delhi, India, 1992).}
\REF\ferrara{S. Ferrara, editor, \it Supersymmetry \rm (World Scientific,
Singapore, 1987).}
\REF\tasione{M.T. Grisaru, in {\it TASI Lectures in Elementary Particle
Physics 1984} (TASI Publications, Ann Arbor, MI, 1984) p.~232;
D.R.T. Jones, {\it op. cit.} p.~284; G.L. Kane, {\it op. cit.} p.~326;
P.C. West, {\it op. cit.} p.~365.}
\REF\tasithree{S. Raby, in {\it From the Planck Scale to the Weak Scale:
Toward a Theory of the Universe}, Proceedings of the 1986 Theoretical
Advanced Study Institute, University of California, Santa Cruz,
edited by H.E. Haber (World Scientific, Singapore, 1987) p.~63.}
\REF\tasifour{G.G. Ross, in {\it The Santa Fe TASI-87}, Proceedings of
the 1987 Theoretical Advanced Study Institute, Santa Fe, NM,
edited by R. Slansky and G. West (World Scientific, Singapore, 1988)
p.~628; M. Wise, {\it op. cit.} p.~787; L.J. Hall,
{\it op. cit.} p.~812.}
\REF\tasifive{M. Dine, in {\it Particles, Strings and Supernovae},
Proceedings of the 1988 Theoretical Advanced Study Institute, Brown
University, Providence, RI, edited by A. Jevicki and C.-I. Tan
(World Scientific, Singapore, 1989), p.~653.}
\REF\tasiseven{H.-P. Nilles, in {\it Testing the Standard Model},
Proceedings of the 1990 Theoretical Advanced Study Institute,
Boulder, CO, edited by M. Cveti\v c and P. Langacker (World Scientific,
Singapore, 1991) p.~633.}
\REF\tasieight{J. Bagger, in {\it Perspectives in the Standard Model},
Proceedings of the 1991 Theoretical Advanced Study Institute,
Boulder, CO, edited by R.K. Ellis, C.T. Hill, and J.D. Lykken
(World Scientific, Singapore, 1992) p.~31.}
In the late 1970s, Fayet was
the first to seriously pursue supersymmetric field theoretic models of
elementary particles at low energies\refmark\fayet.
However, the subject received
its major boost when it was realized that supersymmetry could
in principle be used to explain the origin of the electroweak
scale\refmark{\wittensusy-\fish}.
Many useful pedagogical review articles\refmark{\oraifrev-\xtata}
and textbooks\refmark{\wess-\misra} on supersymmetry
have appeared over the last two decades.  In addition,
a nice collection of some of the classic papers in supersymmetry and
supergravity can be found in ref.~[\ferrara].  Therefore, I will
not consistently cite the original literature in the first two lectures,
since a comprehensive list of references can already be found in the
previously referenced
reviews and textbooks.  In addition, lecture notes
on supersymmetry have also appeared in a quite a few previous
TASIs\refmark{\tasione-\tasieight}.
The student will find many of them valuable as he or she
begins to delve into this field of study.  In these lectures, there
will necessarily be some repetition from lectures past.  Nevertheless,
I hope I can contribute some new insights to a new generation of students
who may be privileged to witness
the true (experimental) birth of supersymmetry in the coming decade.
\endpage

\chapter{%
How to Build a Low-Energy Supersymmetric Model}

The goal of this lecture is to define the minimal supersymmetric
extension of the standard model, usually called the Minimal
Supersymmetric Standard Model (MSSM).  Unfortunately, there is no one
uniform definition of the MSSM.  The reason for this will become
apparent as we proceed.  In this lecture, I will endeavor to take a
``low-energy'' point of view, and define the MSSM directly
in terms of the low-energy effective field
theory appropriate to energy
scales of order 1 TeV and below.  Low-energy phenomenology will be used
to constrain some of the supersymmetric parameters in Lecture 2.
I will try to
avoid additional assumptions that depend on physics at a much higher
energy scale, although it will sometimes be more practical to make use
of some theoretical biases regarding the nature of the physics at the
higher energy scales.

\section{How to Create a Supersymmetric Extension of the Standard
Model}

\REF\smtext{J.F. Donoghue, E. Golowich and B.R. Holstein, {\it
Dynamics of the Standard Model} (Cambridge University Press,
Cambridge, England, 1992); F. Halzen and A.D. Martin, {\it Quarks and
Leptons} (John Wiley and Sons, New York, 1984).}
I begin with a brief description of the ingredients of the Standard
Model\refmark\smtext.
The Standard Model is an SU(3)$\times$SU(2)$\times$U(1) gauge
theory, consisting of gauge fields and matter fields.  The matter
fields include the quarks, leptons and Higgs fields.  The gauge group
quantum numbers of the particles of the Standard Model are well
known.  Prior to electroweak symmetry breaking,
the gauge bosons consist of the eight gluons (8,1,0),
three SU(2) gauge bosons $W^\pm$ and $W^3$ (1,3,0), and one
hypercharge gauge boson $B$ (1,1,0), where I have indicated
parenthetically the transformation properties of the gauge bosons under
the gauge group.  (Gauge bosons transform under the adjoint
representation of the gauge group.)  In describing the fermionic
sector of the model, it is convenient to use left-handed fermion fields.
One generation (using the notation of the third generation)
consists of a weak doublet of quarks $(t_L,b_L)$
$(3,2,1/3)$, two weak singlets $t_L^c$ ($3^*,1,-4/3$) and $b_L^c$
$(3^*,1,2/3)$, a weak doublet of leptons $(\nu_L,\tau_L)$
$(1,2,-1)$ and a weak singlet $\tau_L^c$ $(1,1,2)$.  There is no
$\nu^c_L$ state in the Standard model (which is equivalent to the
statement that there is no right-handed neutrino).  The other
two generations of fermions are just replications of the one just listed.
Note that The U(1) hypercharge ($Y$)
is normalized such that the electric charge
is given by $Q=T_3+Y/2$, where $T_3$ is the third component of
weak isospin.  Finally, the Standard Model contains one complex
weak doublet of Higgs scalars $(G^+,(v+\phi^0+iG^0)/\sqrt{2})$
$(1,2,1)$.  Note that I have explicitly exhibited
the fact that the neutral component of the Higgs field acquires a
vacuum expectation value, $v$.  As a result, the Goldstone bosons
$G^\pm$ and $G^0$ are absorbed by the vector bosons, producing three
massive vector bosons $W^\pm$ and $Z$.  The $Z$ and photon are linear
combinations of $W^3$ and $B$ (the corresponding rotation angle is the
weak mixing angle $\theta\ls{W}$); the photon remains massless.
The $\phi^0$ is the remaining scalar degree of freedom left in the
particle spectrum; this is the physical Higgs boson.
All the fermion states listed above are interaction eigenstates,
but not necessarily mass eigenstates.
For the quarks, the $3\times 3$ quark mass matrices are not
diagonal, so the quark mass eigenstates differ from the interaction
eigenstates.  These two bases are related
by the Cabibbo-Kobayashi-Maskawa (CKM) matrix.
In the case of the leptons, the neutrinos of the Standard Model are
exactly massless.  Thus, we are free to define the neutrino fields
in such a way that the electroweak eigenstates and mass eigenstates
coincide.  The origin of the
fermion masses and the CKM matrix is not explained within the Standard
Model.  This is the so-called ``flavor problem'', which remains one of
the major mysteries of particle physics today.   I will not attempt to
address the flavor problem in these lectures.  Adding supersymmetry to
the Standard Model in a minimal fashion does nothing to shed light
on this issue.




Here are the steps required to extend the Standard Model to
a supersymmetric theory.  First,
add a supersymmetric partner to each Standard Model particle
such that
$$
 \Str~{\bf 1} = 0 \,,\eqn\strace
$$
where
$$
 \Str\, [{\cdots}] = \sum_i (-1)^{2J_i} (2J_i +
1)\,C_i\,[{\cdots}]\,.\eqn\strdef
$$
In eq.~\strdef, $C_i$ counts the electric charge and
color degrees of freedom and
$J_i$ is the spin of particle $i$ [but, replace $2J_i+1$ with 2
if particle $i$ is massless with spin $J_i>0$].  For example, for the
$W^\pm$ we take $C=2$, while for the $u$-quark $C=6$ corresponds
to a color factor of 3, counting both $u\ls{L}$ and $u^c\ls{L}$.
The latter implies that in the supersymmetric extension of the
Standard Model, corresponding to each quark (and antiquark) one
must add two (color-triplet) scalar partners,
usually called $\widetilde q\ls{L}$ and $\widetilde q\ls{R}$,
respectively.\foot{Technically speaking, $\tilde q^\ast\ls{R}$ is
the superpartner of $q^c\ls{L}$.}
Note that eq.~\strace\ is equivalent to the statement that a
supersymmetric theory has an equal number of bosonic and fermionic
degrees of freedom.

\REF\gaugeanom{See, \eg, T.-P. Cheng and L.-F. Li, {\it Gauge Theory
of Elementary Particle Physics} (Oxford University Press, Oxford,
England, 1984), pp.~347-8.}
In the procedure above, one must guard against gauge
anomalies\refmark\gaugeanom.
In the Standard Model, the gauge anomalies exactly
cancel.  However, the superpartners of the
Higgs bosons are chiral fermions which contribute to the
SU(2)$\times$U(1) gauge anomaly.  In order to cancel this anomaly, one
must add a second Higgs doublet with quantum numbers $(1,2,-1)$ and its
superpartners to the theory.  Let us
check that the conditions for anomaly cancellation,
$$\Tr\,T_3^2 Y=\Tr\,Y^3=0\eqn\trcondition$$
are satisfied.  For example, the contributions of the
quarks and leptons to $\Tr\,Y^3$ are
$$
  (\Tr\,Y^3)_{\rm SM} = 3\left(\smallfrac1{27}
  + \smallfrac1{27} - \smallfrac{64}{27}
      + \smallfrac8{27} \right) - 1-1+8 = 0\,,\eqn\smcancel
$$
where the color factor of 3 has been included for the quark
contributions.  If we now include
just one doublet of (left-handed) Higgsinos,
$(\widetilde H^+,\widetilde H^0)_L$, with
$T_3 = \pm \half$ and $Y = 1$, one finds
$\Tr\,Y^3 = 2$.\foot{Note that
the only other supersymmetric fermions, the gauginos,
have $Y=0$ and so do not contribute to the anomaly or its
cancellation.}
By including Higgs supermultiplets in pairs with opposite hypercharge,
the contributions of the higgsinos to the gauge anomalies will cancel.

\REF\smbl{S. Weinberg, {\sl Phys. Rev. Lett.} {\bf 43} (1979) 1566;
F. Wilczek, {\sl Phys. Rev. Lett.} {\bf 43} (1979) 1571.}
\REF\susybl{S. Weinberg, {\sl Phys. Rev.} {\bf D26} (1982) 287.}
Second, we must include supersymmetric interactions among the particles
and their superpartners.  One
may be tempted to include {\it all} possible supersymmetric
interaction terms.  However, this is
phenomenologically unacceptable.  For example, it is possible to
introduce dimension-4 supersymmetric interaction terms that violate
baryon number or lepton number.
In the Standard Model, this does not happen.  That is,
the SU(3)$\times$SU(2)$\times$U(1) gauge invariance is sufficient to
guarantee that all terms in the Lagrangian of dimension 4 or less
automatically respect baryon and lepton number\refmark\smbl.
This is no longer
true in the MSSM\refmark\susybl.  However, the set of unwanted
interaction terms can be removed by imposing a discrete
symmetry, as we will see in section 1.5.

\REF\nonrenorm{M.T. Grisaru, W. Siegel and M. Rocek, {\sl Nucl. Phys.}
{\bf B159} (1979) 429.}
Finally,
supersymmetry is clearly not an exact symmetry of the observed
particle spectrum and hence must be broken.  However, the
supersymmetry breaking terms of the Lagrangian cannot be totally
arbitrary.  In order to preserve the desirable
properties of supersymmetry (\eg, cancellation of quadratic
divergences and the non-renormalization theorems\refmark\nonrenorm),
the supersymmetry-breaking terms must be soft.  In particular,
if the supersymmetry-breaking terms are soft,
the cancellation
of quadratic divergences in all $n$-point Green functions
(for $n \geq 1$) is maintained.  Soft-supersymmetry breaking arises in
one of three ways:
\item{(i)}
Spontaneous global supersymmetry breaking (which implies the existence
of a massless spin 1/2 Goldstino)
\item{(ii)}
Spontaneous local supersymmetry breaking (which implies the existence
of a massive spin 3/2 gravitino)
\item{(iii)}
Explicit (but soft) supersymmetry breaking.

\noindent The third choice above seems rather arbitrary.  However,
it turns out that explicit soft supersymmetry breaking in the
low-energy effective theory can be a consequence of spontaneous
supersymmetry breaking in the fundamental high-energy theory.  We will
return to this point in section 1.4.

\REF\lowneutrino{B. de Wit and D.Z. Freedman, {\sl Phys. Rev. Lett.}
{\bf 35} (1975) 827.}
The recipe for constructing a supersymmetric extension of the Standard
Model, as described above, implies that each known particle of the
Standard Model must possess a new superpartner that has yet to be
discovered by experiment.  Is it possible that one of the known
particles is already a superpartner of another Standard Model
particle?  The answer is clearly no!  A particle and its superpartner
must differ by one half unit of spin while having the same
transformation properties under the SU(3)$\times$SU(2)$\times$U(1)
gauge group, as well as under any global symmetry of the low-energy
effective theory.  No two Standard Model particles satisfy this
constraint.  For example, although $(\nu,\ell)\ls{L}$ and one of the
Higgs doublets possess the same gauge quantum numbers, the former has
lepton number 1 while the latter has lepton number 0.  If lepton
number is to be a good symmetry of the low-energy theory (as suggested
by experiment), the Higgs doublet cannot be a superpartner of
one of the left-handed lepton doublets.  A second example involves
the only case in the Standard Model where two particles that differ by
a half a unit of spin are degenerate in mass: the neutrino and photon.
Could this be a supersymmetric multiplet?  Again, the answer is no
since the photon (which is a gauge boson) must transform under the
adjoint representation of the gauge group, while the neutrino
transforms as an SU(2) weak doublet.  Finally, in light of the
enumeration above of the possible forms of supersymmetry breaking, it
is interesting to consider the possibility that the neutrino
is the Goldstone fermion of spontaneously broken supersymmetry.  In
fact, it has been shown that this possibility is
phenomenologically untenable\refmark\lowneutrino.
The reason is as follows.
The Goldstone particle couples to matter with derivative
couplings.  This leads to low-energy theorems that imply
that the Goldstone particle decouples in the limit of zero
momentum.  Neutrino interactions with matter do not exhibit
this property.

\section{The Low-Energy Supersymmetric Spectrum}

\TABLE\mssmspectrum{}
Let us use the recipe outlined in section 1.1 to construct the MSSM.
The first step is to add a second complex SU(2)-doublet Higgs field with
hypercharge $Y=-1$ to the Standard Model.  I shall denote the $Y=-1$
[$Y=+1$] Higgs doublet fields by $H_1^i$ [$H_2^i$], where $i$ is a weak
SU(2) index.  Starting with this slightly augmented version of the
Standard Model, we construct the particle spectrum of the MSSM as
discussed in section 1.1.  The end result is displayed in
table \mssmspectrum.

\midinsert
\def\q{\hskip.5em\relax}
\def\qq{\hskip.75em\relax}
\def\qqq{\hskip1.2em\relax}
\centerline{\bf Table \mssmspectrum.\enskip The MSSM Particle Spectrum}
\medskip
\tablewidth=\hsize
\thicksize=0pt
\def\tstrut{\vrule height 2.8ex depth 0.8ex width 0pt}
\begintable
          &             & Fermionic &&& \nrneg{8pt}
Superfield& Boson Fields& Partners&
             SU(3)$_C$& SU(2)$_L$& U(1)$_Y$ \crneg{-5pt}
\multispan2 \tstrut \us{Gauge Multiplets}\hfil&&&& \nr
$\wh G$&  $g$&     $\wt g$&       8&  0&  0 \nr
$\wh V$&$W^a$&  $\widetilde W^a$& 1& 3& 0 \nr
$\wh V\pri$&     B&      $\widetilde B$&   1& 1& 0 \crneg{-5pt}
\multispan2 \tstrut \us{Matter Multiplets}\hfil & & & & \nrneg{-2pt}
$\ds{\wh L\atop \wh E}$&
leptons $\Bigg\{ \ds{\wt L^j\,=\,(\tilde\nu,e^-)_L  \atop
                 \ds{\wt E\,=\,\tilde e^+_R\hphantom{(\nu,_L)}}}$\hf&
  $\ds{ (\nu,e^-)_L \atop  e^c_L}$&
$\ds{1\atop\vph 1}$& $\ds{2\atop\vph 1}$& $\ds{-1\qq\atop\vph 2}$
\nrneg{-8pt}
$\ds{\wh Q\atop \ds{\wh U \atop \wh D} } $&
  quarks $\left\{\vbox to 24pt{}   \right.
 \ds{ \wt Q^j\,=\,(\tilde u_L,\tilde d_L)
  \atop \ds{\wt U\,=\,\tilde u^*_R\hphantom{,d_L)^f}
  \atop \ds{\wt D\,=\,\tilde d^*_R\hphantom{,d_L)^f}} } }$\hfill&
    $\ds{(u,d)_L \atop\ds{u^c_L \atop d^c_L}}$&
    $\ds{3\atop\ds{\q\vph3^*\atop\q\vph3^*}}$&
     $\ds{2\atop\ds{\vph1\atop \vph1}}$&
     $\ds{1/3 \atop\ds{\vph-4/3\qqq \atop\vph2/3}}$      \nrneg{-8pt}
$\ds{\wh H_1\atop \wh H_2}$&
   Higgs\quad$\Bigg\{ \ds{ H^i_1 \atop H^i_2}$\hfill&
    $\ds{(\wt H^0_1,\wt H^-_1)_L \atop (\wt H^+_2,\wt H^0_2)_L}$&
     $\ds{1\atop\vph1}$& $\ds{2\atop \vph2}$& $\ds{-1\qq\atop \vph1}$
\nrneg{12pt} &&&&&\cr
&&&&&\endtable
\vskip-15pt
\endinsert

\REF\nonminref{H.E. Haber and M. Sher, \PRB D35&87&2206&;
J.L. Hewett and T.G. Rizzo, \PRP 183&89&193&.}
Before proceeding, it is perhaps appropriate to briefly
mention the possibility of non-minimal supersymmetric extensions of
the Standard Model.  It is important to be open to alternative models;
after all there is no direct experimental information presently
available that selects the MSSM particle spectrum as the unique
choice for the low-energy supersymmetric model.\foot{There may be
indirect evidence from the fact that the three gauge coupling
constants taken as running couplings satisfying the renormalization
group equations of the MSSM appear to coincide (or unify) at a
unique high energy scale.  I will discuss the possible significance of
this result in section 2.2.}
Possible non-minimal extensions include\refmark\nonminref:

\goodbreak
\item{(i)} An extended Higgs sector

\noindent
The most popular alternative is to add an SU(2)$\times$U(1)
Higgs singlet field N and its supersymmetric partner.
More complicated models can possess more than one such singlet field.
Other possibilities involve multiple Higgs ``generations''.  That
is, replicate the ($H_1,H_2$) pair, perhaps in analogy with the
structure of the three quark/lepton generations.

\item{(ii)} An extended gauge sector

\noindent
Instead of taking the low-energy gauge group to be that of the
Standard Model, consider models based on extended gauge groups.
Among the possibilities are
SU(3)$\times$SU(2)$\times$U(1)$\times$U(1) with one new $Z^\prime$
gauge boson,
SU(3)$\times$SU(2)$_L\times$SU(2)$_R$ $\times$U(1)$_{B-L}$
with a new $Z^\prime$ and $W_R^\pm$ gauge bosons, \etc\
In the early days of superstring model building, many subgroups
of $E_6$ were considered as viable candidates for an extended
low-energy gauge group.  Starting with some particular extended gauge
model (with an appropriate Higgs sector to insure proper low-energy
electroweak
symmetry breaking), one would proceed according to the recipe of
section 1.1 to construct the supersymmetric extension of the model.

\item{(iii)} New fermion multiplets (at the electroweak scale)

\noindent
For example, E$_6$ models lead to a vector-like isodoublet
lepton $(N,E^-) \oplus (\overline N, E^+)$ which transforms as
$(1,2,-1)\oplus (1,2,1)$ under SU(3)$\times$SU(2)$\times$U(1),
a vector-like isosinglet
down-type quark $D\oplus D^c$ which transforms as $(3,1,-2/3)\oplus
(3,1,2/3)$ and two SU(2)$\times$U(1) singlet neutral
leptons which transform as (1,1,0).

Unless otherwise indicated, for the remainder of these lectures I
will assume that the low-energy
supersymmetric particle spectrum corresponds to
the particle content of the MSSM as indicated in
table~\mssmspectrum.

\section{The Supersymmetric Lagrangian}

\REF\bjdrell{J.D. Bjorken and S.D. Drell, {\it Relativistic Quantum
Mechanics} (McGraw-Hill, New York, 1964).}
Having settled on the particle content of the model, the next step is
to construct the supersymmetric interactions.  Here, the superfield
formalism and two-component notation for fermions are invaluable.  An
excellent pedagogical source for the necessary formalism is given in
refs.~[\wess] and [\brazil].\foot{Note
that ref.~[\wess] chooses a convention for
the metric tensor that is opposite the Bjorken and Drell
choice\refmark\bjdrell\ used in these lectures.}
I will not review the superfield formalism here; a serious student of
supersymmetry would be well advised to master this formalism
before pursuing a more detailed study of this subject.
Nevertheless, the discussion below should be
accessible even to those who are not experts in the superfield
technique.  Basically, all one needs to know is that superfields
consist of a collection of fields, each differing by a half a unit of
spin.  In addition, I will for the most part refrain from using
two-component fermions.  Instead,
I will use four-component fermion notation.
In doing so, I will need to make use of fermions of definite
chirality: $\psi_L\equiv P_L\psi$ and $\psi_R\equiv P_R\psi$ with
$$
P_{R,L}\equiv\half(1\pm\gamma_5)\,.
\eqn\prldef$$
The connection between two-component and four-component notation for
fermions is summarized by the following brief guide and translation
table:
$$
\psi = \pmatrix{\xi\cr\bar\eta\cr}\,,\quad \psi^c=C\,
          \overline\psi\lsup{\sss T}
  = \pmatrix{\eta\cr\bar\xi\cr}\,,\quad
    P_L = \pmatrix{1& 0\cr 0& 0\cr},\quad
    P_R = \pmatrix{0& 0\cr  0& 1\cr}\,,\eqn\twocomps
$$
where $C$ is the charge conjugation matrix and $\psi^c$ is a charge-%
conjugated spinor.  If one introduces the four-vector
$\overline{\sigma}^\mu\equiv({\bf 1},-{\mathbf{\vec{\sigma}}})$,
then
the following results translate between two and four component
notation
$$\eqalignalign{%
  &\overline\psi_1 P_L \psi_2 = \eta_1\xi_2
    &\overline\psi\lsup c_1 P_L \psi_2^c = \xi_1\eta_2 \cr
  &\overline\psi_1 P_R \psi_2 = \bar\eta_2\bar\xi_1
    &\overline\psi_1\lsup c P_R \psi_2^c = \bar\xi_2\bar\eta_1 \cr
  & \overline\psi\lsup c_1 P_L \psi_2 = \xi_1\xi_2
    & \overline\psi_1 P_L \psi_2^c = \eta_1\eta_2 \cr
&\overline\psi_1 P_R \psi^c_2 = \bar\xi_1\bar\xi_2
  &\overline\psi\lsup c_1 P_R \psi_2 = \bar\eta_1\bar\eta_2 \cr
&\overline\psi_1 \gamma^\mu P_L\psi_2 = \bar\xi_1\bar\sigma^\mu\xi_2
  &\overline\psi\lsup c_1 \gamma^\mu P_L\psi_2^c =
                          \bar\eta_1\bar\sigma^\mu\eta_2\cr
&\overline\psi\lsup c_1\gamma^\mu P_R\psi^c_2 = -\bar\xi_2\bar\sigma^\mu
                                       \xi_1
  &\overline\psi_1\gamma^\mu P_R\psi_2 = -\bar\eta_2\bar\sigma^\mu
                                           \eta_1 \,.\cr}\eqn\spinors
$$
Additional results of this type can be found in
Appendix A of ref.~[\haberkane].

The fields in a low-energy supersymmetric model live in one of two
types of superfields.  The matter fields live in a complex
chiral multiplets $\wh A \equiv (A,\psi_L,F)$, where the superfield
is indicated by a symbol with a caret over it, while the corresponding
symbol with no caret indicates the scalar component of the supermultiplet.
In this superfield, $A$ and $F$ complex scalars
while $\psi_L$ is a complex Weyl spinor.
$F$ is called an auxiliary field, since it is a field
with no kinetic energy term.  That is, its equations of motion are
purely algebraic; consequently it can be re-expressed in terms of
other dynamical fields.  The component
fields in $\wh A$ may be either massless or massive fields.
The massless gauge fields are components of a
massless real vector superfield
$\wh V^a \equiv (\lambda_L^a, V^{\mu a},D^a)$, where $a$ labels the
components of the adjoint representations of the gauge group.
In this superfield,
$\lambda_L$ is a complex Weyl spinor, $V^\mu$ is a massless gauge
field, and $D$ is a real auxiliary scalar field.\foot{I am implicitly
using the so-called Wess-Zumino gauge in which all other field
components of $\wh V$ are set to zero.  In this gauge, $\wh V$ has the
important property that $\wh V^a \wh V^b \wh V^c=0$ as well as all
higher powers of $\wh V$.  For further details see, \eg,
ref.~[\brazil].}
In supersymmetric theories, the number of bosonic degrees of freedom
must equal the number of fermionic degrees of
freedom [eq.~\strace].  Let us check
this fact in the two cases above.  We can either perform the counting
off-shell or on-shell.  Off-shell, we count the auxiliary fields as
independent degrees of freedom.  Moreover, off-shell, each complex Weyl
fermion counts as 4 degrees of freedom, while each gauge field counts as
3 degrees of freedom.  Since a complex scalar represents 2 degrees of
freedom, it is easy to check that the number of bosonic and fermionic
degrees of freedom match in both the chiral and vector superfields.
On-shell, one applies the equations of motion.  This removes
the auxiliary fields.  In addition, the massless vector and
complex Weyl fermion each now have two degrees of freedom
corresponding to the number of physical polarization states (while the
counting of physical scalar degrees of freedom is not
changed).  Again, it is easy to check that the bosonic and fermionic
degrees of freedom match in both cases.

\REF\fayettwo{P. Fayet, {\sl Nucl. Phys.} {\bf B237} (1984) 367.}
We have yet to consider
how to construct massive gauge fields (and their
supersymmetric partners) in the superfield formalism.  In an ordinary
non-abelian
gauge theory, massive gauge bosons arise via the Higgs mechanism when
a massless gauge field absorbs a massless Goldstone boson.  Similarly,
a massless vector superfield can absorb a chiral superfield whose
lowest scalar component is the Goldstone boson.  The end result is a
massive vector multiplet that contains the massive gauge field and
its corresponding superpartners.  Further details can be found in
ref.~[\fayettwo].  However, in obtaining an explicit form for
the supersymmetric Lagrangian, it
is sufficient to write the supersymmetric interactions in
terms of the chiral and massless vector superfields.  The Higgs
mechanism can then be invoked to generate the massive gauge bosons in
the standard way.

In order to construct supersymmetric interactions, it is necessary to
learn how to construct an action (\ie, the integral of the
Lagrangian over spacetime) that is invariant under supersymmetric
transformations.  This procedure is well described in the pedagogical
literature\refmark{\wess-\kirsten}
and will not be repeated here.  But the basic results are easy to
describe.  If one multiplies two chiral superfields together, the result
is a new chiral superfield.  If one multiplies a chiral superfield by its
conjugate or multiplies a vector superfield by itself, the result is
a (massive) vector superfield.  One can show that the highest
component of a chiral superfield (the ``$F$-term'') and the highest
component of a vector superfield (the ``$D$-term'') transform as full
divergences under a supersymmetry transformation.  Thus, by
multiplying superfields together and extracting their $F$ and $D$
terms, one is guaranteed to end up with a viable term for a
supersymmetric Lagrangian, since the corresponding term in the action
would be invariant under a supersymmetry transformation.
If we restrict the terms appearing in the Lagrangian
to dimension 4 or less (so that the resulting theory is
renormalizible), then it is easy to generate all possible
supersymmetric terms in the Lagrangian after a few superfield
multiplications.

Suppose we are given some chiral supermultiplets, which contain the
matter fields and massless vector supermultiplets which
contain gauge fields corresponding to a
gauge group $G$ with structure constants $f_{abc}$ and gauge coupling
constant $g$.  If $G$ is a non-simple group then
there is a separate massless vector supermultiplet
and an independent coupling constant for each
simple component or U(1) factor in $G$.  The fermionic
partners of the gauge fields (the {\it gauginos}) were denoted above
by $\lambda_L$.
The gaugino is a real Majorana fermion which shares the same gauge
quantum numbers as its partner gauge boson.  Henceforth, I
shall use four-component fermion notation, in which case the
Majorana gaugino will be denoted by $\widetilde V$.

I now enumerate all possible
supersymmetric interaction terms involving matter fields and gauge
fields.

\goodbreak
\medskip
{\sl 1. Self-interaction of the gauge supermultiplet}

In a non-supersymmetric gauge theory, one must define the gauge field
strength tensor $F^a_{\mu\nu}$.  The kinetic energy term for the gauge
fields is simply $-{1\over 4}F^a_{\mu\nu}F^{a\mu\nu}$.  For non-abelian
gauge groups, this term also yields
the self-couplings of the gauge fields ($V^3$ and $V^4$ interactions);
such interactions are of course absent for abelian gauge fields.
The supersymmetric generalization introduces the chiral spinor
superfield $\wh W^\alpha$
which can be defined in terms of $\wh V$ using the
supersymmetric covariant
derivative.  This superfield contains $F^a_{\mu\nu}$ as a term in
one of its
components.  With the overall normalization
of $\wh W^\alpha$ appropriately
chosen, the gauge and gaugino kinetic energy terms can be found in
$$
{\cal L}_{\rm gauge}=[\wh W^\alpha \wh W_\alpha]_F+{\rm h.c.}\,,
\eqn\susygauge
$$
where the subscript $F$ indicates that the $F$-term
of the resulting product should be extracted.  Details of this
formalism can be found in refs.~[\wess, \brazil].
Here, I shall simply summarize
the results of this procedure.  In the case of a supersymmetric
non-abelian gauge theory, eq.~\susygauge\ also contains interactions
among the gauge bosons and gauginos.  The self-couplings of the gauge
fields are precisely those of a general non-abelian gauge theory.
This is to be expected, since the gauge boson self-interactions
follow from the requirements of gauge invariance alone.
In addition, there is an interaction between the gauge field and the
Majorana gaugino
$${\cal L}_{V\wt V\wt V} = \half i g f_{abc}
\overline{\wt V}\lsup a\gamma^\mu \wt V^b V^c_\mu\,,
\eqn\gauginogauge$$
whose form also follows from gauge invariance.
The purely vector interaction displayed in
eq.~\gauginogauge\ is a consequence of the
Majorana nature of the gaugino which implies that
$f_{abc} \overline{\wt V}\lsup a\gamma^\mu\gamma_5
\wt V^b = 0$ (due to the antisymmetry of $f_{abc}$).

If $G$ is not simple, then $G$ must be a direct product of simple
group and U(1) factors; the coupling $g$ then depends implicitly on
the indices $a$, $b$ and $c$ as they run over the various pieces of
$G$.  In particular, for
the U(1) parts of the gauge group, the corresponding $f_{abc}$
vanish.  This implies, for example, that there
is no interaction between the U(1)-gaugino ($\wt V\pri$)
and the U(1) gauge field $(V\pri_\mu)$.

\goodbreak
\medskip
{\sl 2. Interaction of gauge and matter supermultiplets}

In an ordinary gauge theory, the interaction of the gauge and matter
fields follows from the gauge invariance of the theory.  In
particular, these interactions arise from the kinetic energy terms of
the matter fields, in which the ordinary derivatives have been
replaced by gauge covariant derivatives.  The supersymmetric
generalization of this mechanism is most easily exhibited in the
superfield formalism.  The requisite terms arise from
$$
  \left[ \widehat A^+ e^{2g\wh V} \wh A \right]_D
  = \left[ \wh A^+\wh A + 2g \wh A^+\wh V\wh A
         + 2g^2 \wh A^+\wh V^2\wh A \right]_D\,,\eqn\kinenergy
$$
where the subscript $D$ indicates that the $D$-term of the
corresponding product of superfields should be extracted.
In eq.~\kinenergy, I have
defined $\wh V\equiv \wh V^a T^a$, where $T^a$ are the gauge group
generators in the representation of the matter superfield $\wh A$.
Note that I have used the fact that $(\wh V)^n=0$
for $n \geq 3$ which is an important property of the
massless vector superfield in the Wess-Zumino gauge.  The first term
in eq.~\kinenergy,
$[A^+A]_D$, simply yields the kinetic energy terms of $A$ and
$\psi_L$.  The remaining terms contain the desired
supersymmetric interactions
$$\eqalign{%
{\cal L} =\ &-g T^a_{ij} V^a_\mu \left( \overline\psi_i
            \gamma^\mu P_L \psi_j + iA^*_i \lrarrow\partial\mu
            A_j\right)\crr
      &-\sqrt 2 g T^a_{ij} \left( \overline{\wt V}\lsup a P_L
       \psi_jA^*_i + \overline\psi_j P_R \wt V^a \wt A_i
       \right) \crr
      &+g^2(T^aT^b)_{ij} V^a_\mu V^{\mu b} A^*_i A_j \,,
\cr}\eqn\gaugematter
$$
where $P_{R,L}$ are defined in eq.~\prldef, $i$ and $j$
label the components of the matter
fields, and $a$ and $b$ are adjoint representation indices of the
gauge group.

If the gauge group is a direct product of simple group and U(1)
factors, then we must associate a different gauge coupling with each
subgroup
factor.  Consider an example where $G$ is the product of a simple
group and U(1).  Let $V^\prime$
[$\widetilde{V^\prime}]$ be the U(1) gauge
[gaugino] field.  Then eq.~\gaugematter\ should be modified by
replacing $gT^a_{ij}\ptildeV a$ with
$$
  gT^a_{ij} \ptildeV a + \half g\pri y_i \delta_{ij}
     \ptildeV{\ds\pri} \qquad \hbox{(no\ sum\ over~i)}
   \,,\eqn\uonegaugematter
$$
where $g^\prime$ is the U(1) gauge coupling constant and
$y_i$ is the U(1) quantum number normalized in a similar
way to hypercharge in the Standard Model.

\goodbreak
\medskip
{\sl 3. Self-interaction of the matter supermultiplet}

Here, there are contributions from $D$-terms and $F$-terms.
If the gauge group $G$ contains no U(1) factors, then the
$D$-terms arise from two sources.
The kinetic energy term of the gauge multiplet yields $\half D^aD^a$,
and the term proportional to
$[\wh A^+e^{2g\wh V}\wh A]_D$ in eq.~\kinenergy\
yields $gD^a T^a_{ij} A^*_i A_j$.  Thus,
$$
  {\cal L}_D = \half D^aD^a + gD^a T^a_{ij} A^*_i A_j\,.\eqn\lagdterm
$$
The equations of motion for $D^a$ are trivial and yield
$$
  D^a = -g T^a_{ij} A^*_i A_j\,.\eqn\dterm
$$
Substituting back into eq.~\lagdterm\ yields ${\cal L}_D =
-\half D^aD^a$ with $D^a$ given in eq.~\dterm.
This is a contribution to the scalar potential
$$
V_D=\half D^a D^a\,,\eqn\ptdterm
$$
where $D^a$ is a function of the scalar fields as specified in
eq.~\dterm.

\REF\fayetilbr{P. Fayet and J. Iliopoulos, {\sl Phys. Lett.} {\bf 51B}
(1974) 461.}
If the gauge group contains a U(1) factor, then there is one
additional source for the D-terms.  Namely, the term
$[2\xi\wh V^\prime]_D\equiv\xi D^\prime$
is gauge invariant (and therefore provides
an acceptable term for the supersymmetric Lagrangian) if
$\wh V^\prime$ is a U(1) gauge vector superfield.  The constant
$\xi$ is a new parameter of the model, called the Fayet-Iliopoulos
term\refmark\fayetilbr.  If one includes this additional
term in the analysis above, then $D^\prime$ can be evaluated from
its equations of motion.  Instead of eq.~\dterm, one finds
$$
  D\pri = -\half g\pri y_i A^*_i A_i - \xi\,,\eqn\uonedterm
$$
where the notation follows eq.~\uonegaugematter, although in this
case, there is an implicit sum over the repeated index $i$.  If U(1)
is just one of the group factors in $G$, then we must modify
eq.~\ptdterm\ to
$$
  V_D = \half [D^aD^a+(D^\prime)^2]\,,\eqn\potuonedterm
$$

The $F$-term contribution to the self-interaction of the matter
supermultiplet derives from the term
$$
  [W(\wh A_i)]_F + {\rm h.c.}={\partial W\over\partial A_i}F_i +
{\rm h.c.}\,,\eqn\superpot
$$
where $W$ is an arbitrary gauge-invariant
function of the chiral superfields
$\wh A_i$ (defined via its power series expansion) called the {\it
superpotential}.  It is important to note that $W$
is analytic in the chiral superfields (\ie, $W$ depends on the $\wh A_i$
but not on the $\wh A^*_i$).  Finally, $\partial W/\partial A_i$
means: form $W(A_i)$, as a power series in $A_i$, by replacing the
superfield $\wh A_i$ by its lowest scalar component $A_i$, and then
compute the derivative.

Below eq.~\kinenergy, I noted that the matter kinetic energy terms
arise from $[A^\dagger A]_D$.  This term also produces one additional
term, $F_i^* F_i$.  It is traditional to combine this term with that
of eq.~\superpot\ and call the result the $F$-term contribution to the
scalar Lagrangian
$$
  {\cal L}_F = F^*_iF_i + \left( {\partial W\over\partial A_i}\
               F_i + {\rm h.c.} \right)\,.\eqn\lagfterm
$$
The equations of motion for $F^i$ are trivial and yield
$$
  F^*_i = -{\partial W\over \partial A_i}\,,\qquad\qquad
  F_i = -\left({\partial W\over \partial A_i}\right)^{\!\ds *}\,,
  \eqn\fterm
$$
Substitution back into eq.~\lagfterm\ yields ${\cal L}_F =
-F^*_i F_i$ with $F_i$ and $F_i^*$
given in eq.~\fterm.  This is a contribution to
the scalar potential
$$
  V_F = F^*_i F_i\,,\eqn\potfterm
$$
where $F_i$ and $F_i^*$ are functions of the scalar fields as specified
in eq.~\fterm.  This completes the
specification of the complete scalar potential which is a sum of
$D$-term and $F$-term contributions
$$
V  = V_D+V_F\,,\eqn\scalarpot
$$
where $V_D$ and $V_F$ are to be expressed in terms of the scalar
fields as explained above.

Finally, $[W(\wh A_i)]_F$ also contains fermion mass terms and
Yukawa interactions
$$
  {\cal L}_Y = -\half \left[ {\partial^2 W\over \partial A_i
                  \partial A_j}  \overline\psi\lsup c_i P_L \psi_j +
                  \left( {\partial^2 W\over
                      \partial A_i\partial A_j}\right)^{\!\ds *}
                 \overline\psi_j P_R \psi^c_i \right]\,.
\eqn\susyyukawa
$$
The appearance of charged-conjugated fields in the above equation is
somewhat awkward, but is unavoidable in supersymmetric models when
four-component fermion notation is used [see eq.~\spinors\ for
the translation into two-component notation].

So far, we have not specified the superpotential $W$.  However,
if one imposes the requirement that the interactions
obtained above contain no terms higher
than dimension-4 (so that the theory is renormalizible),
then $W$ is at most cubic in its superfields,
$$
W(\wh A_i) = f_i \wh A_i + m_{ij} \wh A_i\wh A_j +
               \lambda_{ijk} \wh A_i\wh A_j \wh A_k\ .\eqn\wcubic
$$
Each term in $W(\wh A_i)$ must be a gauge invariant product of the
$\wh A_i$.  For example, $f_i=0$ if the corresponding $\wh A_i$
is non-singlet under the gauge group.

\section{Supersymmetry Breaking}

In the previous sections, I outlined the steps needed to
construct a supersymmetric extension of an arbitrary gauge field
theory.  However, to construct a realistic model of low-energy physics,
the supersymmetry must be broken.

\REF\oraifbr{L.~O'Raifeartaigh, {\sl Nucl. Phys.} {\bf B96} (1975) 331.}
First, let us consider theories of global ($N=1$) supersymmetry.  The
defining anticommutation relations satisfied by the supersymmetric
generators is
$$
\{Q_\alpha,\overline Q_{\dot\beta}\}=2\sigma^\mu\hs{\alpha\dot\beta}
P_\mu
\eqn\qq
$$
in two-component notation, where $\sigma^\mu\equiv({\bf
1},{\bf{\vec{\sigma}}})$ and $P_\mu$ is the momentum operator.
Since $H=P_0$, it follows that\refmark\wittensusy\
$$
\VEV{0|H|0} = \fourth \VEV{0|\,\overline Q_1 Q_1 + Q_1\overline Q_1
              + Q_2\overline Q_2 + \overline Q_2 Q_2\,|0}\geq 0\,.
\eqn\vevh
$$
Unbroken supersymmetry implies that $Q_\alpha|0\rangle = 0$.
Thus, $\VEV{0|H|0}= 0$ if and only if supersymmetry
is unbroken, while $\VEV{0|H|0}$ is strictly positive if and only
if supersymmetry is broken.  Since $\VEV{0|H|0} = \VEV{0|V|0}$ with
$$
V = F^*_i F_i + \half [D^aD^a + (D\pri)^2]\eqn\vscalar
$$
as deduced in the previous section,
we conclude that supersymmetry is spontaneously broken if either
$$
\VEV{0|F_i|0} \neq 0 \qquad\qquad\qquad \
\hbox{O'Raifeartaigh\ breaking\refmark\oraifbr}
\eqn\oraif
$$
and/or\foot{One can show that Fayet-Iliopoulos $D$-term breaking is
possible only if the model possesses a nonzero Fayet-Iliopoulos
parameter $\xi\neq 0$.}
$$
\VEV{0|D^a|0} \neq 0 \quad {\rm and/or} \quad
\VEV{0|D\pri|0} \neq0 \quad\qquad
\hbox{Fayet-Iliopoulos~breaking\refmark\fayetilbr}\eqn\fayetil
$$
As a consequence of the global spontaneous supersymmetry
breaking, the particle spectrum contains a massless spin-1/2
Goldstone fermion (called a {\it Goldstino}) which is typically
the fermionic partner of the corresponding
$F$ or $D$ term with non-zero vacuum expectation value.

Examples of each of these mechanisms are illustrated in
refs.~[\wess,\brazil].
However, realistic models of broken global
supersymmetry based on either of
these two mechanisms are very difficult to construct.  Let me explain
why this approach to supersymmetry breaking has been abandoned by
serious model builders.
In unbroken supersymmetry, superpartners have degenerate masses.
In models of spontaneously broken supersymmetry,
the mass degeneracy among superpartners is broken.  Consider
the full supersymmetric Lagrangian and compute the scalar field-dependent
squared masses of all fields.  For simplicity, I will at first omit the
possibility of a Fayet-Iliopoulos term (\ie, $\xi=0$).
It is convenient to separately consider
the contributions from vector, spinor and scalar fields of the model.
The contribution from vector fields arises from the last term of
eq.~\gaugematter.  The result is
$$
({\cal M}^2_V)_{ab} = 2g^2(T_aT_b)_{ij} A^*_i A_j\,.\eqn\mvector
$$
The squared mass matrices of the fermions and scalars can
also be obtained by examining the full supersymmetric Lagrangian.
Now, take the trace of the resulting squared mass matrices.
Since $(T^aT^a)_{ij} = C_2(R)\delta_{ij}$
where $C_2(R)$ is the quadratic Casimir operator in representation
$R$ of the scalar fields, it follows that
$$
\Tr\, {\cal M}^2_V = 2g^2 C(R) A^*_iA_i\,.\eqn\trmvector
$$
\REF\fgp{S. Ferrara, L. Girardello, F. Palumbo, {\sl Phys. Rev.}
{\bf D29} (1979) 403.}
Similarly, one finds\refmark\fgp\
$$\eqalign{%
  \Tr\,{\cal M}^2_F &= 2\left|{\partial^2 W\over
                           \partial A_i \partial A_j}\right|^2
                   + 4g^2 C(R) A^*_iA_i\,, \crr
  \Tr\,{\cal M}^2_S &= 4\left|{\partial^2 W\over
                           \partial A_i \partial A_j}\right|^2
                   + 2g^2 C(R) A^*_iA_i\,. \cr}\eqn\trmfs
$$
{}From eqs.~\trmvector\ and \trmfs, it follows that
$$
 \Str\, {\cal M}^2 \equiv 3\,\Tr\, {\cal M}^2_V + \Tr\, {\cal M}^2_S
       - 2\,\Tr\, {\cal M}^2_F = 0\,.\eqn\msumrule
$$
It should be noted that this results holds for arbitrary values of the
scalar fields!  Thus, this mass-squared sum rule holds even if
supersymmetry is spontaneously broken (by vacuum expectation values of
scalar fields).  The mass-squared sum rule is a tree-level result of
the theory.  Later on in this section, I will consider the
implications of radiative corrections to this result.

\REF\tasiesb{H.E. Haber, in {\it Testing the Standard Model},
Proceedings of the 1990 Theoretical Advanced Study Institute,
Boulder, CO, edited by M. Cveti\v c and P. Langacker (World Scientific,
Singapore, 1991) p.~340.}
The mass-squared sum rule is a consequence of the
cancellation of quadratic divergences in supersymmetry.  To see this,
consider the one-loop effective scalar potential.  From eq.~(2.75) of
my TASI-90 lectures\refmark\tasiesb,
$$
  V^{(1)}(\phi) = {\Lambda^2\over 32\pi^2}\, \Str\, M^2_i(\phi)
     + {1\over 64 \pi^2}\, \Str\, \left\{ M^4_i(\phi)
       \left[ \ln {M^2_i(\phi)\over \Lambda^2} -\half\right]
          \right\}\,,\eqn\oneloopeffpot
$$
where $M_i(\phi)$ is the scalar dependent mass matrix and $\Lambda$ is
an ultraviolet cut-off.
Thus, $\Str\, M^2_i(\phi) = 0$ implies an absence of scalar
field-dependent quadratic divergences in the effective potential.
In general, one can show that even
in spontaneously broken supersymmetry, no field-dependent
quadratic divergences are generated in any Green function.
Thus, spontaneous global supersymmetry breaking is
soft-supersymmetry-breaking, as expected.

The mass-squared sum rule is slightly more general than the form given
in eq.~\msumrule.  In particular, because the $F$-term and $D$-term
contributions to the supersymmetric Lagrangian
are independent, eq.~\msumrule\ holds separately for the gauge sector
and matter sector, respectively.   For example,
$$
\left.\Tr\, {\cal M}^2_S - 2\,\Tr\, {\cal M}^2_F \right|_{\rm matter} = 0
\,.\eqn\msusysumrule
$$
Unfortunately, eq.~\msusysumrule\ leads to a phenomenological disaster.
For example, if this result is applied to the lepton supermultiplet:
$(\widetilde\ell_L,\widetilde\ell_R;\ell)$, it would imply that one of
the sleptons must be lighter than the
corresponding lepton partner, which is clearly unacceptable.

Let us consider possible approaches to supersymmetry breaking which
avoid the mass-squared sum rule problem.

\medskip\noindent
{\sl 1. The Fayet-Iliopoulos mechanism.}

The scalar field-dependent
mass-squared sum rule was derived above under the assumption that
there was no Fayet-Iliopoulos term ($\xi=0$).  If $\xi\neq0$ then
eq.~\msumrule\ is modified to\refmark\softsusy\
$$
  \Str\,{\cal M}^2 = -2g_a D^a\, \Tr\,T^a\,.\eqn\uonemsumrule
$$
Note that $\Tr\,T^a \not= 0$ is possible only if $T^a$ is a
U(1) generator.  Also, we noted in a footnote to eq.~\fayetil\
that $D$-term supersymmetry breaking requires a nonzero $\xi$; \ie,
$\VEV D$ must be proportional to $\xi$.

\REF\raisemass{P. Fayet, {\sl Phys. Lett.} {\bf 84B} (1979) 416.}
\REF\farrar{G.R. Farrar and S. Weinberg, {\sl Phys. Rev.} {\bf D27}
(1983) 2732.}
\REF\anomalycancel{L. Alvarez-Gaume and E. Witten, {\sl Nucl. Phys.}
{\bf B234} (1984) 269.}
\REF\fnprs{W. Fischler, H.-P. Nilles, J. Polchinski, S. Raby and
L. Susskind, {\sl Phys. Rev. Lett.} {\bf 47} (1981) 757.}
In the Standard Model, the trace of the hypercharge U(1) generator is
zero (summing over complete multiplets), so the mass-squared sum rule
does not depend on whether one introduces a Fayet-Iliopoulos term
corresponding to the hypercharge.  Thus, the mass-squared sum rule
problem remains.  If one is prepared to add additional U(1) gauge
group factors to the Standard Model (along with their corresponding
Fayet-Iliopolous factors $\xi$), then it may be possible to circumvent
the mass-squared sum rule problem through the generation of a nonzero
term on the right-hand side of eq.~\uonemsumrule.
By making clever choices for the quantum numbers of the usual Standard
Model fermions under the new U(1) factors, one can perhaps raise the
masses all squarks and sleptons\refmark\raisemass.
However, this approach is
problematical for a number of reasons.  First, by adding new U(1)
factors to the low-energy gauge group, one reintroduces the
possibility of new U(1) gauge anomalies.  It is very difficult to
construct a model of this type in which all gauge anomalies are
absent\refmark\farrar.
Even if one succeeds in cancelling off the gauge anomalies,
one cannot cancel off the mixed gauge and gravitational anomalies.  As
shown in ref.~[\anomalycancel],
the cancellation of anomalies of the latter type
requires that all U(1) generators are traceless.  If this latter
condition is
satisfied then the right-hand-side of eq.~\uonemsumrule\  must vanish,
and the mass-squared sum rule problem remains.  Moreover, consider a
supersymmetric model
which contains a non-zero Fayet-Iliopoulos term and a U(1) generator
which is not traceless.  Then, one can prove that such a model is not
free of quadratic divergences\refmark\fnprs.
In particular, a quadratic divergence
arises in the renormalization of $\xi$.  This is apparent in the
effective potential argument given above.  Using
eq.~\oneloopeffpot, it follows
that a quadratic divergence arises if $\Str\,M_i^2\neq 0$.  From
eq.~\uonemsumrule, this is indeed the case when $\Tr\,T^a\neq 0$ for
some group generator.  Since the cancellation of quadratic divergences
is crucial to the naturalness properties of the
supersymmetric theory, I
shall reject theories with low-energy gauge group generators
that are not traceless.  As a result, the mass-squared sum rule
problem is not alleviated.

\medskip\noindent
{\sl 2. Radiative corrections}

\REF\agpw{L. Alvarez-Gaum\'e, M. Claudson and M.B. Wise,
{\sl Nucl. Phys.} {\bf  B207} (1982) 96;
S. Dimopoulos and S. Raby, {\sl Nucl. Phys.} {\bf B219}
(1983) 479.}
\REF\sugramodel{A.H. Chamseddine, R. Arnowitt and P. Nath,
{\sl Phys. Rev. Lett.} {\bf 49} (1982) 970;
R. Barbieri, S. Ferrara and C.A. Savoy, {\sl Phys. Lett.} {\bf 119B}
(1982) 343; H.-P. Nilles, M. Srednicki and D. Wyler, {\sl Phys. Lett.}
{\bf 120B} (1983) 346; {\bf 124B} (1983) 337;
E. Cremmer, P. Fayet and L. Girardello,
{\sl Phys. Lett.} {\bf 122B} (1983) 41; L. Ib\'a\~nez, {\sl Nucl. Phys.}
{\bf B218} (1982) 514; L. Alvarez-Gaum\'e, J. Polchinski and M.B. Wise,
{\sl Nucl. Phys.} {\bf B221} (1983) 495.}
\REF\hlw{L. Hall, J. Lykken and S. Weinberg, {\sl Phys. Rev.} {\bf D27}
(1983) 2359; S.K. Soni and H.A. Weldon, {\sl Phys. Lett.} {\bf 126B}
(1983) 215.}
\REF\newhouse{P. van Nieuwenhuizen, {\sl Phys. Rep.}
{\bf 68} (1981) 189.}
\REF\superhiggs{S. Deser and B. Zumino, {\sl Phys. Rev. Lett.} {\bf 38}
(1977) 1433; E. Cremmer \etal, {\sl Nucl. Phys.} {\bf B147} (1979) 105.}
The mass-squared sum rules obtained above are tree-level results.  One
can avoid their dire phenomenological consequences if one can generate
supersymmetry breaking in the low-energy sector at the
loop level.  Models of this type been constructed,
although they are quite contrived\refmark\agpw.
Moreover, once it was
realized that radiative corrections could have an important impact on the
effective low-energy broken supersymmetric theory, it became apparent
that gravitational radiative corrections (which arise from Planck scale
physics) can also lead to appreciable contributions to the low-energy
theory\refmark{\sugramodel, \hlw}.
Thus, instead of constructing extremely contrived models of broken
global supersymmetry, the attention of theorists turned to models of
broken local supersymmetry.  A local supersymmetric theory is a theory
of supergravity.  This then leads to the modern approach of realistic
models of low-energy supersymmetry which avoid
the tree-level mass-squared sum rule problem.

\medskip\noindent
{\sl 3. Supergravity}

Consider a globally supersymmetric gauge theory with chiral matter
supermultiplets.  If the supersymmetry is now promoted to a local
supersymmetry, then the model contains Einstein's gravity in the
classical limit\refmark\newhouse.
The procedure of coupling matter to supergravity is
not unique.  For example, one can consider canonical kinetic energy
terms for the matter superfields coupled to supergravity; in this
case, one says that the chiral multiplets are minimally coupled to
supergravity.\foot{More complicated non-minimal couplings can be
constructed.  Since supergravity models, treated as field theories
are inherently non-renormalizible, one cannot use renormalizibility as
a condition to restrict the possible forms of the couplings.}
In supergravity models, the spin-2 graviton is a member of a
supermultiplet along with a spin-3/2 gravitino.  In a model of
unbroken supergravity, both the graviton and gravitino are massless.
If the supergravity is spontaneously broken, the gravitino gains mass
via the super-Higgs mechanism---the Goldstone fermion of broken
supersymmetry is absorbed (``eaten'') by the massless
gravitino\refmark\superhiggs.  The
degrees of freedom are conserved: the Goldstone fermion and massless
gravitino each have two degrees of freedoms, while the massive
gravitino possesses four degrees of freedom.

\REF\cremmer{E. Cremmer, S. Ferrara, L. Girardello and A. van Proeyen,
{\sl Nucl. Phys.} {\bf B212} (1983) 413.}
Cremmer \etal\refmark\cremmer\
derived the appropriate sum rule for
spontaneously broken supergravity.  Given $N$ chiral
multiplets minimally coupled to supergravity,
eq.~\uonemsumrule\ is modified to
$$
  \Str\, {\cal M}^2 = (N-1)\,(2m^2_{3/2} - \kappa^2D^aD^a)
                    - 2g_aD_a\,\Tr\, T^a\,,\eqn\sugrauonemsumrule
$$
where $m_{3/2}$ is the gravitino mass and $\kappa\equiv (8\pi G_N)^{1/2}
= (8\pi)^{1/2} /M_P$ (where $G_N$ is Newton's constant).
For the reasons explained above, I will not make use
of the Fayet-Iliopoulos mechanism.  In this case, I can take
$\VEV D = 0$ so that the particle masses then satisfy
$$
  \Str\,{\cal M}^2 = 2(N-1) m^2_{3/2}\,.\eqn\sugramsumrule
$$
If $m_{3/2} \simeq {\cal O}(100$~GeV) or larger, then
supersymmetric particle masses will be raised above their
Standard Model partner masses as required by
the present experimental limits on supersymmetry.
Thus, the mass-squared sum rule problem of global
supersymmetry is no longer a problem in supergravity models if the
gravitino mass is sufficiently heavy.

To be complete, it should be noted that the mass-squared sum rules
just quoted are tree-level results.  In ``no-scale'' supergravity
models, $m_{3/2}$ is undetermined at tree-level\refmark\nanop.
This means that in
effect, the gravitino remains massless at tree-level and only acquires
mass via radiative corrections.  In this case, the right-hand side of
eq.~\sugramsumrule\ is once again zero (at tree-level), although in
this case, gravitational radiative corrections are important and can
be used to make the superparticle spectrum phenomenologically viable.
Given that supergravity theories are not renormalizible, it
is not clear how to carry out this procedure in practice.

In light of the above results, it appears that supergravity presents
the most compelling framework for constructing a viable model of
broken supersymmetry.  But, we are interested in low-energy
supersymmetry, where superparticle masses lie in the range of roughly
100 GeV to 1 TeV.  Thus, we must arrange to have $m_{3/2}$ (or the
effective supersymmetry-breaking
scale in the low-energy theory) to lie within
this mass range.  This is a very strong constraint on the supergravity
theory.  For example, one might have guessed that the most natural
supersymmetry-breaking scale in a theory of broken supergravity
is the Planck mass.  However, in this case, the resulting
theory would fail to solve the hierarchy and naturalness
problems of the Standard Model.

\REF\gauginocondensate{%
H.-P. Nilles, {\sl Int. J. Mod. Phys.} {\bf A5} (1990) 4199.}
\REF\cosmoconstant{S. Weinberg, {\sl Rev. Mod. Phys.}
{\bf 61} (1989) 1.}
\REF\stringsusy{L.E. Ib\'a\~nez and D. Lust, {\sl Nucl. Phys.} {\bf B382}
(1992) 305; V.S. Kaplunovsky and J. Louis, CERN-TH-6809-93 (1993).}
The origin of supersymmetry breaking is perhaps the
most pressing theoretical problem in fundamental theories of
supersymmetry.  To discuss these issues in detail would
require another course of lectures.  Instead, I will briefly outline
the most common scenario for producing low-energy supersymmetry from a
more fundamental broken supergravity model.  This scenario has been
called the {\it hidden sector}$\,$ scenario\refmark\hlw.
In this scenario, one posits
two sectors of fields.  One sector (called the ``visible'' sector)
contains all the fields of the
Standard Model (and perhaps additional heavy
fields in a grand unified model
of the strong and electroweak forces).  A second ``hidden'' sector
contains fields which lead to the breaking of supersymmetry at some
large scale $\Lambda_{\rm SUSY}$.  One assumes that none of the fields
in the hidden sector carry quantum numbers of the visible sector.
Thus, the two sectors are nearly decoupled; they
communicate only by weak gravitational interactions.  Thus, the
visible sector only finds out about supersymmetry breaking through its
very weak gravitational
coupling to the hidden sector.  In the visible sector, the
effective scale of supersymmetry breaking (denoted by $\msusy$)
is therefore much smaller than $\Lambda_{\rm SUSY}$.  A typical result
is
$$
\msusy\simeq{\Lambda_{\rm SUSY}^n \over M_P^{n-1}}\,,\eqn\msusybreak
$$
depending on the mechanism for supersymmetry breaking in the hidden
sector.  Two popular models for the breaking mechanism is the Polonyi
model (where $n=2$) based on $F$-type breaking in the hidden sector
(\ie, some $F$-term acquires a vacuum expectation value), and gaugino
condensate models\refmark\gauginocondensate\
(where $n=3$).  In both cases, $\Lambda_{\rm SUSY}$
can be quite large, above $10^{10}$~GeV, while still producing
$\msusy$ of order 1 TeV or less.  Indeed, in these scenarios,
supersymmetry has the potential for solving the hierarchy and
naturalness problems described in the Introduction.  Unfortunately,
there is one unsolved fine-tuning problem (not previously mentioned)
which remains, called the cosmological constant
problem\refmark\cosmoconstant.  Once we take
gravity into account in our particle theories, the vacuum energy
density ($\Lambda_0$)
is a physical quality which in principle is calculable in a
fundamental theory of gravity.  Theoretically, the vacuum energy
density would naively expected to be of order $\mpl^4$, but this is
not our universe.  (A universe with such a large vacuum energy density
would have a lifetime of order the Planck time, $\hbar/M_P c^2\simeq
10^{-43}$~sec!)  Thus, based on the fact that the universe has
endured over 10 billion years (and looks very flat at large scales),
$\Lambda_0/\mpl^4<10^{-121}$.  This is the mother of all fine-tuning
and naturalness problems!  The extent of the fine-tuning problem is
slightly alleviated in broken supersymmetric models.  But at best,
$\mpl^4$ is replaced by $\msusy^4$.  In models of spontaneously
supergravity breaking, there is some freedom that allows a
fine-tuning of parameters to set the cosmological constant to zero.
All model builders must do this in order to have a
theoretically consistent framework.  Whether the cosmological
constant problem must be solved before a theoretically viable model of
supersymmetry breaking can be established is not known.
However, even if the two problems are logically distinct,
a compelling and theoretically
consistent model of supersymmetry breaking has yet to be
obtained.\foot{Perhaps this goal is unattainable without a better
understanding of superstring theory and how the ``low-energy''
effective supergravity theory emerges at the Planck
scale\refmark\stringsusy.}
Thus, for the moment, we shall be content with the overall
scenario described above, leaving the details to future theoretical work.

{}From now on, let us assume that the hidden sector scenario has been
successfully implemented.  Then, if one considers the resulting
spontaneously broken supergravity theory and takes
$M_P \to \infty$, then at low-energies of order $\msusy$, the
theory has the following structure.  The effective low-energy theory
at $\msusy$ is a supersymmetric gauge theory coupled to chiral
supermultiplets, with explicit soft-supersymmetry-breaking terms and
a massive gravitino of order $\msusy$.\foot{In models where the
gravitino mass is generated by radiative corrections, it is
possible to end up with a gravitino whose mass is not connected to
$\msusy$.  In this case, the factors of $m_{3/2}$ that appear in
eq.~(1.41) would be replaced by some other parameter of order
$\msusy$.  Since the gravitino will play almost no role in these
lectures, I will not pursue these alternative possibilities any
further.}
The explicit supersymmetry breaking terms are of a very
particular type\refmark\hlw:
\item{(i)} Majorana mass terms for gauginos
$$
  -{\cal L}_{\rm mass} = \half \left[ M_3\, \overline{\widetilde g}
   \,\widetilde g + M_2 \overline{\widetilde W} \lsup a
                                   {\widetilde W^a}
        + M_1 \overline{\widetilde B} \widetilde B \right]\,.
\eqn\gauginomass
$$
\item{(ii)} The scalar potential is modified by terms proportional to
the gravitino mass
$$
  V = V_F + V_D + m_{3/2}\left[h(A_i)+{\rm h.c.}\right]
       + m^2_{3/2}\,S_{ij} A^*_iA_j\,.\eqn\softpot
$$

\noindent
In eq.~\softpot, $h$ is a cubic polynomial analytic in
the scalar fields $A_i$, and $S_{ij}$ are model-dependent constants.
In the hidden sector scenario of supergravity breaking with minimally
coupled chiral supermultiplets, $S_{ij}=\delta_{ij}$.  In addition,
$h(A_i)$ takes the following form
$$
  h(A_i) = \widetilde A\, W_3(A_i) + \widetilde B\, W_2 (A_i)\,,
\eqn\softsuperpotential
$$
where $W_3$ and $W_2$ are
the cubic terms and the quadratic terms of the superpotential,
respectively, and  $\widetilde A$ and $\widetilde B$ are
model-dependent parameters.  For example, in the simplest Polonyi
models, $\widetilde B = \widetilde A-1$, although this relation
does not hold in more general models.

The supergravity induced soft terms described above should be
regarded as terms of an effective Lagrangian evaluated near
the Planck scale.  The renormalization group is then used to evolve
these terms down to the electroweak scale.  In conclusion,
explicit supersymmetry breaking (SSB) in the low-energy effective
broken supersymmetric model at the energy scale $\msusy$ is
parametrized by:
Majorana mass terms for the gauginos given in eq.~\gauginomass\ and
the following addition to the scalar potential
$$
   V_{\rm SSB} = M^2_i A^*_i A_i + \sum_n A_n\, W_{3n} (A_i)
             + \sum_n B_n\, W_{2n} (A_i)\,,\eqn\deltavee
$$
where each separate term in $W_3 \equiv \sum\limits_{n}W_{3n}$
and $W_2 \equiv \sum\limits W_{2n}$ is weighted by a different
$A$ and $B$ parameter.  Note that I have absorbed the factor of
$m_{3/2}$ into the definitions of $A_n$ and $B_n$.

Finally, we note that all the explicit supersymmetry breaking terms
obtained above are {\it soft}.  To verify this, one must demonstrate
that no new quadratic divergences are generated by the explicitly
broken supersymmetric theory.  To this end, let
us check the one-loop effective potential by computing
$\Str\, M^2_i(\phi)$.  Indeed, an explicit calculation shows that
$\Str\, M^2_i(\phi)$ is non-zero and
proportional to $m^2_{3/2}$; however, it is independent of
$\phi$.  Constant (field-independent) terms in the effective potential
can be dropped, since they only affect the cosmological constant
term which must be fine-tuned to zero anyway.  Thus, the analysis of
the one-loop effective potential is consistent with the claim that no
new (field-dependent) quadratic divergences are generated.  Thus,
spontaneously broken supergravity yields soft-%
supersymmetry breaking in the effective low energy theory.

\REF\hallrandall{L.J. Hall and L. Randall, {\sl Phys. Rev. Lett.}
{\bf 65} (1990) 2939.}
One might be tempted to conclude that all explicit supersymmetry-breaking
terms of dimension 3 or less are soft.  This is not correct, as was
shown by Girardello and Grisaru\refmark\softsusy.  Two counter-examples
are: (i) explicit mass terms for matter fermions, and (ii) cubic
scalar interactions that mix $A_i$ and $A^*_i$.  Both these terms
generate terms in $\Str\,M^2_i(\phi)$ that are linear in the scalar fields
(\ie, tadpoles)
that are neutral under the gauge group.  Thus, the effective potential
analysis demonstrates that these terms can generate new nontrivial
quadratic divergences.  (Ref.~\softsusy\ also makes use of a clever
technique involving a spurion superfield to simulate supersymmetry
breaking and reaches the same conclusions.)  However, in
the absence of completely gauge neutral chiral multiplets,
gauge invariance forbids tadpoles, in which case it appears
that even the two cases just cited do not generate new
nontrivial quadratic divergences\refmark\hallrandall.
Nevertheless, such terms do not seem to emerge from any
theory of spontaneous supergravity breaking.

\section{The Definition of the MSSM}

After this rather long journey into the theory of supersymmetry and
its breaking, we are ready to formally define the MSSM.  The particle
content has already been displayed in section 1.2.
Thus, we must address the question of supersymmetric
interactions and soft-supersymmetry-breaking terms.
I shall proceed as much as possible from the
low-energy point of view.  That is, I will accept a
general form for the soft-supersymmetry-breaking without inquiring
into its origins.

\REF\hhg{J.F. Gunion, H.E. Haber, G. Kane and S. Dawson,
{\it The Higgs Hunter's Guide} (Addison-Wesley Publishing Company,
Reading, MA, 1990) [Erratum: SCIPP-92/58 (1992)].}
The supersymmetric interactions are known once we
specify the superpotential.
The most general gauge-invariant superpotential has the following
form:
$$
  W = W_R + W_{NR}\,.\eqn\generalw
$$
First, $W_R$ is given by
$$
  W_R = \epsilon_{ij} \left[ h_\tau \widehat H^i_1 \widehat L^j
         \widehat E + h_b\widehat H^i_1 \widehat Q^j\widehat D
         - h_t \widehat H^i_2 \widehat Q^j\widehat U
         - \mu \widehat H^i_1 \widehat H^j_2 \right]\,.\eqn\wrparity
$$
Here we use $\epsilon_{ij}$ to combine two SU(2)$_L$ doublets
[where $\epsilon_{ij}=-\epsilon_{ji}$ with $\epsilon_{12}=1$].
The parameters introduced above are the Yukawa coupling matrices
$h_\tau$, $h_b$ and $h_t$ (these are $3\times 3$ matrices
in generation space)
and the Higgs superfield mass parameter, $\mu$.
Generation labels have been suppressed.  If we were to ignore the
first two light generations of quarks and leptons, then
$h_\tau, h_b$ and $h_t$ would be proportional to the corresponding
fermion masses\foot{The signs in eq.~\wrparity\ have been chosen so that
$h_\tau$, $h_b$, and $h_t$ (in the one generation case) are positive.
The sign of $\mu$ is conventional.}
$$
  h_\tau = {gm_\tau\over \sqrt 2 m_W \cos\beta},\qquad\quad
  h_b = {gm_b\over \sqrt 2 m_W \cos\beta},\qquad\quad
  h_t = {g m_t\over \sqrt 2 m_W \sin\beta}\,,\eqn\yukawas
$$
where $\tan\beta \equiv \VEV{H^0_2}/\VEV{H^0_1}$ is the ratio of
neutral Higgs vacuum expectation values.  The generation structure of
the Yukawa coupling matrices corresponds to that of the
two-Higgs-doublet extension of the Standard Model\refmark\hhg.
The MSSM sheds no light on the origin of the flavor dependence
of the model.

Second, $W_{NR}$ is given by
$$
  W_{NR} = \epsilon_{ij} \left[ \lambda_L \wh L^i\wh L^j \wh E +
           \lambda\pri_L \wh L^i \wh Q^j\wh D -
           \mu\pri \wh L^i\wh H^j_2 \right] + \lambda_B \wh U\wh D\wh D
\,,\eqn\norparity
$$
where generation labels are again suppressed.  For example, the
flavor indices on $\lambda_L$ must be such that it is
antisymmetric upon interchange of $\wh L^i$ and $\wh L^j$.
One quickly observes that the terms in $W_{NR}$ violate either
baryon number (B) or lepton number (L).  Specifically,
$$ \eqalignalign{%
  \wh L\wh L\wh E&,\ \wh L\wh Q\wh D,\ \wh L\wh H\qquad\qquad
        &\Delta L\neq 0 \,,\cr
   &\wh U\wh D\wh D   \qquad\qquad     &\Delta B\neq 0 \,.\cr}
\eqn\blviolation
$$
In the MSSM, one sets $W_{NR} = 0$.
In this regard, the MSSM is not as elegant as the Standard Model.
Recall that if one imposes SU(3) $\times$ SU(2) $\times$ U(1) on all
possible Standard Model
interactions, one finds that all terms of dimension 4 or less
automatically preserve $B$ and $L$.  Clearly, this is no longer the
case in the supersymmetric extension of the Standard Model.
How does one impose $W_{NR} = 0$ in the MSSM?
Here, we introduce a discrete symmetry to do our dirty work.
There are two equivalent descriptions:
\REF\matterparity{S. Dimopoulos, S. Raby and F. Wilczek,
{\sl Phys. Lett.} {\bf 112B} (1982) 133.}
\REF\weirdrparity{L.J. Hall, {\sl Mod. Phys. Lett.} {\bf A5} (1990) 467.}
\REF\hallrandalltwo{%
L.J. Hall and L. Randall, {\sl Nucl. Phys.} {\bf B352} (1991) 289.}
\REF\massivegluinos{P. Fayet, {\sl Phys. Lett.} {\bf 78B} (1978) 417.}
\REF\mnmssm{J. Ellis, J.F. Gunion, H.E. Haber, L. Roszkowski, and
F. Zwirner, {\sl Phys. Rev.} {\bf D39} (1989) 844.}
\item{(i)}
Matter parity\refmark\matterparity\

\noindent
The MSSM does not distinguish between Higgs and quark/lepton superfields.
We define a discrete matter parity under which all quark/lepton
superfields are odd while the Higgs superfields are even.

\goodbreak
\item {(ii)} R-parity\refmark\fayet\

\noindent
In superspace, a chiral superfield is
$$
  \wh A = A(x) + \sqrt 2 \theta\psi(x) + \theta\theta
F(x)\,.\eqn\phihat
$$
Under a continuous U(1)$_R$ symmetry, $\theta\to e^{i\alpha}\theta$
and $\wh A \to e^{in\alpha}\wh A$.  In this case, the superfield
$\wh A$ has $R=n$.  This means that the component fields have
$R(A) = n$, $R(\psi)=n-1$, and
$R(F)=n-2$.  The superpotential $W$ must have $R(W)=2$ in order
that the theory conserve U(1)$_R$, since $[W]_F\,+$ h.c.
appearing in the supersymmetric Lagrangian must be neutral under
$R$-parity.  Thus, in order to set $W_{NR}=0$, one may
choose
$$\eqalignalign{%
  R&= 1 \qquad\quad  &\hbox{for}\; \wh H_1,\wh H_2\,, \cr
  R&= \half\qquad &\hbox{for}\; \wh L, \wh E,\wh Q,\wh U,\wh D\,. \cr }
\eqn\rparitychoice
$$
In fact, the imposition of the full continuous
U(1)$_R$ symmetry is too restrictive.  Consider the gauge supermultiplet,
$\wh V$.  Since $\wh V$ is real, we must have $R(\wh V) = 0$ which means
that $R(V_\mu) = R(D) = 0$ and $R(\lambda) = 1$.  That is, the U(1)$_R$
symmetry forbids a nonzero Majorana mass for the gaugino.
In the MSSM, Majorana mass terms for the gaugino are generated by
the soft-supersymmetry-breaking terms.  Since these terms are quadratic
in $\lambda$, the Majorana gaugino mass terms break the continuous
U(1)$_R$ symmetry down to a discrete
$Z_2$ symmetry called $R$-parity.  It is easy to check that
the $R$-parity quantum number is given by
$$
  R = (-1)^{3(B-L) + 2S}\eqn\ztwo
$$
for particles of spin $S$.  By imposing $R$-parity invariance,
on the low-energy supersymmetric model, $W_{NR} = 0$.\foot{Interesting
alternative supersymmetric models exist in which the $R$-parity
symmetry described above is modified.  Among such models are
$R$-parity-violating models, and models that promote the $Z_2$ $R$-parity
of the MSSM to a larger discrete symmetry group\refmark\weirdrparity\
or even to the full
continuous U(1)$_R$ symmetry\refmark\hallrandalltwo.
In the latter case, one must introduce
new color octet fermions to mix with the gluinos\refmark\massivegluinos,
in which case
U(1)$_R$-symmetric massive color-octet Majorana fermions are permitted.
Such alternatives lie beyond the scope of these lectures.}




Before introducing the soft-supersymmetry-breaking terms, the model
we have just defined is a supersymmetric
SU(3)$\times$SU(2)$\times$U(1) gauge theory with three generations of
quarks and leptons, two complex Higgs doublets and their
supersymmetric partners.  It is interesting to note that the model as
it stands now does not spontaneously break the SU(2)$\times$U(1)
electroweak gauge group.  This is easily demonstrated by examining the
Higgs boson contributions to the scalar potential [see
eqs.~\potuonedterm\ and \potfterm].  Explicitly,
$$
  V_{\rm Higgs} = |\mu|^2\left(|H_1|^2+|H_2|^2\right) + \eighth(g^2+g'^2)
      \left( |H_1|^2-|H_2|^2\right)^2 + \half g^2
        |H^*_1H_2|^2\,,\eqn\susyhiggspot
$$
which implies that $V_{\rm Higgs} \geq 0$.
Thus the minimum of the potential is $H_1=H_2=0$ and we conclude that
the electroweak symmetry group is unbroken.  One should not be alarmed
by this result, since we have not yet added the soft-supersymmetry-%
breaking terms.  However, it may be theoretically  useful to have a
supersymmetric SU(2)$_L\times$U(1)$_Y$ model where the gauge symmetry
breaks down to U(1)$_{\rm EM}$ while maintaining the supersymmetry
intact.  Although the MSSM does not allow this (as shown
above), one can construct an exact supersymmetric model
with tree-level SU(2)$_L\times$U(1)$_Y$ breaking by adding
a chiral Higgs superfield, $\wh N$, which is neutral under the gauge
group.  This allows additional terms in the superpotential of the
form\foot{Note that I have omitted
a term in $\delta W$ proportional to
$\wh N^2$, since such a term can be removed by a linear shift in the
definition of the $\wh N$ superfield.}
$$
  \delta W = \lambda \epsilon_{ij} \wh H^i_1 \wh H^j_2 \wh N
               - r\wh N - \third k\wh N^3\,.\eqn\singletpot
$$
Regions of $(\lambda,r,k$) parameter space exist where the new
Higgs potential has its minimum away from $H_1=H_2=0$, thereby
breaking SU(2)$_L\times$ U(1)$_Y$ down to U(1)$_{\rm EM}$.  This
model, along with the attendant soft-supersymmetry-breaking terms has
been called the minimal non-minimal
supersymmetric model (MNMSSM)\refmark\mnmssm.
I will not consider this model further in these lectures.

In order to complete the description of the MSSM, one must examine the
possible soft-supersymmetry breaking terms.  Following the rules
described in the previous section, I enumerate all possible soft terms
below that respect $R$-parity:
$$\eqalign{%
  V_{\rm soft} =\ &m^2_1 |H_1|^2 + m^2_2|H_2|^2 - m^2_{12}
                 (\epsilon_{ij} H^i_1H^j_2 + \hbox{h.c.}) \crr
 &+M^2_{\widetilde Q}\,\left[\tilde t^*_L\tilde t_L
 + \tilde b^*_L\tilde b_L\right] +
       M^2_{\widetilde U}\,\tilde t^*_R\tilde t_R
       + M^2_{\widetilde D}\,\tilde b^*_R\tilde b_R \crr
 &+M^2_{\widetilde L}\,\left[\tilde\nu^*\tilde\nu
 +\tilde \tau^*_L\tilde \tau_L\right]
      + M^2_{\widetilde E}\,\tilde \tau^*_R\tilde \tau_R \crr
 &+{g\over \sqrt 2 m\ls W}\, \epsilon_{ij}\, \left[
     {m_\tau A_\tau\over \cos\beta} H^i_1\, \tilde\ell^j_L\tau^*_R
     + {m_bA_b\over\cos\beta} H^i_1\, \tilde q^j_L \tilde b^*_R
      - {m_tA_t\over \sin\beta} H^i_2\,\tilde q^jt^*_R \right] \crr
 &+\half \left[ M_3\, \overline{\widetilde g}
   \,\widetilde g + M_2 \overline{\widetilde W} \lsup a
                                   {\widetilde W^a}
  + M_1 \overline{\widetilde B} \widetilde B \right]\,,\cr}\eqn\vsoftmssm
$$
where $\tilde\ell_L \equiv \left({\tilde \nu\atop \tilde e_L} \right)$
and $\tilde q_L \equiv \left({\tilde t_L\atop \tilde b_L}\right)$.
As before, the generational labels are suppressed and third
generation notation is used.\foot{More precisely, the
soft-supersymmetry breaking parameters of the squark and slepton
sectors (as well as the corresponding fermion masses) in
eq.~\vsoftmssm\ are 3$\times$3 matrices.}
If one strictly follows
the rules of the previous section, then $m_{12}^2\equiv B\mu$ where
$\mu$ is the Higgs superfield mass parameter.

We have now assembled all the pieces of the MSSM.  To proceed, we
next examine the mass eigenstates of the theory.  This is not a
completely trivial analysis, since any set of particles of
a given spin, $B$, $L$ and SU(3)$_C\ \times$ U(1)$_{EM}$
quantum numbers can mix.
Thus, one must diagonalize mass matrices to obtain the mass
eigenstates and the corresponding mass eigenvalues.  Let us examine
the various sectors of the model.

\medskip
{\sl 1.~~Squarks}

\noindent
In principle, I must diagonalize $6\times 6$ matrices corresponding
to the basis $(\widetilde q_{iL}, \widetilde q_{iR})$
where $i=1,2,3$ are the generation
labels.  The one generation case (using the notation of the third
family) is given below
$$\eqalign{ \def\crrr{\noalign{\vskip12pt}}
M^2_{\tilde t} &= \left[ \matrix{%
    M^2_{\widetilde Q}+ m^2_t+ m^2_Z(\half-e_u s^2_W)\cos2\beta
      & m_t(A_t-\mu\cot\beta) \cr
    m_t(A_t-\mu\cot\beta)
      &M^2_{\widetilde U}+ m^2_t+ m^2_Z e_u s^2_W\cos2\beta \cr}
      \right]  \crrr
M^2_{\tilde b} &= \left[ \matrix{%
    M^2_{\widetilde Q}+ m^2_b- m^2_Z(\half+e_d s^2_W)\cos2\beta
      & m_b(A_b-\mu\tan\beta) \cr
    m_b(A_b-\mu\tan\beta)
      &M^2_{\widetilde D}+ m^2_b+ m^2_Z e_d s^2_W\cos2\beta \cr }
      \right]\,,\cr
}\eqn\squarkmatrix$$
where $e_u=2/3$, $e_d=-1/3$, $s^2_W\equiv\sin^2\theta_W$ and
$\tan\beta \equiv \VEV{H^0_2} \big/ \VEV{H^0_1}$.

\medskip
{\sl 2.~~Sleptons}

\noindent
$$\eqalign{%
  M^2_{\tilde\nu} &= M^2_{\widetilde L} + \half m^2_Z\cos2\beta \crrr
  M^2_{\tilde \tau}  &= \left[ \matrix{%
    M^2_{\widetilde L}+ m^2_\tau- m^2_Z(\half- s^2_W)\cos2\beta
      & m_\tau(A_\tau-\mu\tan\beta) \cr
    m_\tau(A_\tau-\mu\tan\beta)
      &M^2_{\widetilde E}+ m^2_\tau- m^2_Z s^2_W\cos2\beta \cr}
      \right]  \cr
}\eqn\sleptonmatrix$$
The above results indicate that  $\tilde f_L$--$\tilde f_R$ mixing
is unimportant for $m_f \ll M_{\rm SUSY}$ where $M_{\rm SUSY}$
characterizes the scale of the supersymmetry-breaking terms.  This
suggests that only $\tilde t_L$--$\tilde t_R$ mixing is likely to be
phenomenologically relevant.  (If $\tan\beta \gg 1$,
then $\tilde b_L$--$\tilde b_R$ mixing may be non-negligible.)

\endpage
\medskip
\REF\ghia{See, \eg,
Appendix A of ref.~\ghi\ and Appendix C of ref.~[\haberkane].}
\REF\ghi{J.F. Gunion and H.E. Haber, {\sl Nucl. Phys.} {\bf B272} (1986)
1 [Erratum: SCIPP-92/59 (1992)].}
{\sl 3.~~Charginos}

\noindent
The $\wt W^\pm$ and $\wt H^\pm$ can mix; the mass eigenstates are
called {\it charginos}.  In the $\wt W^+$--$\wt H^+$ basis, the
chargino mass matrix is\refmark\ghia\
$$
  X = \pmatrix{ M_2 &\sqrt 2 m\ls W \sin\beta \cr
       \sqrt 2 m\ls W \cos\beta &\mu \cr}\,.\eqn\charginomatrix
$$
In general, two unitary $2\times 2$ matrices $U$ and $V$ are required
to diagonalize the chargino mass-squared matrix
$$
{\cal M}_{\wt\chi^+}^2=VX^\dagger XV^{-1}=U^\ast XX^\dagger
(U^\ast)^{-1}\,.\eqn\chiplusdiag
$$
The two mass eigenstates are denoted by $\widetilde\chi^+_1$
and $\widetilde\chi^+_2$ with corresponding squared masses
$$\eqalign{%
  M^2_{\wt\chi^+_{1,2}} = \half \Big\{
  M^2+\mu^2 + 2m^2_W \mp
    \big [ &(M^2-\mu^2)^2 + 4m^4_W\cos^2
    2\beta                 \cr
 &+4m^2_W (M^2+\mu^2 + 2M\mu \sin2\beta) \left.
    \big]^{1/2} \right\} \,.\cr}
\eqn\physrepeq
$$
By convention, $M_{\wt\chi^+_1}\leq M_{\wt\chi^+_2}$.  The mixing matrix
elements $U_{ij}$ and $V_{ij}$ will appear in the chargino Feynman rules.

\medskip
{\sl 4.~~Neutralinos}

\noindent
The neutral gauginos and higgsinos can mix; the mass eigenstates are
called {\it neutralinos}.
In the $\wt B$--$\wt W^3$--$\wt H^0_1$--$\wt H^0_2$ basis, the
neutralino Majorana mass matrix is\refmark\ghia\
$$
  Y = \pmatrix{%
    M_1 &0  &-m_Zc_\beta s_W &m_Zs_\beta s_W \cr
    0   &M_2 &m_Z c_\beta c\ls W &-m_Z s_\beta c\ls W \cr
    -m_Z c_\beta s\ls W &m_Z c_\beta c\ls W &0 &-\mu \cr
    m_Z s_\beta s\ls W &-m_Z s_\beta c\ls W &-\mu &0 \cr }\,,
\eqn\neutralinomatrix
$$
where $s_\beta = \sin\beta$, $c_\beta=\cos\beta$, \etc\  A $4\times 4$
unitary matrix $Z$ is required to diagonalize the neutralino mass
matrix
$$
{\cal M}_{\wt\chi^0}=Z^\ast YZ^{-1}\,,\eqn\chizerodiag
$$
where the diagonal elements of ${\cal M}_{\wt\chi^0}$ can be either
positive or negative.
The four mass eigenstates are denoted by $\widetilde\chi^0_i$
($i=1,2,3,4$), with corresponding mass eigenvalues $\eta_i
M_{\wt\chi^0_i}$.
The physical neutralino masses are defined to be
positive, with $M_{\wt\chi^0_1}\leq\dots\leq M_{\wt\chi^0_4}^{\vph}$.
The sign of the mass eigenvalue ($\eta_i=\pm1$) is physically
relevant and
corresponds to the CP quantum number of the Majorana neutralino state.
The mixing matrix elements $Z_{ij}$ will appear in the neutralino
Feynman rules.

It is sometimes useful to define the interaction eigenstates that
correspond to the supersymmetric partners of the $\gamma$ and $Z$
$$\eqalign{%
  \wt \gamma &= \cos\theta_W \wt B + \sin\theta_W \wt W^3\,, \cr
  \wt Z &= -\sin\theta_W \wt B + \cos\theta_W \wt W^3\,. \cr
}\eqn\photinozino$$
However, these will not in general be mass eigenstates.  For
example, $\wt\gamma$ is a mass eigenstate only if $M_1=M_2$.  More
generally, $\widetilde\gamma$ is an (approximate) mass eigenstate if
$M_1,M_2 \ll m_Z$, in which case $M_{\wt\gamma}\simeq
M_1\cos^2\theta_W + M_2\sin^2\theta_W$.

For a general set of neutralino parameters, the mass eigenstates are
complicated mixtures of the interaction eigenstates.  Two limiting
cases are useful:

\item{(i)}  $M=M\pri=\mu=0$
$$\eqalignalign{%
\wt \chi^0_1 &= \tilde\gamma,\hskip6cm  &M_{\wt \chi^0_1} = 0 \cr
\wt \chi^0_2 &= \wt H_1\sin\beta+\wt H_2\cos\beta,
                   &M_{\wt \chi^0_2}=0\cr
\wt \chi^0_3 &= \sqrt{\half} \left[\wt Z+\wt H_1\cos\beta-
             \wt H_2\sin\beta\right]  &M_{\wt \chi^0_3} = m\ls Z \cr
\wt \chi^0_4 &= \sqrt{\half} \left[-\wt Z+\wt H_1\cos\beta-
             \wt H_2\sin\beta\right], &M_{\wt \chi^0_4} = m\ls Z\,. \cr}
\eqn\limitone
$$
\item{(ii)} $M,M\pri,\mu\gg m\ls Z$
$$
  \wt \chi^0_i = \left[ \wt B,\,\wt W_3, \,{\sqrt{\half}}(\wt H_1-\wt H_2),
            \,{\sqrt{\half}}(\wt H_1+\wt H_2)\right]\,,\eqn\limittwo
$$
with masses $|M\pri|$, $|M|$, $|\mu|$, and $|\mu|$ respectively.

\medskip
{\sl 5.~~Higgs sector}\refmark\hhg\

\noindent
Including the soft-supersymmetry-breaking terms, the Higgs
potential now reads
$$\eqalign{%
  V_{\rm Higgs} = \ &m^2_{1H} |H_1|^2 + m^2_{2H} |H_2|^2
                 - m^2_{12} (\epsilon_{ij} H^i_1 H^j_2 + h.c.)\crr
                 &+\eighth(g^2+g'^2) \left(|H_1|^2-|H_2|^2\right)^2
                    +\half g^2|H^*_1H_2|^2\,,\cr
}\eqn\mssmhiggspot$$
where $m^2_{iH}\equiv|\mu|^2+m^2_i\quad (i=1,2)$.
The parameters $m^2_i$ ($i=1,2$) are real and can be either
positive or negative.  Now we see how the inclusion
of soft-supersymmetry-breaking terms permits SU(2)$_L\times$U(1)$_Y$
breaking. The parameter region in which SU(2)$_L\times$U(1)$_Y$
breaks down to U(1)$_{EM}$ is
$$\eqalignalign{%
&m^2_{1H} + m^2_{2H} \geq 2|m_{12}^2|\quad
&({\rm required\ for\ stability\
       of}\ V_{\rm Higgs}\ {\rm at\ large\ scalar\ fields}) \cr
&|m_{12}^2|^2 > m^2_{1H} m^2_{2H}
&({\rm required\ for\ SU(2)}_L\times{\rm U(1})_Y\ {\rm breaking})\,.\cr}
\eqn\higgsconditions
$$
Let us proceed on the assumption that these conditions are satisfied.%
\foot{How these conditions are satisfied in the
supergravity scenarios discussed in the previous section is a very
interesting story.  Typically, these conditions are not satisfied
at $M_P$.  However, after renormalization group scaling, one discovers
that these conditions are satisfied at the electroweak scale.  This
picture is called the radiative symmetry breaking scenario, which will
be discussed in Lecture 2.}

The MSSM Higgs potential [eq.~\mssmhiggspot]
is analyzed in detail in my TASI-90 lectures\refmark\tasiesb.
Starting with eight real scalar degrees of freedom, three are
Goldstone modes absorbed by the $W^\pm$ and $Z$.  The
remaining five degrees of freedom yield the physical Higgs bosons
of the model.  Note that $V_{\rm Higgs}$ automatically
conserves CP, since any complex phase in $m^2_{12}$ can be
absorbed into the definition of the Higgs fields.
The following nomenclature is used for the physical Higgs bosons:
$$\eqalignalign{%
 &H^\pm \phantom{,h^0}\qquad &\hbox{charged\ Higgs\ boson\ pair} \cr
 &A^0 \phantom{,h^0}\qquad   &\hbox{CP-odd\ neutral\ Higgs\ boson}\cr
 &H^0,h^0      &\hbox{CP-even\ neutral\ Higgs\ boson}\,.\cr}
\eqn\higgsspectrum
$$
The three parameters $m^2_{1H},\ m^2_{2H}$ and $m^2_{12}$ of the
Higgs potential can re-expressed in terms of the two Higgs vacuum
expectation values, $v_i\equiv\VEV{H_i^0}$ and one physical Higgs
mass.  We are free to choose the phases of the Higgs fields such
that $v_1$ and $v_2$ are positive.   Then, $m_{12}^2$ must be positive,
in which case it follows from eq.~\mssmhiggspot\ that
$$
  m^2_A = {m^2_{12}\over \sin\beta \cos\beta}\,.
\eqn\monetwoeq
$$
It is convenient to choose $m_A$ as one of the physical input
parameters.
 Note that $\mw^2=\mz^2\cos^2\theta_W=\fourth
g^2(v_1^2+v_2^2)$, which fixes the magnitude $v_1^2+v_2^2=(246~{\rm
GeV})^2$.  This leaves two parameters which fix the Higgs sector
masses and couplings: $\mha$ and $\tan\beta\equiv v_2/v_1$.
The charged Higgs mass is given by
$$
  m^2_{H^\pm} = m^2_W +m^2_A\,.\eqn\chhiggs
$$
The neutral CP-even Higgs bosons, $H^0$ and $h^0$ are obtained by
diagonalizing a $2\times 2$ mass matrix, which in the $H_1$--$H_2$
basis is given by
$$
  {\cal M}^2 = \pmatrix{%
   \mha^2 s^2_\beta + m^2_Z c^2_\beta &-(\mha^2+m^2_Z)s_\beta c_\beta
   \cr
  -(\mha^2+m^2_Z)s_\beta c_\beta &\mha^2 c^2_\beta+m^2_Zs^2_\beta \cr }
\, .\eqn\twotimestwo
$$
The eigenstates are
$$\eqalign{%
  H^0 &= (\sqrt 2\,\Re\,H^0_1-v_1) \cos\alpha + (\sqrt2\,\Re\,H^0_2-v_2)
              \sin\alpha   \cr
  h^0 &= -(\sqrt 2\,\Re\,H^0_1-v_1)\sin\alpha + (\sqrt2\,\Re\,H^0_2-v_2)
             \cos\alpha \cr
}\eqn\cpevenhiggs$$
where the mixing angle $\alpha$ is given by
$$
  \cos 2\alpha = -\cos2\beta \left( {\mha^2-m^2_Z\over
                   m^2_{H^0}-m^2_{h^0}}\right),
\qquad\quad \sin2\alpha = -\sin2\beta \left({m^2_{H^0}+m^2_{h^0} \over
                   m^2_{H^0}-m^2_{h^0}}\right)\,.\eqn\alphadef
$$
The corresponding CP-even Higgs mass eigenvalues are
$$
  m^2_{H^0,\,h^0} = \half \left(\mha^2+m^2_Z \pm
       \sqrt{(\mha^2+m^2_Z)^2 - 4m^2_Z \mha^2 \cos^2 2\beta}\right)\,,
\eqn\cpevenhiggsmass
$$
where by definition, $\mhl\leq\mhh$.

All mass relations that have been presented in this section are tree-%
level results which are modified by radiative corrections.  This will
be discussed in more detail in Lecture 3.
For the moment, I shall record the following inequalities that follow
{}from the tree-level mass formulae above
$$\eqalign{%
  &\mhl\leq\mha\,, \cr
  &m_{h^0} \leq m|\cos2\beta|\leq\mz\,,
\qquad {\rm with}\ m\equiv \min(\mz, \mha)\cr
  &m_{H^0} \geq m\ls Z \,,\cr
  &m_{H^\pm} \geq m_W\,. \cr }\eqn\inequalities
$$

The Higgs couplings to gauge bosons are fixed by the SU(2)$\times$%
U(1) gauge invariance.  The Higgs couplings to quarks and leptons are
obtained from eq.~\susyyukawa, where $W=W_R$ is given in
eq.~\wrparity.  Note that this leads to a very specific pattern
of tree-level Higgs-fermion Yukawa couplings in which $H_1$
couples only to down-type fermions ($\tau,b$), while $H_2$ couples
only to up-type fermions ($\nu_\tau,t$).  A comprehensive list of all
MSSM Higgs interactions can be found in Appendix A of ref.~[\hhg].

\goodbreak
\medskip
{\sl 6.~~Quarks, Leptons, and Gauge Bosons}

\noindent
The identification of quark, lepton and gauge boson eigenstates
and the corresponding masses follows the usual Standard
Model analysis.
One constructs the quark mass matrix and
extracts the CKM angles.  The neutrinos are massless, so that the
charged lepton interaction eigenstates and mass eigenstates coincide.
The $Z$ and $\gamma$ are eigenstates of a $2\times 2$ neutral gauge
boson mass matrix.  The Standard Model Higgs boson is replaced by the
Higgs sector described above.  This
completes the enumeration of all the mass eigenstates of the
MSSM.

\REF\rosiek{J. Rosiek, {\sl Phys. Rev.} {\bf D41} (1990) 3464.}
With the MSSM mass eigenstates in hand, and a complete list of
supersymmetric interactions and the soft-supersymmetry-breaking terms,
it is straightforward to compute all the Feynman rules of the MSSM.
These can be found in refs.~[\haberkane], [\ghi], and [\rosiek].

\endpage

\chapter{%
What Do We Know About the Supersymmetric \hfill\break \hbox to 13pt{}
Parameters?}

In this lecture, I will discuss the
constraints on the parameter space of
the low-energy supersymmetric model.  To do this in complete
generality is not practical.  First, one can construct an arbitrary
number of supersymmetric extensions of the Standard Model.
Unfortunately, time and
space limitations forbid a comprehensive survey of all non-minimal
low-energy supersymmetric models.
It is perhaps prudent to focus at first on the MSSM,
although a few comments regarding the alternate phenomenology of
$R$-parity violating theories will be given at the end of this
lecture.   Second, even within the MSSM framework, the number of
new parameters is quite large.  Experimental data constrains some
of the parameters, and various theoretical assumptions can further
reduce the parameter freedom.  Imposing theoretical biases on the
structure of the MSSM is not necessarily bad, as long as one is aware
of the assumptions made and the possibility that some
of these assumptions could be incorrect.

\section{The MSSM Parameter Space}

Based on the results of Lecture 1, the parameters of the MSSM are as
follows.  In the supersymmetry-conserving sector, one has the usual
Standard Model parameters: the SU(3)$\times$SU(2)$\times$U(1) gauge
couplings ($g_s$, $g$ and $g\pri$ respectively), and the
Higgs-fermion Yukawa coupling matrices ($h_t$, $h_b$, and $h_\tau$).
One further parameter is the Higgs superfield mass parameter,
$\mu$.  Other possible parameters that could have arisen in the
superpotential have been eliminated under the assumption of
$R$-parity invariance.  The remaining parameters of the MSSM arise
{}from soft-supersymmetry breaking terms.  These include three gaugino
Majorana mass parameters ($M_3$, $M_2$ and $M_1$),
diagonal squark and slepton mass squared-mass matrices in the
$\tilde f_L$--$\tilde f_R$ basis
($M_{\wt Q}^2$, $M_{\wt U}^2$, $M_{\wt D}^2$, $M_{\wt L}^2$,
and $M_{\wt E}^2$), Higgs-squark-squark
and Higgs-slepton-slepton trilinear interaction terms (which are
proportional to the matrix $A$-parameters $A_t$, $A_b$ and $A_\tau$),
and the Higgs sector mass parameters ($m_{1H}^2$, $m_{2H}^2$ and
$m_{12}^2$).  Electroweak symmetry breaking is driven by the
soft-supersymmetry breaking Higgs mass parameters.  As a result,
$m_{1H}^2$, $m_{2H}^2$ and $m_{12}^2$ can be re-expressed in terms of
two Higgs vacuum expectation values ($v_1$ and $v_2$) and one
physical Higgs mass (usually chosen to be the mass of the CP-odd
Higgs scalar, $\mha$).  Note that $v_1^2+v_2^2=(246~\gev)^2$ is fixed
by the $W$ (or $Z$) mass while the ratio $\tanb\equiv v_2/v_1$ is a
free parameter of the model.  In a model with only one generation of
quarks, leptons and their superpartners, the above list contains 14
new parameters.  In addition, the $A$-parameters and gaugino mass
parameters may contain nontrivial complex phases that cannot be
removed by field redefinitions.  These provide potential new sources
of CP-violation.  In the full three-generation model, the number
of new parameters is substantially greater since the squark and
slepton mass-squared parameters and the $A$-parameters are now
$3\times 3$ matrices.  The possibility of intergenerational
mixing can lead to additional complications.

To make predictions and to perform phenomenological analyses
with 14 or more free parameters is impractical.
Therefore, most practitioners of low-energy supersymmetry
make assumptions to reduce the parameter freedom.
Different model assumptions can lead to different ground rules.
This leads to the following quandary---what is the real MSSM?
Of course, there is no unambiguous answer to this question.
However, it is important to be aware of the underlying assumptions of
any starting point.  When a theorist or phenomenologist states:
``The MSSM predicts that $\ldots$'', one should carefully check the
underlying assumptions.  Below, I will
present a  brief guide to theoretical prejudices and experimental
data used to constrain the MSSM parameter space.

\section{Theoretical Constraints and Biases}

In a completely phenomenological treatment of low-energy
supersymmetry, the MSSM parameters would be considered completely
free parameters to be determined from experimental data.  Once the
supersymmetric parameters are measured, one could then attempt to
extract information about the underlying fundamental physics
which is associated
with energy scales much higher than those directly probed by
colliders.  Since the conjecture of low-energy
supersymmetry is motivated by an attempt to embed low-energy physics
in a more fundamental framework, it seems appropriate to exploit this
motivation in constraining the low-energy
supersymmetric parameters.  This leads to a number of theoretical
considerations that provide some arguments for constraining the
parameter freedom of the MSSM.
\REF\barbig{R. Barbieri and G.F. Giudice, {\sl Nucl. Phys.}
{\bf B306} (1988) 63.}
\REF\rossrob{G.G. Ross and R.G. Roberts, {\sl Nucl. Phys.}
{\bf B377} (1992) 571.}

\smallskip\pointbegin
{\sl What is the maximal size of the dimensionful parameters of the MSSM?}

This is the polite version of the common complaint of
experimentalists---``when are you theorists going to
give up on supersymmetry?''
Roughly speaking, we do not want the dimensionful supersymmetric
parameters (\ie, $\mu$ and the soft-supersymmetry-%
breaking mass parameters)
to be much larger than 1 TeV; otherwise we
lose the ``naturalness" of the model.  But can one
be more precise?  Indeed, one can invent a method for measuring
how much fine-tuning goes into setting the hierarchy between
the electroweak scale and the supersymmetry breaking scale.  Two recent
examples of such a measure are given in refs.~[\barbig] and [\rossrob].
In each case, a dimensionless ratio (let us call it $x$)
is constructed that measures the amount of fine-tuning.
A model of low-energy supersymmetry is deemed acceptable if $x$ is
restricted to lie within a certain range (corresponding to, say, moderate
fine-tuning).  The problem with such a condition is obvious.
How does one establish a reasonable allowed range for $x$?  Without
detailed knowledge of the fundamental theory at high energies
(including a precise understanding of the underlying mechanism of
supersymmetry breaking), there is no way to do this. If one regards
$x$ as a one-dimensional parametrization of low-energy supersymmetry
models, then this problem can be understood as a lack of information
as to the nature of the probability measure in $x$.  Different
theorists can propose different conditions and reach somewhat
different conclusions.  My personal belief is that the conditions of
refs.~[\barbig] and [\rossrob]\ are too strict, but at the present state
of knowledge, the decision on a reasonable range for $x$ must be left
to the aesthetic sensibilities of the physicist.  To paraphrase a
famous former supreme court judge (when asked to define what is
``obscene''), I cannot formally define a measure that tells me when
the supersymmetric mass parameters are too large, but I know it when
I see it.

\smallskip\point
{\sl What are the limits on $\tan\beta$ [and on $m_t$]?}

Above, we considered the question of
theoretical constraints on the maximal value of
dimensionful parameters of the MSSM.  Here, we consider the
possibility of determining upper (and lower) bounds for dimensionless
parameters of the MSSM.  Since the gauge coupling constants
and the Higgs-fermion Yukawa couplings (except for $h_t$) are
known, it is sufficient to focus on two unknown dimensionless
parameters: $h_t$ and $\tanb$.  From eq.~\yukawas, one can
equally well consider the two independent unknown parameters
to be $\mt$ and $\tanb$.  Clearly, once the top quark is
discovered (perhaps in 1993 or 1994 at Fermilab), the
reduction of parameter uncertainty will have an impact
on the future development of low-energy supersymmetry
model building and phenomenology.

\REF\rgecollection{T.P. Cheng, E. Eichten and L.-F. Li, {\sl Phys. Rev.}
{\bf D9} (1974) 2259; K. Inoue, A. Kakuto, and Y. Nakano,
{\sl Prog. Theor. Phys.} {\bf 63} (1980) 234; H. Komatsu,
{\sl Prog. Theor. Phys.} {\bf 67} (1982) 1177; K. Inoue, A. Kakuto,
H. Komatsu and S. Takeshita,
{\sl Prog. Theor. Phys.} {\bf 67} (1982) 1889; {\bf 68} (1982) 927
[E: {\bf70} (1983) 330] {\bf 71} (1984) 413; C. Hill, C.N. Leung and
S. Rao, {\sl Nucl. Phys.} {\bf B262} (1985) 517.}
\REF\bagger{J. Bagger, S. Dimopoulos and  E. Masso, {\sl Phys. Lett.}
{\bf 156B} (1985) 357; \sl Phys. Rev. Lett. \bf 55\rm (1985) 920.}
\REF\russ{G.M. Asatryan, A.N. Ioannisyan and S.G. Matinyan,
{\sl Sov. J. Nucl. Phys.} {\bf 53} (1991) 371 [{\sl Yad. Fiz.} {\bf 53},
(1991) 592];
M. Carena, T.E. Clark, C.E.M. Wagner, W.A. Bardeen and K. Sasaki,
{\sl Nucl. Phys.} {\bf B369} (1992) 33.}
\REF\handz{H.E. Haber and F. Zwirner, unpublished.}
\REF\cdflimit{F. Abe \etal\ [CDF Collaboration], {\sl Phys. Rev. Lett.}
{\bf 68} (1992) 447.}
Upper bounds on dimensionless parameters cannot be deduced on the
basis of the low-energy theory alone.  But once one is willing to
consider the Standard Model (and its MSSM extension) as the low-energy
effective theory of a more fundamental high-energy theory, one
can make progress.  The weakest bounds are obtained as follows.
Consider the theory of particle physics at a scale $\Lambda$,
where $\Lambda$ is allowed to vary from the electroweak scale
to the Planck scale.  In the MSSM, the first scale  of new physics
encountered is the mass scale that characterizes the scale of low-energy
supersymmetry breaking.  This is the mass scale which determines the
sizes of all the dimensionful supersymmetric parameters discussed
above.  Let $\msusy$ represent this mass scale.  Admittedly, this is
an oversimplification, since it neglects the variation among the many
dimensionful supersymmetric parameters.  However, the approximation
of one overall scale of low-energy supersymmetry is sufficient for our
purposes here.  Above $\msusy$, I assume that no new physics enters
until the scale $\Lambda$ is reached.  $\Lambda$ could be the
Planck mass, a grand unification scale ($10^{16}~\gev$), or perhaps
some intermediate scale where physics not yet imagined enters.
In all cases, it must be true that all dimensionless coupling constants
(the guage and Higgs-fermion Yukawa couplings) remain finite
at all energy scales below $\Lambda$.  The theoretical tools needed to
implement this condition are the renormalization group equations (RGEs).

The one-loop renormalization group equations (RGEs)
take the form
$$
{dp_i\over dt}=\beta_i(p_1,p_2,..)\,,\qquad
\hbox{ where}~t\equiv\ln\,\mu^2\,,\eqn\rges$$
where $\mu$ is the energy scale, and
the parameters $p_i$ stand for
the squared Yukawa couplings $h_f^2$
($f=t$, $b$ and $\tau$; the two lighter generations can be neglected)
and the squared gauge couplings $g_j^2$ ($j=$3, 2, 1) corresponding to
SU(3)$\times$SU(2)$\times$U(1) respectively.  The $h_f$ are defined
in eq.~\yukawas, and I assume a Higgs-fermion coupling pattern
as specified by the MSSM.  The $g_j$ are
normalized such that they are equal at the grand unification
scale.  It is also convenient to define
$$    g_s\equiv g_3,\qquad
g\equiv g\ls2\,,\qquad g\pri\equiv \sqrt{\threefifths} g\ls1\,,
\eqn\gaugecoups$$
where $g$ and $g\pri$ are normalized in the usual way for low-energy
electroweak physics, {\it i.e.} $\tan\theta_W=g\pri/g$.
In addition, let
 $N_g=3$ be the number of generations and
$N_H$  the number of low-energy scalar doublets.
In the MSSM, $N_H=2$ for
$\mu>\msusy$.  For values of $\mu<\msusy$, $N_H=2$ if $\mha\simeq
{\cal O}(\mz)$ and $N_H=1$ if $\mha\simeq{\cal O}(\msusy)$.

\REF\langtasi{P. Langacker, TASI-92 Lectures, in These Proceedings.}
\TABLE\rgeone{}
\FIG\tanlima{%
The region of $\tanb$--$\mt$ parameter space in which
all running Higgs-fermion Yukawa couplings remain finite at all
energy scales, $\mu$, from $\mz$ to
$\Lambda=10^{16}$~GeV\refmark\handz.
Non-supersymmetric two-Higgs-doublet (one-loop)
renormalization group equations \break
(RGEs) are used for $\mz\leq\mu\leq\msusy$ and the RGEs of the
minimal supersymmetric model are used for $\msusy\leq\mu\leq\Lambda$
(see table \rgeone).  Five different values of $\msusy$ are shown;
the allowed parameter space lies below the respective curves.}
\FIG\tanlimb{%
The region of $\tanb$--$\mt$ parameter space in which
all running Higgs-fermion Yukawa couplings remain finite at all
energy scales from $\mz$ to
$\Lambda=100$ TeV.  See caption to fig.~\tanlima.}
In Table~\rgeone, I list the $\beta$-functions
appropriate for the Standard Model
and the MSSM\refmark{\rgecollection,\bagger}.
The latter are used for energy scales between $\mz$
and $\msusy$, and the former for energy scales between $\msusy$ and
$\Lambda$.  By requiring that $h_t$, $h_b$, and $h_\tau$ remain finite
at all energy scales up to $\Lambda$\refmark{\bagger-\handz},
one obtains an allowed region
of $\mt$ {\it vs.}~$\tanb$ as a function of $\Lambda$.
The results are shown in figs.~\tanlima\ and
\tanlimb\ for two different choices of $\Lambda$\refmark\handz.
The allowed region of parameter space lies below the curves shown.
For example, if there is no new physics (other than perhaps minimal
supersymmetry) below the grand unification scale of $10^{16}$~GeV,
then based on the CDF limit\refmark\cdflimit\ of $\mt>91$~GeV, one
would conclude that $0.5\lsim\tanb\lsim 50$.  The lower [upper] limit on
$\tanb$ arises from the constraint on the growth of the Higgs-top
[-bottom] quark Yukawa coupling.  The lower limit on
$\tanb$ becomes even sharper if the top-quark mass is heavier.
Remarkably, the limits on $\tanb$ do not get substantially weaker
for $\Lambda$ as low as 100 TeV.  In contrast,
the limits on $\mt$ get
significantly weaker as $\Lambda$ is taken smaller.  However,
one can use other constraints based on precision
electroweak measurements to conclude that $\mt<197~\gev$ at 95\% CL%
\refmark\langtasi.
It is amusing that these results are about the same as those shown
in fig.~\tanlima, for $\msusy=1$~TeV.

\REF\twoloop{V. Barger, M.S. Berger and P. Ohmann, {\sl Phys. Rev.}
{\bf D47} (1993) 1093.}
\pageinsert
\centerline{\bf
Table \rgeone. One-loop Renormalization Group Equations}
\centerline{\bf for Yukawa and Gauge Coupling Constants}
\bigskip
\hrule
\vskip6pt
1. $\mu>\msusy$
$$\eqalign{%
\beta_{h_t^2}&={h_t^2\over16\pi^2}\left[6
h_t^2+ h_b^2-\sixteenthirds g_3^2-3g^2-\thirteenninths g'^2\right]\crr
\beta_{h_b^2}&={h_b^2\over16\pi^2}\left[6 h_b^2+h_t^2
+h_{\tau}^2-\sixteenthirds g_3^2-3g^2-\sevenninths g'^2\right] \crr
\beta_{h_{\tau}^2}&={h_{\tau}^2\over16\pi^2}\left[4
h_{\tau}^2+3 h_b^2-3g^2-3g'^2\right]\crr
\beta_{g'^2}&={g'^4\over48\pi^2}\Big[10N_g+\threehalf N_H\Big]\crr
\beta_{g^2}&={g^4\over48\pi^2}\Big[6N_g+\threehalf N_H-18\Big]\crr
\beta_{g_3^2}&={g_3^4\over48\pi^2}\Big[6N_g-27\Big]   \cr}
$$
\hrule
\vskip6pt
2. $\mu<\msusy$
$$\eqalign{%
\beta_{h_t^2}      &={h_t^2\over16\pi^2}\big[\ninehalf
h_t^2+\half h_b^2-8g_3^2-\ninefourth
g^2-\seventeentwelfth g'^2\big]     \crr
\beta_{h_b^2}      &={h_b^2\over 16\pi^2}\big[\ninehalf h_b^2+\half
h_t^2+h_{\tau}^2-8g_3^2-\ninefourth g^2-\fivetwelfth
g'^2\big]     \crr
\beta_{h_{\tau}^2} &={h_{\tau}^2\over16\pi^2}\big[\fivehalf
h_{\tau}^2+3h_b^2-\ninefourth g^2-\fifteenfourth
g'^2\big]     \crr
\beta_{g'^2}  &={g'^4 \over48\pi^2}\Big[\twentythirds N_g+\half
N_H\Big]    \crr
\beta_{g^2}   &={g^4\over48\pi^2}\Big[4N_g+\half N_H-22\Big]  \crr
\beta_{g_3^2} &={g_3^4\over48\pi^2}\Big[4N_g-33\Big]   \cr}
$$
\hrule
\vfill
\endinsert

How reliable are the results just quoted?  First, they are based
only on one-loop RGEs.  If two-loop RGEs are used, then the results
quoted above will change only very slightly\refmark\twoloop.
The reason is that as
the energy scale increases, the Higgs-quark Yukawa coupling ($h_f$)
increases
slowly at first.  As the energy scale approaches the Landau pole (where
$h_f\to\infty$), $h_f$ begins to increase more rapidly.  In fact
the results shown in figs.~\tanlima\ and \tanlimb\ are very insensitive
to whether one defines $\Lambda$ to be the Landau pole or the point at
which $h_f^2/4\pi\sim{\cal O}(1)$.  Thus, these results should not be
very sensitive to the regime in which $h_f$ is large enough such that
higher terms in the RGE matter.  Of course, once $h_f$ enters the
non-perturbative regime, the analysis just presented breaks
down.  However, presumably, new physics enters once this point is
reached, so the results of figs.~\tanlima\ and \tanlimb\ (which assume
that no new physics enters below the scale $\Lambda$) remain valid.

\REF\tanbt{See \eg, G.F. Giudice and G. Ridolfi, {\sl Z. Phys.}
{\bf C41} (1988) 447; M. Olechowski and S. Pokorski, {\sl
Phys. Lett.} {\bf B214} (1988) 393; M. Drees and M.M. Nojiri,
{\sl Nucl. Phys.} {\bf B369} (1992) 54.}
Note that in the above analysis, no assumption has been made about
the nature of the physics above $\Lambda$.  (For example, no assumption
regarding grand unification has been made.)   Additional assumptions
about the physics at the high-energy scale can reduce the allowed
parameter space and strengthen the resulting bounds.
For example, in models of spontaneously broken
supergravity, one obtains an effective Planck scale
field theory consisting of a supersymmetric model with
soft-supersymmetry-breaking terms, as described in Lecture 1.
Renormalization group evolution is required to derive the appropriate
set of low-energy parameters at the electroweak scale.  In the RGE
analysis, one can also find an allowed region of $\mt$ {\it vs.}~%
$\tanb$, as a function of the Planck scale input parameters.  However
the resulting constraints can be stronger since the form of the Planck
scale theory is not completely arbitrary.
The range of allowed $\tanb$ values is found to be somewhat
smaller; roughly\refmark\tanbt\
$$1\lsim\tanb\lsim{m_t\over m_b}\,.
\eqn\tanbinequality$$
Values of $\tanb\leq 1$ are disallowed since such values are
incompatible with electroweak symmetry breaking driven by the
soft-supersymmetry-breaking Higgs mass-squared
terms at the electroweak scale.

\smallskip\point
{\sl How can low-energy phenomenology restrict the MSSM parameters?}

\REF\dipole{%
J. Ellis, S. Ferrara and D.V. Nanopoulos,
{\sl Phys. Lett.} {\bf B114} (1982) 231;
W. Buchm\"uller and D. Wyler, {\sl Phys. Lett.} {\bf B121} (1983) 321;
J. Polchinski and M. B. Wise, {\sl Phys. Lett.} {\bf B125} (1983) 393;
F. del Aguila, M.B. Gavela, J.A. Grifols, and A. M\'endez,
{\sl Phys. Lett.} {\bf B126} (1983) 71;
D.V. Nanopoulos and M. Srednicki, {\sl Phys. Lett.} {\bf B128} (1983) 61;
A.I. Sanda, {\sl Phys. Rev.} {\bf D32} (1985) 2992.}
\REF\fcncsusy{J. Ellis and D.V. Nanopoulos, {\sl Phys. Lett.}
{\bf 110B} (1982) 44; T. Inami and C.S. Lim, {\sl Nucl. Phys.}
{\bf B207} (1982) 533;
M.J. Duncan and J. Trampetic, {\sl Phys. Lett.} {\bf B134} (1984) 439;
F. Gabbiani and A. Masiero, {\sl Nucl. Phys.} {\bf B322} (1989) 235;
for a recent re-analysis and more complete guide to the
literature, see
J.S. Hagelin, S. Kelley and T. Tanaka, MIU-THP-92-59 (1992).}
\REF\flavorsusy{L.J. Hall, V.A. Kostelecky and S. Raby,
{\sl Nucl. Phys.} {\bf B267} (1986) 415.}
\REF\fermimass{S. Dimopoulos, L.J. Hall and S. Raby,
{\sl Phys. Rev. Lett.} {\bf 68} (1992) 1984; {\sl Phys. Rev.} {\bf D45}
(1992) 4192; {\bf D46} (1992) 4793.}
There is no experimental evidence to date for the existence of
supersymmetric particles or interactions.  This fact imposes
constraints on any low-energy supersymmetric model in a number of ways.
First, one can perform direct searches for supersymmetric particle
production in accelerator experiments.  Current limits on
supersymmetric particle masses will be briefly summarized in section 2.3.
These mass limits in turn impose restrictions on the basic supersymmetric
parameters.  Second, all CP-violating phenomena observed to date are
consistent with a single complex phase in the CKM quark mixing matrix.
I noted briefly in section 2.1 that the MSSM introduces new parameters
with (possible) complex phases.  If these phases were ${\cal O}(1)$,
then CP-violating effects induced by squark, slepton and/or gaugino
exchange would result in
an observable electric dipole moment for the neutron and electron
that is substantially above the present experimental
limits\refmark\dipole.
Either the corresponding supersymmetric particle masses are
very large (above 1~TeV) or the corresponding phases are small
($\lsim 10^{-2}$).
Third, it is
an experimental fact that
flavor changing neutral current (FCNC)
processes are greatly suppressed.
The most stringent constraints derive from the properties of
$K^0$--$\bar K^0$ mixing (and to a lesser extent from
$B^0$--$\bar B^0$ mixing).
These results place very stringent limits on the matrix structure of
the squark mass parameters\refmark\fcncsusy.
To a very good approximation, the diagonal
squared-masses of squarks (in the $\tilde q_L$--$\tilde q_R$ basis)
of a given electric
charge must be degenerate.  These last two results present a
formidable challenge
to model builders attempting to derive the low-energy MSSM parameters
{}from more fundamental physics at a higher energy scale.
One successful approach is the standard low-energy
supergravity scenario, where
the soft-supersymmetry breaking squark
mass parameters are flavor independent and all
$A$-parameters and gaugino masses are real at the Planck scale.
Note that the renormalization group
evolution is sensitive to $h_b,$ and $h_t$, which yields
squark mass parameters with some non-trivial flavor
dependence at the electroweak scale.  However, the magnitude of such
flavor dependence is consistent with present experimental constraints
on FCNC interactions.

Although the required degeneracy of squark and slepton masses
can be arranged, the MSSM provides no explanation for its
origin.  Moreover,
the MSSM provides no fresh
insights into the structure of quark (and lepton) masses and mixing.
Depending on the structure of the physics at high energies,
supersymmetry could even generate a real ``flavor problem"---\ie,
large flavor changing neutral currents at the electroweak scale
in conflict with experimental observations\refmark\flavorsusy.
This could occur, for
example, if there were some flavor dependence
at an intermediate energy scale.
Integrating out the physics at that intermediate
scale can produce soft-dimension-two squark mass terms with
non-trivial flavor structure.  Thus, present limits on FCNCs can
impose severe constraints on the nature of the physics at scales
well above 1~TeV.  On the other hand,
the possibility of non-trivial
flavor structure at high scales may ultimately
provide the origin of the quark and lepton masses and mixing angles
observed at low-energies\refmark\fermimass.

\smallskip\point
{\sl Limits on the $A$-parameters [and $\mu$]}

\REF\aparms{J.F. Gunion, H.E. Haber and M. Sher,
{\sl Nucl. Phys.} {\bf B306} (1988) 1.}
\REF\kanequiros{M. Quiros, G.L. Kane and
H.E. Haber, {\sl Nucl. Phys.} {\bf B273} (1986) 333.}
{}From the mass matrices for the top and bottom squarks presented
in eq.~\squarkmatrix, one notes an off-diagonal term in
the squark squared mass-matrices
$$
  M^2_{\rm LR} = \cases{%
            m_b(A_b-\mu\tan\beta),&
            for $\tilde b_L$--$\tilde b_R$ mixing, \cr
            m_t(A_t-\mu\cot\beta),&
            for $\tilde t_L$--$\tilde t_R$ mixing. \cr
}\eqn\msqlr$$
If $M^2_{\rm LR}$ is too large, the smaller of the two
squark squared-mass eigenvalues can be driven negative.
This would imply that the SU(3)$_C\times$U(1)$_{EM}$
ground state was not a minimum of the scalar potential.
The true ground state would not conserve both color and EM;
clearly this possibility must be rejected.  As an example,
if $M_{\wt Q} \simeq M_{\wt U} \simeq M_{\wt D}$ and
$A_t \gg \mu\cot\beta$, then one finds that the
squared-mass of the lightest top-quark
is positive if $A_t \lsim 3.5 M_Q$.  Even if all squark
and slepton squared-masses are positive, this does not necessarily
imply that the SU(3)$_C\times$U(1)$_{EM}$ ground state is a {\it global}
minimum.  To insure that the global minimum preserves both color and EM,
one must examine the complete scalar potential (including  Higgs bosons,
squarks, and sleptons). By searching in a general way for all
color and/or EM violating local minima, one can
check whether these minima lie
above the desired SU(3)$_C\times$U(1)$_{EM}$ preserving ground state.
The resulting constraint also leads to
some restrictions (typically upper bounds) on combinations
of $A$-parameters\refmark\aparms.
Roughly speaking, the $A$-parameters should be less
than $c\cdot{\cal O}(M_Q)$ where $c$ is a number of order 1
($c \approx 3$ is a typical value).

\smallskip\point
{\sl Are there theoretically compelling constraints on Planck scale
parameters?}

\noindent
In section 1.4, I described how the spontaneous breaking of supergravity
can lead to an effective field theory at the Planck scale that is a
broken supersymmetric model where the supersymmetry breaking is
parametrized by all possible soft-supersymmetry breaking terms.
Until the origin of the supersymmetry breaking of the fundamental
theory is understood, one must regard the Planck-scale
soft-supersymmetry breaking terms as phenomenological input parameters
to the model.  However, perhaps one can do better.   By introducing
(hopefully theoretically well motivated) constraints to the Planck scale
parameters, one can reduce the parameter freedom of the resulting
low-energy supersymmetric model.  Here are some of the most common
Planck scale assumptions.
\smallskip
\item{(i)}
Grand unification of gauge coupling constants
and gaugino Majorana mass terms
$$\eqalign{%
g_1(M_X) &= g_2(M_X) = g_3(M_X) = g_U\,, \cr
M_1(M_X) &= M_2(M_X) = M_3(M_X) = m_{1/2}\,.  \cr }\eqn\gunif
$$
where $M_X \lsim M_P$ is the grand unification scale.  Note that the
$g_i$ evaluated at $\mz$ are related to the low energy gauge
couplings as specified in eq.~\gaugecoups.
The relation between $g_1$ and $g_2$
is consistent with the canonical normalization of the
generators $\Tr\,T^aT^b = \half \delta^{ab}$ at the grand unified scale.

\noindent
The renormalization group (RG) scaling of the $g_i$ are given in
Table~\rgeone.  The RG scaling of the $M_i$ are also
simple since the ratios
$M_i/g_i$ are independent of scale.  Thus,
$$
  M_i (Q^2) = m_{1/2}\left[{g^2_i (Q^2) \over g^2_U} \right]\,.
\eqn\mgaugino
$$
In terms of the gaugino mass parameters at the electroweak scale,
$$
  {M_1(m^2_Z)\over M_2 (m^2_Z) } =
  {g^2_1(m^2_Z)\over g^2_2 (m^2_Z) } = {5\over 3}
    {g'^2 \over g^2}\,.\eqn\gutmassrelation
$$
That is, the low-energy gaugino mass parameters satisfy
$$
  M_3 = {g^2_s\over g^2} M_2,\qquad M_1 = {5\over 3}
        {g'^2\over g^2} M_2\,,\eqn\gauginomassrelation
$$
which reduces three independent gaugino mass parameters to one.
Since the gluino
mass is equal to $|M_3|$, it is common to let the gluino mass be the
free parameter of the model.  The other two gaugino mass parameters
are then determined.  The assumption that the gaugino mass parameters
satisfy eq.~\gauginomassrelation\ is so prevalent in the
literature that it is usually taken for granted without
explicit acknowledgement.  Effectively, eq.~\gauginomassrelation\
has been incorporated into the definition of the MSSM.  Nevertheless,
there are examples of models in which eq.~\gauginomassrelation\
does not hold\refmark\kanequiros.
\smallskip
\item{(ii)}
Universal diagonal scalar masses
$$\eqalign{%
  &M^2_{\wt Q}(M_P) = M^2_{\wt U}(M_P)
  = M^2_{\wt D}(M_P) = M^2_{\wt L}(M_P)
  = M^2_{\wt E} (M_P) = m_0^2 {\bf 1} \,,\crr
  &m^2_1(M_P) = m^2_2(M_P) = m_0^2 \,. \cr}\eqn\plancksqmasses
$$

\noindent
In Lecture 1, I indicated that in spontaneously broken
supergravity models, universal scalar masses at the Planck scale
automatically
arise in models with canonical kinetic energy terms.  However,
there is no strong theoretical reason to limit the kinetic energy
terms in this manner.  For example, in superstring-motivated
models, the kinetic energy terms are expected in general to be
of a more complicated form\refmark\stringsusy,
resulting in deviations from
eq.~\plancksqmasses.  On the other hand,
phenomenological considerations at the low-energy scale (in
particular, the observed suppression of FCNCs) suggest that
deviations from eq.~\plancksqmasses\ had better be small.
\smallskip
\item{(iii)}
Universal $A$-parameters
$$
  A_t(M^2_P) = A_b(M^2_P) = A_t(M^2_P) = A_0 {\bf 1}\,.\eqn\planckaparm
$$

\noindent
 As above, this assumption is automatically satisfied only in
 the simplest models.

\item{(iv)}
The $\mu$ and $B$-parameters
$$\eqalign{%
   \mu(M^2_P) &= \mu_0\,, \cr
   B(M^2_P) &= B_0 \,,\cr}\eqn\muandbparm
$$
where $m^2_{12} \equiv \mu B$.

\noindent
Some model builders go further and
impose an additional relation between $A_0$ and $B_0$.  For example, in
the simplest Polonyi model,  $B_0 = A_0 - 1$.  However, given that
the fundamental mechanism of supersymmetry breaking is far from
being understood, it is prudent to regard $A_0$ and $B_0$ as
independent parameters.

\smallskip
\item{(v)}
Unification of Higgs-fermion Yukawa couplings
$$h_b(M_X) = h_\tau(M_X)\,.\eqn\mbmtau$$

\REF\gutyukawas{B. Ananthanarayan, G. Lazarides and Q. Shafi,
{\sl Phys. Rev.} {\bf D44} (1991) 1613; S. Kelley, J.L. Lopez and
D.V. Nanopoulos, {\sl Phys. Lett.} {\bf B274} (1992) 387; J. Ellis,
S. Kelley and D.V. Nanopoulos, {\sl Nucl. Phys.} {\bf B373} (1992) 55.}
\noindent
This relation corresponds to the
well known prediction in SU(5) grand unified models of $m_b = m_\tau$
at $M_X$, which yields $m_b \simeq 3m_\tau$ at the electroweak scale.
Analogous relations involving Yukawa couplings from the first two
generations
do not work.  There are ways to alter the corresponding relations
for the other two generations by introducing additional Higgs multiplets
and arranging appropriate generational dependent couplings at the
grand unification scale.  A recent example of such an attempt can be
found in ref.~[\fermimass].  On the other hand, in superstring
models, relations such as eq.~\mbmtau\ may not hold.  It will
quite interesting to determine the phenomenological viability of
eq.~\mbmtau.  A slightly more radical unification condition
$h_b(M_X) = h_\tau(M_X) = h_t(M_X)$ has also been
studied\refmark{\gutyukawas,\twoloop}.
This relation emerges naturally in certain SO(10) grand unified
models, and leads to a low-energy value of $\tan\beta$
near its upper limit, $\tan\beta \simeq m_t/m_b$.

\REF\zwirneretal{G. Gamberini, G. Ridolfi and F. Zwirner,
{\sl Nucl. Phys.} {\bf B331} (1990) 331.}
In summary, what I have just described is an approach,
sometimes called the minimal low
energy supergravity model (MLES)\refmark\sugramodel,
in which the Planck scale
parameters (which play the role of initial conditions for the
RGEs that determine the low-energy MSSM parameters) are completely
determined by the following
$$
  g_U, m_{1/2}, m_0, \mu_0, A_0, B_0, h_t(M_X), h_b(M_X), h_\tau(M_X)\,.
\eqn\mlesparms
$$
In particular, the five supersymmetric input parameters, $m_{1/2}$,
$m_0$, $\mu_0$, $A_0$ and $B_0$ are sufficient to predict all
low-energy MSSM parameters listed at the beginning of this lecture!
In order to carry out this program, one must use the full set of
RGEs to evaluate the MSSM parameters at the electroweak scale.
In addition, one must make sure that electroweak symmetry breaking
takes place at the correct energy scale.
In the MLES, it is easy to check that SU(2)$\times$U(1)
breaking does not occur at tree-level (\ie, using Planck
scale parameters).  Namely,
$$\eqalign{%
  m^2_{1H} &= m^2_{2H} = m^2_0 + \mu^2_0\,, \cr
  m^2_{12} &= \mu_0B_0\,, \cr}\eqn\planckhiggsmasses
$$
which does not simultaneously satisfy the two requirements specified in
eq.~\higgsconditions.  However, when RG-evolution is employed the
resulting low-energy values for $m_{1H}^2$, $m_{2H}^2$ and $m_{12}^2$
do satisfy eq.~\higgsconditions, and SU(2)$_L\times$U(1)$_Y$
symmetry breaking is triggered.  This is an example of
radiative symmetry breaking generated by summing leading-%
logarithmic corrections to the scalar potential.
The basic mechanism underlying the electroweak
symmetry breaking can be traced to the
following diagrams that contribute to the radiative
corrections to $m^2_{2H}$.
\vskip12pt
\vbox{%
\plotpicture{\hsize}{3cm}{tasidiag.topdraw}
}
\vskip6pt\noindent
These diagrams would cancel in the supersymmetric limit.
But because supersymmetry is broken an incomplete cancellation takes
place.  The result is
$$
  \delta m^2_{2H} \simeq {-3\phantom{1}\over 16\pi^2}\ h^2_t M^2_Q \ln
    \left({M^2_P\over M^2_Q} \right) \,.\eqn\deltamtwoh
$$
Thus, for large $h_t$, it is possible to drive $m^2_{2H}$
negative and trigger SU(2)$_L\times$U(1)$_Y$ breaking.
Moreover, since the parameter $m_{2H}^2$ evolves logarithmically,
it is possible to arrange an exponential hierarchy between the
Planck scale and the electroweak scale.  To carry out all the
details requires a full numerical RGE analysis, along with a
careful treatment
of supersymmetric thresholds as one passes below the masses of
the various supersymmetric particles.  The question of where one
should stop the RG evolution is a subtle one.
See ref.~[\zwirneretal] for a detailed discussion of these points.

Based on the discussion just presented, it is easy to see why
$\tan\beta > 1$ in the MLES approach.\foot{The MLES parameters [eq.~%
\mlesparms] and the conditions of radiative symmetry
breaking determine the low-energy values of $\mz$ and $\tan\beta$.}
Starting from the Higgs potential [eq.~\mssmhiggspot]
one can easily work out the minimum conditions:
$$\eqalign{%
  m^2_{1H} &= m^2_{12}{v_2\over v_1} + \eighth (g^2+g'^2)
               (v^2_2-v^2_1)\,. \cr
  m^2_{2H} &= m^2_{12}{v_1\over v_2} - \eighth (g^2+g'^2)
               (v^2_2-v^2_1)\,. \cr}\eqn\minconditions
$$
Eliminating $m^2_{12}$ from these equations, and solving
for $\tan\beta = v_2/v_1$, one finds
$$
  \tan^2\beta = {m^2_{1H} + \half m^2_Z \over
                 m^2_{2H} + \half m^2_Z}\,,\eqn\tansqbeta
$$
where $m^2_Z = \frac14(g^2+g'^2)(v^2_1+v^2_2)$.  In the MSSM, the
$b$-quark couples to $H_1$ and the $t$-quark couples to $H_2$.
Assuming $h_t>h_b$, it follows that $m_{2H}^2$
is driven negative in the RG-evolution as explained above.
In particular, $m^2_{2H} < m^2_{1H}$ at the low-energy scale,
which implies [using eq.~\tansqbeta] that $\tanb>1$.  On the other
hand, if $h_t<h_b$ then $m_{1H}^2$ would be driven negative first,
and one would conclude that
$m^2_{1H} < m^2_{2H}$ at the low-energy scale.  Applying
eq.~\tansqbeta\ would lead to $\tanb<1$ which contradicts the
initial assumption of $h_t<h_b$.  Thus, the radiative symmetry
breaking mechanism is possible only if $h_t>h_b$ which implies that
$\tanb<m_t/m_b$.  Thus we have demonstrated the result previously
quoted in eq.~\tanbinequality.

\smallskip\point
{\sl What are the implications of coupling constant unification for
the MSSM?}

\REF\amaldiold{U. Amaldi \etal, {\sl Phys. Rev.} {\bf D36} (1987) 1385.}
\REF\drtj{M.B. Einhorn and D.R.T. Jones, \NP B196&82&475&.}
\REF\susyprotonfive{N. Sakai and T. Yanagida, {\sl Nucl. Phys.}
{\bf B197} (1982) 533;
J. Ellis, D.V. Nanopoulos and S. Rudaz, \NP B202&82&43&.}
\REF\nathsusyproton{R. Arnowitt and P. Nath, {\sl Phys. Rev. Lett.}
{\bf 69} (1992) 725; {\sl Phys. Lett.} {\bf B287} (1992) 89;
NUB-TH-3056/92 (1992);
J. Hisano, H. Murayama and T. Yanagida, Tohoku preprint TU-400 (1992).}
\REF\susygrand{%
U. Amaldi, W. de Boer and H. Furstenau, {\sl Phys. Lett.} {\bf B260}
(1991) 447; U. Amaldi \etal, {\sl Phys. Lett.} {\bf B281} (1992) 374.}
\REF\ekn{J. Ellis, S. Kelly and D.V. Nanopoulos, {\sl Phys. Lett.}
{\bf B287} (1992) 95; {\sl Nucl. Phys.} {\bf B373} (1992) 55.}
\REF\zich{F. Anselmo, L. Cifarelli, A. Peterman, and A. Zichichi,
{\sl Nuovo Cim.} {\bf 104A} (1992) 1817; {\bf 105A} (1992) 581;
{\bf 105A} (1992) 1201.}
Grand unified theory (GUT)
extensions of the Standard Model were studied
extensively in the 1970s\refmark\gutproton.
During the 1980s, experimental
information of two types contributed to the demise of the
simplest grand unification models based on SU(5) [or SO(10) with
no intermediate gauge breaking scale].  First, no proton decay
($p\to e^+\pi^0$ was the expected dominant mode) was observed
at the rates predicted by the model.  Second, experimental
measurements of $\sin^2\theta_W$ disagreed with the GUT predictions.
In the latter case, the disagreement was only of order 10\%; however,
as time progressed and the experimental measurements of $\sin^2\theta_W$
became more accurate, the disagreement between theory and experiment
became more significant.  Amaldi and co-workers\refmark\amaldiold\
presented a comprehensive analysis of neutral current parameters
in 1987 (predating the precision electroweak measurements at LEP).
Moreover, their work
provided compelling evidence for ruling out the simplest GUT
models.  In that same paper, Amaldi \etal\ noted, almost parenthetically,
that the prediction of the minimal GUT extension of the MSSM for
$\sin^2\theta_W$ was in very good agreement with the experimentally
measured value.\foot{This result had been long appreciated
by theorists; see \eg, ref.~[\drtj].}
Moreover, it had been already noted in the literature
that the experimental limits on proton decay
were less problematical for supersymmetric GUT
models\refmark{\matterparity,\susybl,\drtj-\nathsusyproton}.

With precision electroweak LEP data now available, the viability of
the various GUT models can be reanalyzed.  Not surprisingly, the
conclusions have not changed.  A new analysis by
Amaldi \etal\refmark\susygrand\
inspired a number of groups to take a further look at the
unification of coupling constants in the Standard Model and
its extensions\refmark{\ekn,\zich}.
Instead of predicting $\sin^2\theta_W$ from some
specific GUT model, these new analyses begin with the determination
of the low-energy gauge couplings (and other low-energy parameters)
{}from experiment.  The most complete analyses then employ two-loop
renormalization group equations of the Standard Model and the MSSM
and ask whether the three gauge couplings meet at some high energy
scale.  The coupling constant evolution in the MSSM depends on the
precise value of the supersymmetric masses.  In first approximation,
one can assume that all supersymmetric particle masses are roughly
degenerate of order $\msusy$.  Then, the question of coupling
constant unification is studied as a function of $\msusy$.
The end result of the RGE analysis confirms the results of
ref.~[\amaldiold]:
the three gauge coupling constants ($g_1$, $g_2$ and $g_3$)
do not unify in the Standard Model, whereas in the MSSM the three
gauge coupling constants meet at a single point if
$M_{\rm SUSY} \simeq {\cal O}(1~{\rm TeV})$.  To quote a specific
example, the analysis of ref.~[\zich] finds that:
$$\eqalign{%
&\msusy=10^{3.4\pm0.9\pm0.4}~\gev\,,\cr
&M_X=10^{15.8\pm0.3\pm0.1}~\gev\,,\cr
&\alpha_U^{-1}=26.3\pm 1.9\pm1.0\,,\cr}
\eqn\susygutvalues$$
where $\alpha_U\equiv g^2_U/4\pi$.

The announcement of the result that $\msusy \sim {\cal O}(1$ TeV)
created a great stir
in the particle physics community.  But, before being carried
away by all the hype, consider the reactions of the optimist,
the pessimist, and the cynic.  The optimist says: the unification
of coupling constants is the first experimental verification of
the low-energy supersymmetric scenario.  The pessimist says:
the unification of coupling constants only rules out the simplest
GUT extensions of the Standard Model.  It may imply new physics
at any scale between the weak scale and the GUT scale and says
nothing about TeV scale physics.  The cynic says: in the GUT
extension of low-energy supersymmetry, there are three unknown
parameters: $g_U$ (the unified coupling constant at the GUT scale),
$M_X$ (the GUT scale or unification point) and $\msusy$.  Thus the
RGEs for $g_1$, $g_2$ and $g_3$ provide three equations and three
unknowns.  A unique solution is essentially guaranteed, so the
unification of coupling constants is no surprise at all.
The optimist clearly overstates the (experimental) case for
supersymmetry.  On the other hand, the pessimist admits that
the unification of couplings implies that the desert hypothesis
of no new physics between the electroweak scale and the GUT scale is
incorrect.  New physics must enter somewhere between $\mz$ and
$M_X$.  This is an exciting result!  Clearly, low-energy supersymmetry
is one possible model for such new physics.  Although there is no
guarantee that the new physics is associated with the TeV scale,
the arguments based on the hierarchy and naturalness problems of the
Standard Model strongly suggest that new TeV scale physics must exist.
The simplest possible scenario would be one in which this TeV scale
physics also accounts for the unification of couplings.  Finally, the
cynic's remarks that the unification of couplings is guaranteed
is technically true (if we ignore the effects of  supersymmetric
thresholds).  Nevertheless, in solving the RGEs for $\msusy$ and
$M_X$, there was no guarantee that the coupling constant unification
that emerges would be
consistent with sensible values of these parameters.  Perhaps,
the results quoted in eq.~\susygutvalues\ are coincidental.
Nevertheless, the fact that such values correspond precisely to the
expected range of a successful GUT extension of low-energy supersymmetry
may be more than coincidental and should not be simply dismissed.

If we are to take the unification of coupling constants as a strong
hint for low-energy supersymmetry, we must address a number of
key questions.  How significant a result
is the unification of coupling constants for low-energy
supersymmetry?  How reliable is the result that the
coupling constants unify?  After all the experimental value of
$\alpha_s \equiv g^2_s/4\pi$ is not known very precisely.
Can one really determine $M_{\rm SUSY}$?  Surely, in a realistic
MSSM, the supersymmetric spectrum is complicated, with multiple
mass thresholds.  How important are the unknown thresholds
to the results of
the analysis?  Finally, the unification of coupling constants
has been achieved with no mention of a particular GUT.  Does
the physics at the GUT scale modify the point of unification?
What about proton decay in such models?

\REF\threshes{R. Barbieri and L.J. Hall, {\sl Phys. Rev. Lett.}
{\bf 68} (1992) 752;
L.J. Hall and U. Sarid, {\sl Phys. Rev. Lett.} {\bf 70} (1993) 2673;
P. Langacker and N. Polonsky, {\sl Phys. Rev.} {\bf D47} (1993) 4028.}
I shall not answer these questions fully here.  But the following
remarks may be useful.  First, the
uncertainty in $\alpha_s$ is definitely a concern.  The
analysis of ref.~\susygrand\ took
$\alpha_s = 0.113 \pm 0.005$.  More recent data from LEP
suggest a somewhat higher value.  The experimental value for
$\alpha_s$ also depends on the particular analysis used to
extract it.  For example, the value of $\alpha_s$ obtained from
jet event topologies is somewhat lower than the value deduced
{}from $\Gamma(Z\to$ hadrons).   Anselmo \etal\refmark\zich\ considered
the maximum possible range of uncertainty for $\alpha_s$ and
examined its impact on the output value of $\msusy$.  They quote
$$
  10^{0.5 \pm 0.5} < M_{\rm SUSY} < 10^{5\pm 1}\ \gev\eqn\msusyrange
$$
for $0.102\pm 0.008 \leq \alpha_s \leq 0.134\pm 0.008$.

\REF\pdg{K. Hisaka \etal\ [Particle Data Group], \sl Phys. Rev. \bf
D45 \rm  (1992) S1.}
The above result suggests that $\msusy$ is not really well fixed
by the coupling constant unification analysis.  Additional
uncertainties also limit how well one can determine $\msusy$.
These include unknown threshold effects at the GUT scale
as well as the effects of a realistic low-energy supersymmetric
spectrum with multiple thresholds between, say, 100~GeV and
1~TeV\refmark\threshes.
Recently, there have been some attempts to include
the effects of a  realistic set of low-energy supersymmetric
thresholds.  For example, Ross and Roberts\refmark\rossrob\
present an ambitious
analysis in the context of the MLES model.  They begin with Planck
scale boundary conditions, and use RGE evolution to determine the
MSSM low-energy parameters.  They then check to see whether the
resulting low-energy gauge couplings unify when scaled back up to
the GUT scale (which is required to lie below the Planck scale).
By iterating this procedure, they insure that the end result is
a low-energy MSSM model whose gauge coupling constants unify
at a reasonable energy scale.  By such a procedure, they
automatically include the effects of the low-energy supersymmetric
thresholds.

\REF\paul{P. Langacker, in {\it Electroweak Physics Beyond the
Standard Model}, Proceedings of the International Workshop on Electroweak
Interaction Beyond the Standard Model, Valencia, Spain, Oct 2-5, 1991,
edited by  J.W.F. Valle and J. Velasco (World Scientific, Singapore,
1992) p.~75.}
Finally, consider the coupling constant unification
in the context of grand unification.  The proton is unstable
in such grand unified models,
so one must check that a specific GUT is not
in conflict with the experimental limit,
$\tau(p\to e^+\pi^0) > 5.5 \times 10^{32}$~yr\refmark\pdg.
I have already
noted that in the simplest (nonsupersymmetric) SU(5) grand
unified model, the predicted proton lifetime is too short.  The
reason for this is related to the value of $M_X$.  In the original
SU(5) models\refmark\gutproton,
typical values for $M_X$ (based on approximate
coupling constant unification) were around
$5\times 10^{14}$~GeV.  The proton
decays via the exchange of superheavy gauge bosons whose mass is
proportional to $M_X$.  The theoretically predicted rate for
proton decay scales as $1/M_X^4$.  In the non-supersymmetric GUT model,
$\tau(p\to e^+\pi^0)\simeq  10^{31\pm 1}$yr, which is clearly
ruled out.  In the GUT model extension of low-energy supersymmetry,
$M_X$ is approximately two orders of magnitude larger than in the
case of the original SU(5) model.  As a result, in the supersymmetric
model, $\tau(p\to e^+\pi^0)\gsim 10^{36}\ {\rm yr}$,
which is clearly not in conflict with the present data.
However, in the supersymmetric model, genuinely new diagrams
contribute to proton decay\refmark{\susybl,\susyprotonfive};
one example is shown in the figure below.
\vskip15pt
\vbox{%
\plotpicture{\hsize}{3cm}{dimen5.topdraw}
}
\vskip12pt
The loop diagram on the left yields an effective dimension-5 operator.
Its contribution to the proton decay rate
scales only as $1/M_X^2$ and could lead to a partial decay rate
larger than the present experimental limit.  In the case of the
diagrams shown above, extra factors of light fermion masses
due to the higgsino couplings to the quarks play an
important role in limiting the size of these diagrams to an acceptable
level.  The diagrams above provide a mechanism for
$p\to \bar\nu K^+$, which will be a dominant decay
channel in the supersymmetric model.
Present experimental limits are
$\tau(p\to \bar\nu K^+) > 10^{32}$~yr\refmark\pdg.
Interesting constraints on MLES model parameters
based on the requirement that the
predicted proton partial decay rates
are not in conflict with present experimental limits
have been recently obtainly in ref.~[\nathsusyproton].

\TABLE\sinthetaw{}
Finally, it is worth noting that the unification of coupling
constants places a number of interesting restrictions on
non-minimal extensions of the MSSM (which I abbreviate by
NMSSM).  Returning to the more traditional approach, I compare
the predictions for $\sin^2\theta_W$ in the SU(5) grand unified
model in the case of the Standard Model (SM) with $N_H=1$ or 2
light Higgs doublets, and in the case of the MSSM (with $N_H=2$)
or a non-minimal extension (NMSSM) with $N_H=4$\refmark\drtj.
In the computation
of $\sin^2\theta_W$, which is defined in the $\overline{MS}$-scheme
and evaluated at $\mz$, I have taken $\alpha_s = 0.118\pm 0.007$
and $\msusy=1$~TeV.  The results, taken from ref.~[\paul], are shown in
Table~\sinthetaw\
and should be compared with
$\sin^2\theta_W(m_Z) = 0.2328\pm 0.0007$ based on a recent analysis of
LEP data\refmark\langtasi.
Clearly, the MSSM is the only model in the above list that is
compatible with the experimental data.

\midinsert
\centerline{\bf Table \sinthetaw.\enspace
Predictions for $\bold{\sin^2\theta_W}$\refmark\paul}
\vskip6pt
\thicksize=0pt
\tablewidth=12cm
\begintable
Model&  $N_H$&  $\sin^2\theta_W(m_Z)$ \crneg{-5pt}
SM \hf&  1&    $0.2106\pm 0.0020$ \nrneg{5pt}
SM \hf&  2&    $0.2147\pm 0.0020$ \nrneg{5pt}
MSSM \hf&  2&  $0.2320\pm 0.0017$ \nrneg{5pt}
NMSSM \hf&  4& $0.2530\pm 0.0015$ \cr
&& \endtable
\endinsert

\section{Experimental Searches for Supersymmetry}

One of the important
goals of the present and
next generation of colliders (Tevatron, LEP, LEP-II, LHC, SSC and
the next $\epem$ linear collider [NLC])
is to unambiguously answer
the question---does low energy supersymmetry exist?  If the
answer is yes, then experiments at those colliders will begin
to measure the parameters of low-energy supersymmetry.  One will
be able to check directly whether the low-energy supersymmetric
model can be characterized as an MSSM, or whether additional
non-minimal structure exists.  Furthermore, one will be
able to test a variety of theoretical assumptions and prejudices
outlined in the previous section.  It will be exciting to determine
the implications of the low-energy supersymmetric parameters for
the underlying fundamental theory in order to extract information
about the physics at very high energy scales.

\TABLE\facilities{}
\midinsert
\centerline{\bf Table \facilities.\enskip Present and Future Facilities}
\smallskip
\def\tstrut{\vrule height 2.8ex depth 1.0 ex width 0pt}
\thicksize=0pt
\tablewidth=\hsize
\begintable
&&&& Dates of \nrneg{8pt}
Machine& Type& $\sqrt s$~(TeV)& $\cal L$ (pb$^{-1}$/year)& Operation \cr
Fermilab\hf&&&& \nrneg{8pt}
Tevatron\hf& $p\bar p$& 1.8& 10--100& now---1999\hf\nr
LEP\hf&    $\epem$&    0.1& ($10^6$--$10^7$ $Z$'s)& now---1994\hf \nr
LEP-II\hf& $\epem$&    0.2&  500&     1995--1999 \hf\nr
SSC\hf&    $pp$&        40&  10,000&  start: 2000 \hf\nr
LHC\hf&    $pp$&    10--17&  $10^5$&  start: 2000 (?)\hf \nr
NLC\hf& $\epem$&  0.5&  $2\times 10^4$&   start: 2005 (?)\hf\cr
&&&& \endtable
\vskip-12pt
\endinsert

\REF\kolb{E.W. Kolb and M.S. Turner, {\it The Early Universe}
(Addison-Wesley Publishing Company, Reading, MA, 1990).}
\REF\dark{S. Kelley, J. L. Lopez, D.V. Nanopoulos, H. Pois and K. Yuan,
{\sl Phys. Rev.} {\bf D47} (1993) 2461; R.G. Roberts and L. Roszkowski,
 RAL-93-003 (1993).}
\REF\wimp{D.O. Caldwell, {\sl J. Phys.} {\bf G17} Suppl. (1991) S325.}

In this section, I shall only touch on highlights of
the phenomenology of supersymmetry.  For further details, see
refs.~[\haberkane] and [\xtata].
The conservation of $R$-parity in the
MSSM plays a key role in determining
the phenomenology of supersymmetric particle production and
detection at present and future colliders.
$R$-parity conservation implies that all Standard Model
particles (including all Higgs bosons)
have $R=1$, while their supersymmetric partners have $R=-1$
[see eq.~\ztwo].  Four important consequences that are
relevant for phenomenology follow.
\smallskip
\item{(i)}
The lightest supersymmetric particle (LSP) is stable.

\noindent
The LSP is almost certainly color and
electrically neutral.  This leaves two favored candidates:
the lightest neutralino state ($\widetilde\chi^0_1$) or the lightest
sneutrino ($\tilde\nu$).\foot{%
If the gravitino ($\tilde g_{3/2}$) is the LSP, then the second
lightest supersymmetric particle will be very long-lived.  In collider
experiments, we can forget about the gravitino; the second lightest
supersymmetric particle will play the same role as the LSP does in
the more conventional scenario.}
The LSP also plays a very
important role in cosmology---it is a prime candidate for the dark
matter\refmark\kolb.
There is evidence that non-baryonic dark matter
must exist throughout the universe.  Moreover, theorists strongly
believe that the total energy density of the universe is precisely
equal to the critical density---the dividing line between an open and
closed universe.  Baryonic matter accounts for no more
than about 10\% of
critical density.  So, it seems plausible to assert that the
energy density of the dark matter is sufficient in order
that $\Omega\equiv
\rho/\rho_c=1$.  Neutrinos (if they had mass) and the LSP are two
possible dark matter candidates.  If one imposes the requirement that
the contribution of the LSP to $\Omega$ is a significant fraction of 1,
then one finds additional constraints on the parameters of
the MSSM\refmark\dark.
Moreover, the properties of the LSP must be compatible with the negative
results of dark matter detection experiments\refmark\wimp.
Due to the coherent
interactions of the $\wt\nu$ with high-$Z$ material, the non-observation
of dark matter candidates rules
out $\Omega_{\tilde\nu} \sim {\cal O}(1)$.
On the other hand, $\widetilde \chi^0_1$ remains a viable dark
matter candidate. Its interactions with matter are incoherent
(since $\wt\chi^0_1$ is a Majorana fermion), so it would not have been
directly observed by existing dark matter searches.  Cosmologically
interesting regions in MSSM parameter space  exist where
$\Omega_{\widetilde \chi^0_1} \sim {\cal O}(1)$.

\goodbreak
\smallskip
\item{(ii)}
Supersymmetric particles are produced in pairs at colliders.

\noindent
In collider experiments, a collision is initiated by the
scattering of two $R$-even particles (\eg, $\epem$,
$gg$, $gq$, $qq$, $q\bar q$, \etc).
Thus, the final state must be $R$-even, which implies that only
an even number of $R$-odd particles can be produced in a collision.
\smallskip
\item{(iii)}
All supersymmetric particles (excluding the LSP) are unstable
and decay.

\noindent
When an $R$-odd particle decays, the
decay products must contain an odd number of $R$-odd particles.
After the decay of
all unstable particles in a given decay chain,
one is left with Standard Model ($R$-even) particles plus an odd
number of LSP's.  Schematically, the end result of the decay of
an unstable supersymmetric particle $\widetilde X$ is
$$
  \widetilde X \to Y + (2n+1) \widetilde \chi\ls{\rm LSP}
   \qquad n = 0,1,2,\dots\eqn\susydecay
$$
where $Y$ consists of $R$-even particles.
Since $\widetilde X$ is heavy, it will decay rapidly
(even if its decay amplitude is of weak interaction strength),
leaving no gaps in collider detectors.
\smallskip
\item{(iv)}
The LSP interacts very weakly with matter.

\noindent
The LSP interacts with matter via the exchange of a heavy virtual
supersymmetric particle, as required by $R$-parity invariance,
since at each interaction vertex an even number of $R$-odd
particles must appear.
Thus, in collider detectors the LSP behaves like a neutrino.

As a result of the four basic properties just enumerated, it
follows that one or more LSP's will always be
produced in the decay chain of any unstable supersymmetric
particle.  The LSP then escapes the collider detector without
being directly observed.  Thus, one of the hallmark
signatures of supersymmetric particle production is the ``observation''
of missing (transverse) energy.\foot{In hadronic colliders, where the
total energy of a given hard collision is not known, it is not possible
to detect energy that escapes in a direction longitudinal to the beam
direction.  Transverse missing energy can be inferred from
the detection of an imbalance in the total
momentum transverse to the beams.}
To prove that a missing transverse energy signal was evidence for
physics beyond the Standard Model, one would have to prove that the
missing energy could not be accounted for by neutrinos or fake
missing energy generated by mismeasurement.

\TABLE\susylimits{}
\REF\cdfsusy{F. Abe \etal\ [CDF Collaboration], {\sl Phys. Rev. Lett.}
{\bf 69} (1992) 3439.}
At hadron colliders, the most important supersymmetric particle
searches involve the production of gluinos and squarks.  Since these are
colored (strongly interacting) particles, they are
produced with the largest cross sections.  In contrast, at
$\epem$ colliders,
all particles with non-zero electroweak charge are produced
with comparable cross sections (of order 1 ``unit" of $R$).  These
include squarks, sleptons, sneutrinos, charginos and neutralinos.  The
properties of the various signatures depend in detail on the
various decay branching ratios.  The missing transverse energy
signature is only one of many possible distinguishing features
of the supersymmetric particle decay chains.
Present bounds on supersymmetric particle masses have been obtained
{}from a variety of experiments at $e^+e^-$ and  $p\bar p$ colliders.
No signal has yet been found above Standard Model backgrounds.
A comprehensive list of supersymmetric particle searches has been
compiled by the Particle Data Group\refmark\pdg.
More recently, new limits
on squark and gluino masses (obtained at the Tevatron) have
been published by the CDF Collaboration\refmark\cdfsusy.

\REF\barnetthaber{R.M. Barnett, J.F. Gunion,
and H.E. Haber, {\sl Phys. Rev. Lett.} {\bf 60} (1988) 401;
{\sl Phys. Rev.} {\bf D37} (1988) 1892; H. Baer, X. Tata, and
J. Woodside, {\sl Phys. Rev. Lett.} {\bf 63} (1989) 352;
{\sl Phys. Rev.} {\bf D41} (1990) 906; {\bf D44} (1991) 207.}
A word of caution is in order when considering the published
limits on squark and gluino masses at $p\bar p$ colliders.
First, mass limits are sometimes given assuming
that the $\tilde q$ and/or $\tilde g$ decays exclusively consist of
{\it direct} decays into $\wt\chi^0_1$
(plus hadronic jets).  In this case, $\wt\chi^0_1$ (which
is assumed to be the LSP) may carry off
substantial ``missing'' transverse energy which provides the
experimental signature for these events.  However, one must also
consider the possibility of $\tilde g$ and $\tilde q$
``cascade" decays.  These are decays
in which the $\tilde g$ and/or $\tilde q$
first decays into a heavier neutralino or chargino state,
which then subsequently decays and produces a $\wt\chi^0_1$ at the
end of the decay chain.
In cascade decays, the missing transverse energy is
on average softer than in direct decays to $\wt\chi^0_1$. However,
the probability of cascade decays increases as the squark
and gluino masses
become larger\refmark\barnetthaber.
As a result, Tevatron limits on
squark and gluino masses based on an assumption of no
cascade decays are overestimates of the true experimental
mass limits.  Second, limits on squark masses are obtained
at the $p\bar p$ colliders under the assumption of
five flavors of mass-degenerate $\tilde q\ls L$
and $\tilde q\ls R$.  Omitted from this list are $\tilde t_L$
and $\tilde t_R$, since $\tilde t_L$--$\tilde t_R$ mixing
may significantly lower the mass of the lightest $\tilde t$
eigenstate.  A mass limit (from $p\bar p$ collider data)
for the $\tilde t$, assuming that it is lighter than all other
squarks, has not been published.

The present limits on
supersymmetric particles is summarized in Table~\susylimits.
These mass limits are still far away
{}from the ${\cal O}(1~{\rm TeV})$ upper limit expected from
naturalness arguments.  Only the future colliders (LHC, SSC, NLC and
beyond) will have sufficient energy to fully probe the TeV region.
These machines will either discover evidence for supersymmetric
particles or rule out the low-energy supersymmetry scenario.

\REF\brahm{D.E. Brahm, L.J. Hall and S.D.H. Hsu,
{\sl Phys. Rev.} {\bf D42} (1990) 1860.}
\REF\hidaka{K. Hidaka, {\sl Phys. Rev.} {\bf D44} (1991) 927.}
\pageinsert
\vskip12pt
        \parskip=0pt  \parindent=0pt   \itemsize=18pt
\centerline{\bf Table \susylimits.
Present Limits on Supersymmetric Particles$^*$}
\vskip8pt
\def\tstrut{\vrule height 3.0ex depth 1.1 ex width 0pt}
\thicksize=0pt
\tablewidth=\hsize
\parasize=7cm
\begintable
& \quad Bound on&&        \nrneg{8pt}
& \quad Particle Mass&&        \nrneg{8pt}
Particle& \quad(GeV)&& Source \crneg{-5pt}
$\widetilde \chi^0_1$&  18.4 & & \nrneg{16pt}
&&\hf \vtop{\hbox{$\left\{\vbox to 1.4cm{}\right.$}} \hskip-2pc&
                                       \nrneg{6.0pc}
&&&             \para{Based on the LEP non-observation of
              $\wt\chi^0_i$ and $\wt\chi_1^+$, CDF non-observation
              of $\tilde g$, and the assumption of gaugino mass
              unification.$^{\dag}$} \nrneg{1.8cm}
$\wt \chi^0_2$&  45& &                        \nr
$\widetilde \chi^0_3$&  70&&                  \nr
$\widetilde \chi^0_4$& 108&&                  \nr

$\wt \chi^\pm_1$&   45.2 &&   LEP \hf \nr
$\wt \chi^\pm_2$&   99 && See neutralino mass limits above.\hf\nr

$\tilde\nu$& 41 &&\para{Assumes the $ \tilde\nu$ decays are invisible
                       (otherwise $M_{\tilde\nu} <$ 32 GeV).
                       Based on LEP measurement of $\Gamma(Z\to$
                       invisible final states).} \nr
$\tilde e$&  45 &&\para{LEP; assumes  $M_{\wt \chi^0_1}<41$ GeV%
                        $^{\ddag}$ }\nr
$\tilde\mu$&  45 && \para{LEP; assumes  $M_{\wt \chi^0_1}
                        < 41$ GeV$^{\ddag}$ }\nr
$\tilde \tau$& 45 &&\para{LEP; assumes  $M_{\wt \chi^0_1}
                        < 38$ GeV$^{\ddag}$ }\nr
$\tilde q$&  45 && LEP; assumes $M_{\wt \chi^0_1} < 20$ GeV\hf \nr
     &  74 && UA2 (any $M_{\tilde g})^\star$\hf \nr
     & $\sim$95\9 && CDF ($M_{\tilde q} < M_{\tilde g})^\star$\hf \nr
$\tilde g$&  79 && UA2 ($M_{\tilde g} < M_{\tilde q})$\hf \nr
          &  $\sim$95\9 && CDF \hf\crneg{6pt}
&&&\endtable
\item{%
^*} All limits (except the CDF limits on squark and gluino
     masses) are taken from ref.~[\pdg].
\item{%
^{\dag}} Theoretical analysis of ref.~[\hidaka], based on LEP
     non-observation of $Z^0\to \wt\chi^+_1\wt\chi^-_1,$ $\wt\chi^0_1
     \wt\chi^0_2$, the LEP measurements of $\Gamma(Z^0)$ and
     $\Gamma(Z\to$ invisible final states),
     and the CDF Collaboration limit on $M_{\tilde g}$.
     In addition, eq.~\gauginomassrelation\
     relating gaugino Majorana mass parameters is assumed.
\item{%
^{\ddag}} Assumes $M_{\tilde\ell_L} = M_{\tilde\ell_R}$
          and $\tilde\ell\to \ell\,\wt\chi\lsup0_1$.
      If $\tilde\ell$ is stable, then  $M_{\tilde\ell}>40$ GeV.
\item{%
^\star} Assumes five flavors of mass-degenerate
     $\tilde q_L,\tilde q_R$.
\vfill
\endinsert

\section{Alternative Low-Energy Supersymmetry Phenomenologies}

In Lecture 2, I have focused primarily on the parameters of the
MSSM and its phenomenology.  The assumption of $R$-parity
invariance played a central role in establishing the importance
of the missing-energy signature for the MSSM.  However, it is important
to be open to alternatives.  Thus, in this section, I briefly
consider the implications of $R$-parity violating low-energy
supersymmetry.  The $R$-parity violation can be either
spontaneous or explicit.

\REF\valle{J.C. Romao, C.A. Santos and J.W.F. Valle,
{\sl Phys. Lett.} {\bf B288} (1992) 311.}
\REF\merlo{S. Dimopoulos and L.J. Hall, {\sl Phys. Lett.} {\bf B207}
(1987) 210;
S. Dimopoulos, R. Esmailzadeh, L.J. Hall, J.-P. Merlo and G.D. Starkman,
{\sl Phys. Rev.} {\bf D41} (1990) 2099.}
\REF\herbie{H. Dreiner and G.G. Ross,
{\sl Nucl. Phys.} {\bf B365} (1991) 597.}

Spontaneously broken $R$-parity can be
achieved within the context of the MSSM!
Simply arrange the parameters of the scalar potential
so that $\VEV{\tilde\nu_\tau} \neq 0$.  This will
spontaneously break lepton number and R-parity.  For example,
if the supersymmetric parameters are such that
$$
  M^2_{\tilde\nu} = M^2_L + \half m^2_Z\cos2\beta < 0\,,\eqn\snumass
$$
then the sneutrino field will acquire a vacuum expectation value.
Phenomenological limits on lepton number violation imply that
$\VEV{\tilde\nu_\tau} \lsim 10$ GeV,
for $M_{\rm SUSY} \lsim 1$ TeV\refmark\brahm.
However, such a model would possess a Majoron (the Goldstone
boson of spontaneous lepton violation) and an associated
light scalar which would have been detected in $Z$ decay.
Thus, the model just described is not phenomenologically viable.
Realistic models of spontaneous R-parity violation
necessarily involve a non-minimal extension of the
MSSM\refmark\valle.

\TABLE\rparitycouplings{}
Explicitly broken $R$-parity models can be obtained by setting
$W_{NR} \neq 0$ [see eq.~\norparity],
One can set bounds on $\lambda_L,\lambda'_L, \mu'$ and $\lambda_B$
based on B and L violation limits in a variety of Standard Model
processes\refmark{\merlo,\herbie}.
For example, an analysis of ref.~\herbie\ yields upper limits for
$\lambda_L,\lambda'_L$ and $\lambda_B$ scaled by
[$M_{\tilde f}/100$ GeV] given in Table~\rparitycouplings.

\midinsert
\centerline{\bf Table \rparitycouplings.\enspace
Bounds on $\bold R$-parity Violating Couplings\refmark\herbie}
\def\tstrut{\vrule height 2.5ex depth 1.0ex width 0pt}
\thicksize=0pt
\tablewidth=13cm
\begintable
&&&&& \crneg{-3pt}
\multispan2 \tstrut \hf $\widehat L_a \widehat L_b \widehat E_c$\hf&
\multispan2 \tstrut \hf $\widehat L_a \widehat Q_b \widehat D_c$\hf&
\multispan2 \tstrut \hf $\widehat U_a \widehat D_b \widehat D_c$\hf \cr
abc& $\lambda_L <$&  abc&  $\lambda'_L <$&  abc& $\lambda_b<$ \crneg{-5pt}
121&   0.04&         111&  0.01&            112& $10^{-9}$ \nr
122&   0.04&         112&  0.03&  & \nr
123&   0.04&         113&  0.03&  & \nr
131&   0.10&         121&  0.26&  & \nr
132&   0.10&         122&  0.45&  & \nr
133&   0.10&         123&  0.26&  & \nr
231&   0.09&         131&  0.26&  & \nr
232&   0.09&         133&  0.01&  & \nr
233&   0.09&         211&  0.09&  & \nr
   &       &         212&  0.09&  & \nr
   &       &         213&  0.09&  & \nr
   &       &         221&  0.22&  & \nr
   &       &         231&  0.22&  & \cr
&&&&& \endtable
\endinsert

The phenomenology of $R$-parity violating models is radically
different from that of the MSSM.  In contrast to the MSSM,
models of $R$-parity violation exhibit the following features:
(i) the LSP in unstable; (ii)  superpartners can be singly produced; and
(iii) there is a potential for observable $L$ (or $B$) violation in
high-energy collisions.
In particular, the missing energy signature is no longer a
necessary feature of supersymmetric particle production.
See, \eg, ref.~[\merlo] for a detailed treatment of this alternative
phenomenology.
\endpage
\chapter{%
Radiative Corrections to Tree-level Low-Energy \hfill\break
\hbox to 13pt{} Supersymmetry}

\REF\georginatural{H. Georgi and A. Pais, {\sl Phys. Rev.} {\bf D10}
(1974) 539.}
In this lecture, I will examine the influence of radiative corrections
on the MSSM.  Precision electroweak measurements at LEP now permit us
to test the Standard Model to roughly one part in a thousand.  No
deviations from the Standard Model have yet been observed; this
already imposes interesting constraints on models of physics beyond the
Standard Model.  Radiative corrections can also affect the {\it
natural} relations of the MSSM.  These are relations among MSSM
observables that are modified by finite (and
therefore calculable) radiative corrections\refmark\georginatural.
Naively, one might
expect the radiative corrections to natural relations to be small.
However, I shall show that there are very large corrections to
relations among observables in the MSSM Higgs sector.  These results
have an significant impact on the direct searches for Higgs
bosons at present and future colliders.

\section{MSSM Contributions to Precision Electroweak
Measurements}

\REF\decouple{T. Appelquist and J. Carazzone, {\sl Phys. Rev.} {\bf
D11} (1975) 2856; B. Ovrut and H. Schnitzer, {\sl Phys. Rev.} {\bf
D22} (1980) 2518; Y. Kazama and Y.-P. Yao, {\sl Phys. Rev.} {\bf D25}
(1982) 1605; J.C. Collins, {\it Renormalization} (Cambridge University
Press, Cambridge, 1984), Chapter 8.}
\REF\veltman{M. Veltman, {\sl Nucl. Phys.} {\bf B123} (1977) 89;
M.B. Einhorn, D.R.T. Jones, and M. Veltman, {\sl Nucl. Phys.}
{\bf B191} (1981) 146.}
\REF\nondecouple{J.C. Collins, F. Wilczek and A. Zee, {\sl Phys. Rev.}
{\bf D18} (1978) 242.}
\REF\doug{%
D. Toussaint, {\sl Phys. Rev.} {\bf D18} (1978) 1626.}
\REF\cfh{M. Chanowitz, M.
Furman and I. Hinchliffe, {\sl Phys. Lett.} {\bf 78B}
(1978) 285; {\sl Nucl. Phys.} {\bf B153} (1979) 402.}
\REF\sirlin{%
W.J. Marciano and A. Sirlin, {\sl Phys. Rev.} {\bf D22} (1980) 2695
[E: {\bf D31} (1985) 213].}
\REF\abj{G. Altarelli, R. Barbieri, and S. Jadach, {\sl Nucl. Phys.}
{\bf B369} (1992) 3 [E: {\bf B376} (1992) 444].}
\REF\lang{P. Langacker and M. Luo, {\sl Phys. Rev.} {\bf D44} (1991)
817; J. Ellis, G.L. Fogli
and E. Lisi, {\sl Phys. Lett.} {\bf B274} (1992) 456;
D. Schaile, {\sl Z. Phys.}
{\bf C54} (1992) 387; P. Renton, {\sl Z. Phys.} {\bf C56} (1992) 355.}
\REF\screen{M. Veltman, {\sl Acta Phys. Pol.} {\bf B8} (1977) 475;
M.B. Einhorn and J. Wudka, {\sl Phys. Rev.} {\bf D39} (1989) 2758.}
It may be possible to detect deviations from the Standard Model in
precision electroweak measurements due to the effects of the virtual
exchange of particles associated with new physics beyond the Standard
Model.  Typically, the intrinsic mass scale of the new physics
(call it $M$) lies sufficiently above $\mz$ such that we can formally
integrate out the effects of the new ``heavy physics''.  The end
result is that there are two contributions to the radiative
corrections to the energy scale $\mz$.  The first
contribution represents pure Standard Model physics, while the second
contribution reflects the remnant of the heavy physics that has been
integrated out.
The decoupling theorem\refmark\decouple\
implies that radiative corrections to
electroweak observables from such new physics
should be of ${\cal O}(g^2\mz^2/M^2)$.
That is, the virtual effects due to new physics should
formally decouple as
$M\to\infty$.  However, in spontaneously broken gauge theories, an
exception to the decoupling theorem
arises\refmark{\veltman--\doug}.
Suppose we wish to
consider the virtual effects of a certain particle whose mass $m$
is proportional to a dimensionless coupling of the theory.
In this case,
the virtual effects due to such a particle do {\it not} decouple in the
large $m$ limit.
Two examples in the Standard Model are the top quark, whose mass
is proportional to a Higgs-quark Yukawa coupling, and the Higgs boson,
whose mass is proportional to the Higgs self-coupling.

The classic example of non-decoupling can be seen in the $\rho$-parameter
of electroweak physics\refmark{\veltman,\cfh}.
If one defines $\rho\equiv\mw^2/\mz^2\cos^2\theta_W$, then at tree-level
$\rho=1$ in any SU(2)$\times$U(1) model whose
Higgs sector consists entirely of weak scalar doublets (\eg, the
Standard Model and the MSSM).  Since $\rho=1$ is a ``natural'' relation
in such models, the deviation of $\rho$ from one is calculable when
radiative corrections are incorporated.  To proceed, one must
carefully define the observables of the model in order to establish
a useful one-loop definition for $\rho$.  For example, in ref.~[\sirlin],
$\rho\equiv\rho\ls{NC}$
is defined as the ratio of the neutral current
neutrino-nucleon inclusive cross-section relative to the corresponding
charged current cross-section, normalized to 1 at tree-level
[see eq.~(3.23)].
A recent theoretical analysis of LEP data in ref.~[\abj] gives
$\rho= 0.9995\pm 0.0051$.
In the Standard Model, the deviation of
$\rho$ from 1 is predicted to be
quite small unless there exist particles with
nontrivial electroweak quantum numbers that are substantially
heavier than the $Z$ boson.
It is therefore convenient to define
$$\rho\equiv \rho\ls{\rm RSM}+\delta\rho\,,\eqn\refsm$$
where $\rho\ls{\rm RSM}\simeq 1$ is the $\rho$ parameter in a
``reference Standard Model'' (RSM) in which the radiative corrections
to $\rho$ are very small.
In order to exhibit the explicit
dependence of $\rho$ on the top quark and Higgs masses, let us
choose a RSM in which $\mt=\mb$ and $\mhsm=\mz$
(where $\mhsm$ is the mass of the Standard Model Higgs boson).
Then, in the Standard Model,
assuming $\mhsm\gg\mz$\refmark{\veltman,\cfh,\sirlin}
$$
  \delta\rho \simeq{g^2N_c\over 32\pi^2 m^2_W}F(\mt^2,\mb^2)
          -{3 g^2 \over 64 \pi^{2} c\ls{W}^2}
\left[s^2\ls W\ln \biggl(
 {\mhsm^2 \over \mz^2} \biggr)-{c\ls{W}^4\over s\ls{W}^2}\ln\left(
{\mw^2\over\mz^2}\right)-1\right]\,,
\eqn\rhocorrection$$
where $N_c=3$,
$s\ls{W}\equiv\sin\theta_W$, $c\ls{W}\equiv\cos\theta_W$, and the
function $F$ is defined by
$$
F(m_1^2,m_2^2)\equiv\half(m_1^2+m_2^2)-{m_1^2 m_2^2\over m_1^2-
m_2^2}\,\ln\left({m_1^2\over m_2^2}\right)\,.\eqn\effdef$$
In deriving eq.~\rhocorrection,
$\mhsm$ is assumed to be much larger than $\mz$.
As advertised, the decoupling theorem is not respected in the
limit of large top quark and/or Higgs mass.  The quadratic dependence
of $\delta\rho$ on $m_t$ yields a useful constraint on the top quark
mass.\foot{Radiative corrections to other electroweak observables
also depend nontrivially on $m_t$ (\eg, see refs.~[\abj--\lang]).
For example, a recent analysis by Langacker\refmark\langtasi\
yields $\mt<197$~GeV at $95\%$~CL.}
In contrast, the dependence of $\delta\rho$
on $m_{\phi^0}$ is only logarithmic and is therefore not very useful in
constraining the Higgs mass.\foot{This is an example of Veltman's
screening theorem\refmark\screen.}
\REF\oblique{B.W. Lynn, M.E. Peskin and R.G. Stuart, in {\it Physics at
LEP}, edited by J. Ellis and R. Peccei, CERN Yellow Report CERN-86-02
(1986) p.~90;
B. Lynn and D.C. Kennedy, {\sl Nucl. Phys.} {\bf B321}
(1989) 83; {\bf B322} (1989) 1; D.C. Kennedy, {\sl Nucl. Phys.} {\bf
B351} (1991) 81; M.E. Peskin, in {\it Physics at the 100~GeV Mass
Scale}, Proceedings of the 1989 SLAC Summer Institute, edited by
E.C. Brennan, SLAC-Report-361 (1990) p.~71.}

The contributions of physics beyond the Standard Model to electroweak
observables occur primarily through virtual loop corrections to
gauge boson propagators, sometimes called ``oblique''
corrections\refmark\oblique.  As an example, this is typically a
good approximation in the case
of virtual Higgs boson corrections, since the Higgs coupling to
light fermions is suppressed by a factor $m_f/\mw$.  However,
there are a number of cases where one-loop vertex and box
corrections involving the
coupling of charged Higgs bosons to $t\bar b$ are not especially small.
I will consider this possibility briefly in section 3.2.
In this section, I shall work in the oblique approximation and assume
that the radiative corrections to an electroweak observable of
interest are dominated by
virtual heavy particle corrections to gauge boson
propagators.

\REF\tasiradcorr{W.F.L. Hollik, {\sl Fortschr. Phys.} {\bf 38}
(1990) 165;
F. Jegerlehner, in {\it Testing the Standard Model},
Proceedings of the 1990 Theoretical Advanced Study Institute,
Boulder, CO, edited by M. Cveti\v c and P. Langacker (World Scientific,
Singapore, 1991) p.~476.}
\REF\kennedy{D.C. Kennedy, in {\it Perspectives in the Standard Model},
Proceedings of the 1991 Theoretical Advanced Study Institute in
Elementary Particle Physics, Boulder, CO, 3-28 June 1991,
edited by R.K. Ellis, C.T. Hill, and J.D. Lykken (World Scientific,
Singapore, 1992) p.~163.}
\REF\sirlindef{A. Sirlin, {\sl Phys. Rev.} {\bf D22} (1980) 971.}
\REF\sirlinwmass{W.J. Marciano and A. Sirlin, {\sl Phys. Rev.}
{\bf D29} (1984) 945 [E: {\bf D31} (1985) 213.}
Before proceeding with a discussion of the general parametrization
of the oblique corrections, let us recall some basic facts from
the theory of on-shell renormalization of the electroweak
theory\refmark{\tasiradcorr,\kennedy}.
It is sufficient to focus our attention on the renormalization
of the gauge boson propagators. Let
$$i\Pi^{\mu\nu}_{ij}(q)=ig^{\mu\nu}A_{ij}(q^2)+iq^\mu q^\nu
B_{ij}(q^2)\,,\eqn\propagator$$
be the sum of all one-loop Feynman graphs contributing to
the $V_i$--$V_j$ two-point function, where $q$ is the
four-momentum of the vector boson ($V=W,\ Z$ or $\gamma$).
Only the functions $A_{ij}$ are
relevant for the subsequent analysis.  It is convenient to write
$$A_{ij}(q^2)=A_{ij}(0)+q^2F_{ij}(q^2)\,,\eqn\fdef$$
which define the quantities $F_{ij}$.  Gauge invariance
implies that $A_{\gamma\gamma}(0)=0$.
In addition, the sum of new
heavy particle contributions to $A_{Z\gamma}(0)$ also vanishes exactly.
Only gauge boson loops can produce
nonzero contributions to $A_{Z\gamma}(0)$ [in the standard $R$-gauge].
One now defines the gauge boson squared-mass
counterterm, $\delta m_V^2$, by
$$m^2_{V0}=m^2_V-\delta m^2_V\,,\eqn\countert$$
for $V=W$ or $Z$,
where $m_{V0}$ is the bare mass and $m_V$ is the renormalized mass.
The counterterm is fixed by the condition
$$\delta m_V^2= \Re\,A_{VV}(m_W^2)\,.\eqn\onshelcondition$$
This guarantees that the physical gauge boson mass corresponds to the pole
of the renormalized propagator.  Next, we introduce Sirlin's
definition\refmark\sirlindef\
of $\sin^2\theta_W$ in terms of the physical $W$ and $Z$
masses
$$
\sin^2\theta_W = 1-{\mw^2\over \mz^2}\,.\eqn\sirlindeff
$$
In the ``on-shell'' renormalization scheme of Sirlin\refmark\sirlindef,
the input
parameters to the electroweak theory are taken to be $\alpha$ (the
fine-structure constant), $\mz$ and $G_F$.  All other observables
in electroweak physics are predicted in terms of
these three input parameters.

For example, the $W$ mass is one of the predictions of the
electroweak theory,  as I now demonstrate.
Tree-level relations of the model can be regarded as relations among
bare parameters (which are denoted by the subscript 0).  Thus,
$$
  m_{W0}^2 = {\pi\alpha_0\over \sqrt 2 G_{F0}
                \sin^2\theta_{W0}}\,.\eqn\treewmass
$$
It is useful to introduce the following combination of physical
parameters
$$
  \mu \equiv {\pi\alpha\over \sqrt 2 G_F} =
    (37.2802\ {\rm GeV})^2 \,.\eqn\mudeff
$$
Let us introduce renormalized quantities and counterterms as before:
$\alpha_0 = \alpha-
\delta\alpha$, \etc\
Then eq.~\treewmass\ can be used to
derive a formula for the one-loop renormalized $W$ mass parameter
$$
  m^2_W = {\mu^2\over \sin^2\theta_W}
   \left( 1-{\delta\alpha\over \alpha} +
         {\delta G_F\over G_F} +
         {\delta s^2\over s^2}
         + {\delta m^2_W \over m^2_W} \right)\,,\eqn\renormwmass
$$
where $s^2\equiv \sin^2\theta_W$.  One can derive a formula for
$\delta s^2/s^2$ by taking the variation of eq.~\sirlindeff
$$
{\delta s^2\over s^2}={1-s^2\over s^2}\left({\delta\mz^2\over\mz^2}
-{\delta\mw^2\over\mw^2}\right)\,.\eqn\deltasw
$$
Inserting this result into eq.~\renormwmass, one finds
$$
  m^2_W = {\mu^2\over s^2} \left[ 1 - {\delta\alpha\over \alpha}
          + {\delta G_F\over G_F} + \left({1-s^2\over s^2} \right)
           {\delta m^2_Z\over m^2_Z} +
          \left({2s^2-1\over s^2}\right){\delta m^2_W\over m^2_W}
         \right]\,.\eqn\oneloopwmass
$$
The $W$ and $Z$ counterterms are fixed as before by
requiring that the renormalized parameters $\mw^2$ and $\mz^2$
be the physical masses measured in experiment.  In terms of the
notation of eq.~\fdef,
$$\eqalign{%
{\delta m^2_W \over m^2_W} &= {A_{WW}(0)\over m^2_W}
    + F_{WW}(m^2_W)\,, \cr
{\delta m^2_Z \over m^2_Z} &= {A_{ZZ}(0)\over m^2_Z}
    + F_{ZZ}(m^2_Z)\,.\cr }\eqn\physmasses
$$
By inserting eq.~\sirlindeff\ for $s^2$ on the right hand side of
eq.~\oneloopwmass, one obtains an implicit equation for the
physical $W$ mass.
As a result, all infinites must cancel exactly when the right-hand
side of eq.~\oneloopwmass\ (appropriately regularized) is evaluated.

The prediction for the $W$ mass in the Standard Model
is well known\refmark\sirlinwmass.
Here, we are interested in the $W$ mass shift from the Standard Model
prediction due to new physics.  In particular, we need to compute the
new physics contributions to $\delta\alpha/\alpha$ and $\delta
G_F/G_F$ in the oblique approximation.  First, we define
the observable $\alpha$ via the Thomson limit in QED.  That is,
consider
$$
  \lim_{q^2\to 0}\; \left[
  \lower.9cm\hbox{ \vrule width0pt depth0pt height 2.1cm
\getpicture{10cm}{2.0cm}{alpha.topdraw}
             \vrule height0pt depth0pt width 10.2cm }
  \right]
$$
where the unlabeled loop contains the contributions from new
heavy physics.
In equations, the one-loop definition of $\alpha$ (or $e^2$)
reads
$$
  \lim_{q^2\to 0} \; e^2\left({-i\over q^2}\right)
= e^2_0 \left({-i\over q^2}\right)
+ e^2_0 \left({-i\over q^2}\right) i A_{\gamma\gamma}
        (q^2)\left({-i\over q^2}\right)\,,\eqn\thompson
$$
which implies that
$$\eqalign{%
\alpha &= \lim_{q^2\to 0}\; \alpha_0 \left( 1+
          {A_{\gamma\gamma} (q^2) \over q^2}\right) \crr
&= \alpha_0\left[1 + F_{\gamma\gamma}(0)\right] \crr
&=(\alpha-\delta\alpha) \left[ 1+F_{\gamma\gamma}(0) \right]\,. \cr}
\eqn\alphadeff
$$
Thus, in the oblique approximation,
$$
{\delta\alpha\over\alpha}  = F_{\gamma\gamma}(0)\,.\eqn\deltaalpha
$$
Note that in the $q^2\to 0$ limit, $Z$--$\gamma$ mixing
diagrams do not contribute, since new heavy physics does not
contribute to $A_{Z\gamma}(0)$.

The physical value for the Fermi constant $G_F$
is defined through $\mu$-decay. Consider

$$ \lim_{q^2\to 0}\; \left[
  \lower2.3cm\hbox{ \vrule width0pt depth0pt height 4.7cm
\getpicture{10cm}{4.5cm}{g_f.topdraw}
             \vrule height0pt depth0pt width 10cm }
     \right]
$$
\endpage\noindent
where, as above, the unlabeled loop contains contributions
{}from new heavy physics.
A similar calculation as above yields
$$\eqalign{%
  G_F &= \lim_{q^2\to 0}\; G_{F0} \left( 1 +
         {A_{WW}(q^2) \over q^2-m^2_W}\right) \crr
      &= (G_F - \delta G_F) \left( 1 -
         {A_{WW}(0) \over m^2_W}\right) \,.\cr }\eqn\gfdeff
$$
Thus, in the oblique approximation,
$$ {\delta G_F\over G_F} = -{A_{WW}(0)\over m^2_W}\,.\eqn\deltagf$$
We can define a neutral current coupling $G_{NC}$
in analogy with $G_F$.  The physical process of relevance is
$\nu e^-$ elastic scattering.
Then, by a similar computation as above,
one arrives at a result analogous to
eq.~\deltagf
$$
 { \delta G_{NC}\over G_{NC}} = -{A_{ZZ}(0)\over
m^2_Z}\,.\eqn\deltagnc
$$

Combining the results obtained above, eq.~\oneloopwmass\ yields the
following formula for the contribution of new physics to the
$W$ mass (in the oblique approximation)
$$
  m^2_W = {\mu^2\over s^2} (1+\Delta r)\,,\eqn\deltarnum
$$
where
$$\eqalign{%
\Delta r &= -F_{\gamma\gamma}(0) + \left( {1-s^2\over s^2}\right)
   \left[ {A_{ZZ}(0)\over m^2_Z} - {A_{WW}(0)\over m^2_W}\right] \crr
         &\quad + \left( {1-s^2\over s^2}\right) F_{ZZ}(m^2_W)
            + \left( {2s^2-1\over s^2}\right) F_{WW}(m^2_W)\,. \cr}
\eqn\obliquedeltar
$$
Another physical observable of interest is $\rho\ls{NC}$ which was
discussed earlier.  Formally, we may define
$$
  \rho\ls{NC}\equiv {G_{NC}\over G_F} \equiv 1 +
\delta\rho\,.\eqn\rhonc
$$
Taking the variation of this equation yields
\vskip2pt
$$
  \delta\rho = {\delta G_{NC}\over G_{NC}} - {\delta G_F\over G_F}\,.
\eqn\deltarhonc
$$
\vskip2pt\noindent
In the oblique approximation, we find [using eqs.~\deltagf\ and
\deltagnc] that the contribution of new physics to the shift in
$\rho\ls{NC}$ is
$$
  \delta\rho = {A_{WW}(0)\over m^2_W} - {A_{ZZ}(0)\over m^2_Z}\,.
\eqn\rhoshift
$$

\REF\tatsu{M.E. Peskin and T. Takeuchi, {\sl Phys. Rev. Lett.} {\bf 65}
(1990) 964; {\sl Phys. Rev.} {\bf D46} (1992) 381.}
\REF\otherst{G. Altarelli and R. Barbieri, {\sl Phys. Lett.} {\bf B253}
(1990) 161.}
\REF\morest{W.J. Marciano and J.L. Rosner, {\sl Phys. Rev. Lett.}
{\bf 65} (1990) 2963;
D.C. Kennedy and P. Langacker, {\sl Phys. Rev. Lett.} {\bf 65} (1990)
2967 [E: {\bf 66} (1991) 395]; {\sl Phys. Rev.} {\bf D44} (1991) 1591;
D.C. Kennedy, {\sl Phys. Lett.}, {\bf B268} (1991) 86; J.Ellis,
G.L. Fogli, and E. Lisi, {\sl Phys. Lett.} {\bf B285} (1992) 238.}
One can study other electroweak observables in a
similar fashion.  A remarkable fact is that the heavy particle
contributions to the oblique radiative corrections can be summarized in
terms of three numbers called $S$, $T$ and $U$\refmark{%
\tatsu--\morest, \kennedy}.
$T$ is related simply to the $\rho$ parameter
$$\rho-1=\alpha T\,,\eqn\tdef$$
where $\alpha$ is the usual fine structure constant.  To formally
define the three quantities $S$, $T$ and $U$, one proceeds as follows.
The oblique radiative corrections to electroweak observables can be
expressed in terms of the various functions $A_{ij}(0)$ and
$F_{ij}(q^2)$ introduced in eq.~\fdef, where the $q^2$ that appears
depends on the particular observable of interest.
A major simplification takes place if one is interested in the
effects of new heavy physics characterized by a scale $M\gg\mz$.
In this case, since $q^2$ is of order $\mz^2$,  one only makes
an error of ${\cal O}(\mz^2/M^2)$ by neglecting the $q^2$ dependence
of the $F_{ij}$.  Then, one can show that the oblique corrections
to electroweak observables due to heavy physics can be expressed
in terms of three particular
combinations of the $A_{ij}(0)$ and $F_{ij}$
$$\eqalign{\alpha T&\equiv {A\ls{WW}(0)\over\mw^2}-{A\ls{ZZ}(0)\over
\mz^2}\,,\cr
{g^2\over 16\pi c\ls{W}^2}S&\equiv F\ls{ZZ}(\mz^2)
-F\ls{\gamma\gamma}(\mz^2)+\left({2s\ls{W}^2-1\over s\ls{W}
c\ls{W}}\right)F\ls{Z\gamma}(\mz^2)\,,\cr
{g^2\over 16\pi}(S+U)&\equiv  F\ls{WW}(\mw^2)-F\ls{\gamma\gamma}(\mw^2)-
{c\ls{W}\over s\ls{W}}F\ls{Z\gamma}(\mw^2)\,.\cr}
\eqn\studefs$$
Note that the $A_{ij}(0)$ and $F_{ij}$ in the above formulae
are divergent quantities.  Nevertheless, if one includes a complete
set of contributions from a gauge invariant sector, then $S$, $T$,
and $U$ will be finite constants.
The Higgs sector by itself does not constitute a gauge invariant
sector (the gauge bosons must also be included).
In order to obtain finite quantities that solely
reflect the influence of heavy Higgs physics, one must
define $\delta S$, $\delta T$ and $\delta U$ relative to some
reference Standard Model  where $\mhsm$ is fixed to a convenient
value.  For example, if we choose a RSM with
$\mhsm=\mz$, then the
change in $S$ due to a fourth generation of fermions $U$ and $D$ (with
electric charges $e\ls{D}+1$ and $e\ls{D}$ respectively) and a heavy
Higgs boson of mass $\mhsm$ is given by
$$
\delta S\simeq {N_c\over 6\pi}\left[1+(1+2e\ls{D})\ln\left(
{m\ls{D}^2\over m\ls{U}^2}\right)\right]
+{1\over 12\pi}\left[\ln\left({\mhsm^2\over\mz^2}\right)-3\pi\sqrt{3}
+{107\over 6}\right]\,,\eqn\esscorrection$$
where $m\ls{U}$, $m\ls{D}$, $\mhsm\gg\mz$ has been assumed.
Once again, the non-decoupling effects of the heavy physics are
apparent.

\REF\sirlindeltar{A. Sirlin, {\sl Phys. Rev.} {\bf D29} (1984) 89.}
The effects of heavy physics on numerous
electroweak observables are immediately known once
the corresponding
contributions to $S$, $T$ and $U$ have been computed.
For example, consider the shift in the $W$ mass due to new physics.
It is common practice to rewrite the factor $1+\Delta r$ that appears
in eq.~\deltarnum\ as $1/(1-\Delta r)$.  These expressions are
formally equivalent up to one-loop.  However the latter form is more
useful in that it correctly sums the one-loop leading
logs\refmark\sirlindeltar.
Thus, the one-loop prediction for $\mw$ is obtained by solving the
following equation for the $W$ mass
$$\mw^2\left(1-{\mw^2\over\mz^2}\right)=\left({\pi\alpha\over\sqrt{2}
G\ls{F}}\right)^2{1\over 1-\Delta r}\,.\eqn\wmass$$
The contributions of new physics to $\Delta r$ in the oblique
approximation was obtained in eq.~\obliquedeltar.  It is then a
simple matter to re-express this result in terms of $S$, $T$ and $U$:
$$\Delta r={g^2\over 8\pi}\left[S-2c\ls{W}^2 T+\left({2s\ls{W}^2-1\over
2s\ls{W}^2}\right)U\right]\,.\eqn\deltar$$
\vskip6pt\noindent
Other examples can be found in refs.~[\tatsu--\morest].

In computing the contributions
of the MSSM to $S,\ T$ and $U$, it is convenient to consider
separately the contributions of the
various sectors of particles not contained in the Standard Model.
Each sector, when appropriately
defined, yields a finite shift to $S$, $T$ and $U$.
The relevant particle sectors include all supersymmetric
particles and the physical Higgs bosons (beyond the minimal
neutral Higgs scalar of the Standard Model).  Specifically,
we examine the contributions to $S$, $T$ and $U$ from:
\item{A.} The squarks and sleptons.  Note that each squark and each
slepton generation (\ie, summing over both up-type and
down-type superpartners)
contributes separately a finite result to $S$, $T$
and $U$.
\item{B.} The neutralinos and charginos.
\item{C.} The MSSM Higgs sector.

\noindent
The treatment of the MSSM Higgs sector requires some care to
insure a finite contribution to $S$, $T$ and $U$.  As before,
one must first
define the RSM.  Then,
$$\eqalign{S&=S\ls{\rm RSM}+\delta S\,,\cr
T&=T\ls{\rm RSM}+\delta T\,,\cr
U&=U\ls{\rm RSM}+\delta U\,,\cr}
\eqn\stsusy$$
where the MSSM Higgs sector contributions to
$\delta S,\ \delta T$ and $\delta U$
are obtained from eq.~\studefs\
by computing the MSSM Higgs loops contributing to $A_{ij}(0)$ and
$F_{ij}$ (including diagrams
with one virtual Higgs boson and one virtual gauge boson) and
subtracting off the corresponding Higgs loops of the RSM.
In the present case, it is convenient to
define the RSM to be the Standard Model
with the Standard Model Higgs boson mass set
equal to the mass of the lightest CP-even Higgs scalar of the MSSM.
In addition, until $\mt$ is known, the definitions of $\delta S$,
$\delta T$ and $\delta U$ will depend on the value of $\mt$ chosen
for the RSM. Typically, one chooses $\mt=\mz$ (equal to the present
experimental CDF lower bound\refmark\cdflimit)
in order to obtain conservative limits on the possible new physics
contributions to $S$, $T$  and $U$.

\REF\hehth{H.E. Haber, SCIPP preprint in preparation.}
\REF\barbi{R. Barbieri, M. Frigeni, F. Giuliani and H.E. Haber,
{\sl Nucl. Phys.} {\bf B341} (1990) 309; M. Drees and K. Hagiwara,
{\sl Phys. Rev.} {\bf D42} (1990) 1709;
P. Gosdzinsky and J. Sola, {\sl Phys. Lett.} {\bf B254} (1991) 139;
{\sl Mod. Phys. Lett.} {\bf A6} (1991) 1943; M. Drees, K. Hagiwara,
and A. Yamada, {\sl Phys. Rev.} {\bf D45} (1992) 1725;
R. Barbieri, M. Frigeni and F. Caravaglios, {\sl Phys. Lett.}
{\bf B279} (1992) 169; J. Ellis, G.L. Fogli, and E. Lisi,
{\sl Phys. Lett.} {\bf B286} (1992) 85; {\sl Nucl. Phys.} {\bf B393}
(1993) 3.}
\REF\passarino{G. Passarino and M. Veltman, {\sl Nucl. Phys.}
{\bf B160} (1979) 151.}
\REF\higgsrad{S. Bertolini, {\sl Nucl. Phys.} {\bf B272} (1986) 77.}
\REF\oldhiggsrad{R.S. Lytel, {\sl Phys. Rev.} {\bf D22} (1980) 505;
J.-M. Fr\`ere and J.A.M. Vermaseren, {\sl Z. Phys.} {\bf C19} (1983) 63;
W. Hollik, {\sl Z. Phys.} {\bf C32} (1986) 291; {\bf C37} (1988) 569.}
\REF\higgsradtwo{C.D. Froggatt, R.G. Moorhouse, and I.G. Knowles,
{\sl Phys. Rev.} {\bf D45} (1992) 2471; {\sl Nucl. Phys.} {\bf B386}
(1992) 63; T. Inami, C.S. Lim and A. Yamada, {\sl Mod. Phys. Lett.}
{\bf A7} (1992) 2789.}
Some of the results for the MSSM contributions to $S$, $T$ and
$U$ are presented below\refmark\hehth.  The qualitative behavior of these
results is easily summarized.  The contribution of any given
sector to $\delta S$, $\delta T$ and $\delta U$ behaves as
$$\delta S({\rm MSSM})\sim\delta T({\rm MSSM})\sim
{\cal O}\left({\mz^2\over\msusy^2}\right)\,,\eqn\susyst$$
$$\delta U({\rm MSSM})\sim
{\cal O}\left({\mz^4\over\msusy^4}\right)\,,\eqn\susyu$$
in the limit where $\msusy\gg\mz$.  In each sector, $\msusy$
corresponds roughly to the mass parameter that contributes the
dominant part of the corresponding sector masses.\foot{These
masses are: in sector A, the SU(2)$\times$U(1) conserving
diagonal squark and slepton masses ($M_{\widetilde Q}$, $M_{\widetilde U}$,
$M_{\widetilde D}$, $M_{\widetilde L}$
and $M_{\widetilde E}$);
in sector B, the gaugino mass parameters $M_1$ and $M_2$ and the
Higgs superfield mass parameter $\mu$; and in sector C, the
CP-odd Higgs mass, $\mha$.}
Thus, all MSSM contributions
to $S$, $T$ and $U$ vanish at least quadratically in $\msusy$ as the
supersymmetric particle masses become large.  Unlike the
non-supersymmetric examples presented previously,
the effects of the supersymmetric particles (and all Higgs bosons
beyond $\hl$) smoothly decouple; the resulting low-energy effective
theory at the scale $m_Z$ is precisely that of the Standard Model.
This decoupling behavior is easily understood.
As an example, consider the effects of the MSSM Higgs sector.
As indicated above, the sum of
the contributions to $S$, $T$ and $U$ is finite after
subtracting off the contribution of the Standard Model
Higgs boson with $\mhsm=\mhl$.  According to eq.~\cpevenhiggsmass,
the mass of $\hl$ cannot be arbitrarily large---it is
bounded at tree level by $\mz$.  All other Higgs masses can become
large by taking $\mha\gg\mz$.  In this limit, we see
that $\mhpm\simeq\mhh\simeq\mha$ and $\mhl\simeq\mz|\cos2\beta|$.
However, in this limit, the large Higgs masses are due  to the
large value of the mass parameter $m_{12}$ [see eq.~\monetwoeq]
rather than a large Higgs self-coupling (as in the
large Higgs mass limit of the Standard Model).  In particular, the
Higgs self-couplings in the MSSM are gauge couplings which can
never become large.  As a result, the decoupling theorem applies.
Similar arguments can be applied to the other sectors.  Supersymmetric
particle masses can be taken large by increasing the values of
SU(2)$\times$U(1) conserving mass parameters.  Thus, the virtual effects
of heavy supersymmetric particles must decouple.\foot{One can also
show that the couplings of the supersymmetric fermions (charginos
and neutralinos) to the massive gauge bosons become purely
vector-like in the limit of large fermion mass.  Thus, again the
decoupling theorem applies to the heavy virtual supersymmetric
fermion exchanges.}

Let us now turn to some specific calculations\refmark{\barbi,\hehth}.
The only remaining
question is to evaluate the constant of proportionality that
is implicit in eqs.~\susyst\ and \susyu.  In particular, note
that $\delta U\ll \delta S$, $\delta T$,
so I shall focus primarily
on $\delta S$ and $\delta T$ below.  It is important
to emphasize that the $S$, $T$, $U$ formalism is useful only in
the limit where the new physics is sufficiently
heavy as compared to $\mz$.  In the present application,
``sufficiently heavy'' means that it is a good
approximation to keep only the leading
${\cal O}(\mz^2/\msusy^2)$ terms.  Because of various additional
factors such as $1/16\pi^2$
which typically arise in loop-calculations,
I expect the results presented
below to be reasonably accurate for supersymmetric masses above, say,
150 GeV.  For lighter supersymmetric particle masses,
more precise computations are required.\foot{However,
note that the calculation of $T$ (or
the $\rho$ parameter) is meaningful over the entire range of possible
supersymmetric masses.}

Consider first the contributions of the top/bottom squark system.
One can easily diagonalize the
two $2\times 2$ squark mass matrices [eq.~\squarkmatrix]
to find the mass eigenstates
and corresponding mixing angles.  To a good approximation, we can
neglect the mixing between $\widetilde b_L$ and $\widetilde b_R$.
Denote the top-squark mass eigenstates by
$\widetilde t_1$ and $\widetilde t_2$
with corresponding mixing angle $\theta_t$.  Then, in terms of
the function $F$ defined in eq.~\effdef,
$$\eqalign{
\delta\rho(\widetilde t,\widetilde b)=&{g^2 N_c\over 32\pi^2\mw^2}
\left[\cos^2\theta_t F(m^2_{\tilde t_1},m^2_{\tilde b_L})+
\sin^2\theta_t F(m^2_{\tilde t_2},m^2_{\tilde b_L})\right.
\cr &\qquad\qquad\quad\left.-
\sin^2\theta_t\cos^2\theta_t F(m^2_{\tilde t_1},m^2_{\tilde t_2})
\right]\,.\cr}\eqn\deltarhosq$$
$F(m_1^2,m_2^2)$ has the following properties in the
limit of $|m_1^2-m_2^2|\ll m_1^2$, $m_2^2$
$$\eqalign{%
F(m_1^2,m_2^2) &\simeq {(m_1^2-m_2^2)^2\over 6m_2^2}\ , \crr
F(m_1^2,m_3^2)-F(m_2^2,m_3^2) &\simeq (m_1^2-m_2^2)
  \left[{1\over 2}+ {m_3^2 \over m^2_3-m^2_2}
  + {m^4_3\over (m^2_3-m^2_2)^2} \ln\left({m^2_2\over m^2_3}\right)
  \right]\, . \cr}
\eqn\flimit$$
Using these results and explicit formulae for the top-squark masses
and mixing angles obtained from diagonalization of the squark mass
matrix, one ends up with
$$\delta\rho(\widetilde t,\widetilde b)\simeq  {g^2N_c m^4_t K\over
   32\pi^2m^2_W \msusy^2}
\,,\eqn\rhostsb$$
where I have assumed that  $\msusy \gg \mz,\mt$. Here, $\msusy$
is the largest of the supersymmetric mass parameters that appear
in the squark mass matrix, and $K$ is a dimensionless function
of the model parameters.  In particular, $K$ approaches a constant
in the limit where one or more of the various supersymmetric
masses become large, so that eq.~\rhostsb\ exhibits the
expected decoupling behavior.  It may appear that the
correction exhibited in eq.~\rhostsb\ is significantly enhanced by
a factor of $\mt^4$.  However, this result is deceptive, since we
must compare it to the corresponding Standard Model result
[eq.~\rhocorrection].  One sees that the MSSM contribution is
in fact suppressed by $\mt^2/\msusy^2$ relative to the Standard Model
result.  Of course, this is not a suppression if the relevant
supersymmetric mass parameters are small.  For example,
in the supersymmetric limit ($M_{\tilde t_1}=M_{\tilde t_2}=\mt$ and
$M_{\tilde b_L}=\mb$), one finds $\delta\rho(\widetilde t,\widetilde b)=
\delta\rho(t,b)$, in which case $\delta\rho$ would be
twice as large as its predicted Standard Model value.  This
demonstrates that an appreciable MSSM contribution to precision
electroweak measurements is possible in principle if supersymmetric
particle masses are light enough.  However, once these masses are taken
larger than $\mt$, their contributions to one-loop effects diminish
rapidly.

\FIG\stsqk{%
The contribution to the $S$ and $T$ parameters from
the squark and slepton sector of the MSSM as a function
of $M_{\widetilde Q}$.  The squark mass spectrum is determined
by MSSM mass parameters defined in section 1.5.
For simplicity, all soft supersymmetry breaking diagonal
squark (and slepton) mass parameters are taken
equal to $M_{\widetilde Q}$.
In addition, $A=M_{\widetilde Q}$, $\mu=-200$~GeV and
$\tan\beta=2$
are the parameters that determine the strength of the
off-diagonal squark mixing.  The solid curves show the contributions
of three generations of squarks and sleptons, while the dashed curves
show the partial contribution arising from the top/bottom squark
sector alone.}

In fig.~\stsqk, I plot the squark and slepton contributions to
$S$ and $T$ for a typical set of MSSM parameters.  For simplicity,
all diagonal squark soft-supersymmetry breaking parameters
have been taken equal to $M_{\widetilde Q}$.  In addition, I have
included squark mixing (which is appreciable only in the top
squark sector) by taking all $A$-parameters equal to $M_{\widetilde Q}$.
For illustrative purposes, I have chosen
$\mu=-200$~GeV (and $\tan\beta=2$) which enhances somewhat the top
squark mixing.  This choice of parameters leads to a large mass splitting
of the top squark eigenstates (and a rather light top squark) when
$M_{\widetilde Q}$ is near its lower limit as shown in fig.~\stsqk.
Nevertheless, we see that the contributions to $S$ and $T$ from the
squark and slepton sector never exceed 0.1 for the parameters shown.
As suggested above, larger values for $\delta T$ can occur only
for parameter choices approaching the supersymmetric limit.

\FIG\stchi{%
The contribution to the $S$ and $T$ parameters from the
neutralino and chargino sector of the MSSM as a function
of $\mu$ for $\tan\beta=2$.  The four curves shown correspond
to $M=50$, $250$, $500$ and $1000$~GeV [with $M_2\equiv M$ and
$M_1=(5g^{\prime 2}/3g^2)M$].  In (a), curves in the region
of $|\mu|\leq100$~GeV are not shown, since in this region of
parameter space the light chargino mass is of order $m_Z$ (or less).
In (b), $T$ is related to the
$\rho$ parameter via $\delta\rho=\alpha\delta T$
which is an experimental observable over the entire mass parameter
region.}
Consider next the contributions to $S$ and $T$ from the neutralino
and chargino sector, shown in fig.~\stchi.  Again, one never sees
values of $\delta S$ and $\delta T$ larger than 0.1.  I have omitted
plotting $\delta S$ in the region of small $\mu$, since the lightest
chargino mass is $\lsim\mz$ in this region of parameter space.
In this case, it is no
longer true that all particle masses associated with the ``new physics''
lie significantly above $\mz$, so that the assumption that the
radiative corrections are simply parametrized by $S$, $T$ and $U$ breaks
down.

Finally, I consider the contributions to $S$ and $T$ from the MSSM
Higgs sector.  Here one must be careful to define these contributions
relative to the RSM, as discussed above.  In this case, I subtract
out the contribution of the Standard Model Higgs boson whose mass
is chosen to be equal to the mass of the lightest CP-even scalar
($\mhl$).
The results of an exact one-loop
computation of the MSSM Higgs
contributions to $S$, $T$ and $U$ are recorded below.\foot{%
See refs.~[\doug,\higgsrad--\higgsradtwo] for
previous work on radiative corrections in two-Higgs doublet models.}

\noindent
First, for $\delta S$ and $\delta U$ I find
$$\eqalign{
\delta S&=  {1\over\pi\mz^2}\Biggl\{
\sin^2(\beta-\alpha)
{\cal B}_{22}(\mz^2;\mhh^2,\mha^2)-{\cal B}_{22}(\mz^2;\mhpm^2,
\mhpm^2)\cr
& +\cos^2(\beta-\alpha)\biggl[{\cal B}_{22}
(\mz^2;\mhl^2,\mha^2)+{\cal B}_{22}(\mz^2;\mz^2,\mhh^2)
-{\cal B}_{22} (\mz^2;\mz^2,\mhl^2)\cr
&\qquad\qquad -\mz^2{\cal B}_0(\mz^2;\mz^2,\mhh^2)
+\mz^2{\cal B}_0(\mz^2;\mz^2,\mhl^2)\biggr]\Biggr\}\,,\cr
\delta U&= -\delta S+{1\over\pi\mz^2}\Biggl\{
{\cal B}_{22}(\mw^2;\mha^2,\mhpm^2)-2{\cal B}_{22}(\mw^2;\mhpm^2,
\mhpm^2)\crr
& +\sin^2(\beta-\alpha)
{\cal B}_{22}(\mw^2;\mhh^2,\mhpm^2)  \crr
& +\cos^2(\beta\!-\!\alpha)\biggl[{\cal B}_{22}
(\mw^2;\mhl^2,\mhpm^2)\!+\!{\cal B}_{22}(\mw^2;\mw^2,\mhh^2)
\!-\!{\cal B}_{22} (\mw^2;\mw^2,\mhl^2)\cr
&\qquad\qquad -\mw^2{\cal B}_0(\mw^2;\mw^2,\mhh^2)
+\mw^2{\cal B}_0(\mw^2;\mw^2,\mhl^2)\biggr]\Biggr\}\,.\cr}
\eqn\esshiggs$$
The following notation has been introduced for the various
loop integrals
$${\cal B}_{22}(q^2;m_1^2,m_2^2)\equiv B_{22}(q^2;m_1^2,m_2^2)-
B_{22}(0;m_1^2,m_2^2)\,,\eqn\calbtwo$$
$${\cal B}_{0}(q^2;m_1^2,m_2^2)\equiv B_{0}(q^2;m_1^2,m_2^2)-
B_{0}(0;m_1^2,m_2^2)\,,\eqn\calbzero$$
and $B_{22}$ and $B_0$ are defined according to ref.~[\passarino]
(up to an overall sign in some cases since I use the Bjorken and Drell
metric\refmark\bjdrell).  Explicitly,
$$B_{22}(q^2;m_1^2,m_2^2)=\fourth(\Delta+1)\left[m_1^2+m_2^2
-\third q^2\right]-\half\int_0^1\,dx\,X\ln(X-i\epsilon)
\,,\eqn\beetwo$$
\vskip-5pt
$$B_0(q^2;m_1^2,m_2^2)=\Delta-\int_0^1\,dx\,\ln(X-i\epsilon)
\,,\eqn\beezero$$
where
$$X\equiv m_1^2x+m_2^2(1-x)-q^2 x(1-x)
\eqn\xdefinition$$
and $\Delta$ is the regulator of dimensional regularization defined by
$$\Delta={2\over 4-n}+\ln(4\pi)+\gamma\,,
\eqn\deltadef$$
$n$ is the number of space-time
dimensions and $\gamma$ is Euler's constant.  Of course, in the
calculation of physical observables, terms proportional to $\Delta$ must
exactly cancel.  The following two
relations are particularly useful:
$$4B_{22}(0;m_1^2,m_2^2)=F(m_1^2,m_2^2)+A_0(m_1^2)+A_0(m_2^2)\,,
\eqn\bzerotwo$$
$$B_0(0;m_1^2,m_2^2)={A_0(m_1^2)-A_0(m_2^2)\over m_1^2-m_2^2}\,,
\eqn\bzerozero$$
where $F(m^2_1,m^2_2)$ is defined in eq.~\effdef, and
$$A_0(m^2)\equiv m^2(\Delta+1-\ln m^2)\,.\eqn\azero$$
Finally, consider $\delta\rho=\alpha \delta T$.  I find:
$$\eqalign{\delta T=&{1\over 16\pi\mw^2 s\ls{W}^2}\Biggl\{
F(\mhpm^2,\mha^2)+\sin^2(\beta-\alpha)\left[F(\mhpm^2,\mhh^2)-
F(\mha^2,\mhh^2)\right]\crr &\qquad\qquad
+\cos^2(\beta-\alpha)\left[F(\mhpm^2,\mhl^2)
-F(\mha^2,\mhl^2)+F(\mw^2,\mhh^2)\right.\crr &\left.
\qquad\qquad\qquad-F(\mw^2,\mhl^2)
-F(\mz^2,\mhh^2)+F(\mz^2,\mhl^2)\right]\crr & \qquad\qquad
+4\mz^2\left[B_0(0;\mz^2,\mhh^2)-B_0(0;\mz^2,\mhl^2)\right]\crr &
\qquad\qquad
-4\mw^2\left[B_0(0;\mw^2,\mhh^2)-B_0(0;\mw^2,\mhl^2)\right]\Biggr\}
\,,\cr}\eqn\teehiggs$$
where $s\ls{W}\equiv\sin\theta_W$.

\FIG\sthiggs{%
The contribution to the $S$ and $T$ parameters from
the MSSM Higgs sector
relative to the Standard Model with Higgs mass set equal to $\mhl$,
as a function of $\mha$.  Three curves corresponding to
$\tanb= 1$, 2, and 10 are shown.  In (a), the dotted curve corresponds
to the asymptotic prediction [eq.~\val\shiggsapprox] for $\tanb=10$.
In the region of small $\mha$ where the dotted curve departs from the
dashed curve, it is no longer useful to use $S$ in the
parametrization of the radiative
corrections.  In (b), the two dotted curves correspond to
the asymptotic predictions [eq.~\val\thiggsapprox]
for $\tanb=1$ and 10 respectively.}
Note that the above expressions for $\delta S,\ \delta T$ and $\delta U$
are valid for an
arbitrary two-Higgs doublet extension of the Standard
Model\refmark{\higgsrad--\higgsradtwo}.
In the MSSM, tree-level relations exist among the Higgs masses and
angles $\alpha$ and $\beta$ as discussed at the end of Lecture 1.
By virtue of
these relations, the numerical values of $\delta S$ and
$\delta T$ in the MSSM are much smaller than 1.  Numerical
results are shown in fig.~\sthiggs.  To understand why
the numerical values for the MSSM-Higgs contributions to
$\delta S$ and $\delta T$ are particularly small, it is
instructive to evaluate the expressions obtained above in the
limit of $\mha\gg\mz$.  I find\foot{Asymptotically, $\delta U$%
(MSSM--Higgs) = ${\cal O}(m^4_Z/m^4_A)$, which is completely
negligible.}
$$\delta S({\rm MSSM}\!-\!{\rm Higgs})
\simeq{\mz^2(\sin^2 2\beta-2\cos^2\theta_W)\over 24\pi\mha^2}\,,
\eqn\shiggsapprox$$
$$\delta T({\rm MSSM}\!-\!{\rm Higgs})
\simeq{\mz^2(\cos^2\theta_W-\sin^2 2\beta)\over 48\pi\mha^2
\sin^2\theta_W}\,.\eqn\thiggsapprox$$

An analysis of $S$, $T$ and $U$ based on LEP data (assuming a
RSM where $\mt=\mhsm=\mz$) reported in ref.~[\paul] yields:
$\delta S=-0.97\pm 0.67$, $\delta T=-0.18\pm 0.51$ and
$\delta U=0.07\pm 0.97$.
It is hard to imagine that the these quantities
could ever be measured to an accuracy better than 0.1, whereas
the analysis above implies that the contributions of the MSSM to
$S$ and $T$ must lie below 0.1 if all supersymmetric particle masses
are above $\mz$.
One must also consider the possibility of other contributions
to $\delta S$ and $\delta T$.  As long as $\mt$ is not well known,
there will be $\mt$ dependence in these quantities (entering through the
$\mt$ choice of the RSM).  However,
even when $\mt$ is known with some accuracy, it is doubtful
that virtual effects of the MSSM will be detected
via its oblique radiative corrections.

\section{MSSM Radiative Corrections to Processes Involving
$\bold b$ Quarks}

\REF\gapjf{G. Altarelli and P. Franzini, {\sl Z.~Phys.} {\bf C37}
(1988) 271; G.G. Athanasiu, P.J. Franzini and F.J. Gilman,
{\sl Phys. Rev.} {\bf D32} (1985) 3010;
S.L. Glashow and E.E. Jenkins, {\sl Phys. Lett.} {\bf B196} (1987) 233;
F. Hoogeveen and C.N. Leung, {\sl Phys. Rev.} {\bf D37} (1988) 3340;
J.F. Gunion and B. Grzadkowski, {\sl Phys. Lett.} {\bf B243}
(1990) 301.}
\REF\cqgjnn{C.Q. Geng and J.N. Ng, {\sl Phys. Rev.} {\bf D38}
(1988) 2857.}
\REF\stefan{
S. Bertolini, F. Borzumati, A. Masiero and G. Ridolfi,
{\sl Nucl. Phys.} {\bf B353} (1991) 591.}
\REF\buras{A. Buras, P. Krawczyk, M.E. Lautenbacher and C. Salazar,
{\sl Nucl. Phys.} {\bf B337} (1990) 284.}
\REF\hewett{V. Barger, J.L. Hewett and R.J.N. Phillips, {\sl Phys.
Rev.} {\bf D41} (1990) 3421.}
\REF\zbbar{A. Djouadi, G. Girardi, C. Verzegnassi, W. Hollik and
F.M. Renard, {\sl Nucl. Phys.} {\bf B349} (1991) 48; M. Boulware and
D. Finnell, {\sl Phys. Rev.} {\bf D44} (1991) 2054.}
\REF\wise{B. Grinstein and
M.B. Wise, {\sl Phys. Lett.} {\bf B201} (1988) 274.}
\REF\willey{W.-S. Hou and R.S. Willey, {\sl Phys. Lett.} {\bf B202}
(1988) 591; {\sl Nucl. Phys.} {\bf B326} (1989) 54;
T. Rizzo, {\sl Phys. Rev.} {\bf D38} (1988) 820;
X.-G. He, T.D. Nguyen and R.R. Volkas, {\sl Phys. Rev.} {\bf D38}
(1988) 814; M. Ciuchini, {\sl Mod. Phys. Lett.} {\bf A4}
(1989) 1945.}
\REF\joanne{J.L. Hewett, {\sl Phys. Rev. Lett.} {\bf 70} (1993) 1045;
V. Barger, M.S. Berger, and R.J.N. Phillips, {\sl Phys. Rev. Lett.}
{\bf 70} (1993) 1368.}
\REF\joannetwo{J.L. Hewett, Argonne preprint ANL-HEP-PR-93-21 (1993).}
\REF\barbii{R. Barbieri and G.F. Giudice, CERN-TH.6830/93 (1993).}

In
the previous section, I demonstrated that oblique radiative corrections
due to supersymmetric particle exchange are not observable if
all supersymmetric particle masses lie much above $\mz$.  However,
this leaves the possibility that certain non-oblique radiative
corrections might be detectable.  In this section, I shall briefly
address this possibility.

A promising class of non-oblique radiative corrections consists
of processes that involve an external $b$-quark.
Three examples of such processes that have been studied in the
literature are: (i) charged Higgs box diagram
contributions to $B^0$--$\overline{B^0}$
mixing\refmark{\gapjf--\hewett}; (ii) the charged Higgs
vertex correction to $Z\to b\bar b$\refmark\zbbar;
and (iii) the charged Higgs vertex corrections to rare
$b$-decays\refmark{\cqgjnn--\buras,\wise--\joannetwo}\
such as $b\to s\gamma$,
$b\to s\ell^+\ell^-$, $b\to sg$ and $b\to s\nu\bar\nu$.
Here, I shall take the charged Higgs coupling to $t\bar b$
to be of the form that occurs in the MSSM
$$g_{H^- t\bar b}={g\over{2\sqrt{2}\mw}}\
[m_t\cot\beta (1+\gamma_5)+m_b\tan\beta (1-\gamma_5)]\, .
\eqn\hpmqq$$
Of course, in the MSSM
there will also be supersymmetric particle contributions to
all of the one-loop processes mentioned above.
Some of these contributions
(\eg, loops containing top-squarks) could be as important as
the charged Higgs effects.  Depending on the sign of the relative
contributions (which depends in detail on the MSSM parameters),
the overall MSSM contribution to the various rare $b$ decays could
be substantially different from the
charged Higgs effects alone\refmark{\stefan,\zbbar,\barbii}.

Among the rare $b$-decays, the charged Higgs contribution to
$b\to s\gamma$ is perhaps the most promising.  The theoretical
prediction for this rate in the Standard Model is
$BR(B\to K\gamma+X)\simeq 3.6\times 10^{-4}$ ($4.1\times 10^{-4}$),
for $\mt=150$ (200) GeV, where the leading log QCD corrections have been
included\refmark\buras.  Incorporating the
charged Higgs contribution
enhances the $b\to s
\gamma$ branching ratio over the Standard Model expectation.  An
explicit computation shows that the amplitude for $b\to s\gamma$
in the two-Higgs doublet model (omitting supersymmetric particle
contributions) has the following structure
$$\eqalign{%
{\cal M}(b\to s\gamma) &={eg^2m_b^3 V_{tb} V^\ast_{ts}\over\mw^2}\crr
&\quad \times\left[ A_W+(A_{H_1}\cot^2\beta+A_{H_2}) \left.
\times\cases{1,&$\mhpm\ll\mt$\cr
\mt^2/\mhpm^2\,,&$\mt\ll\mhpm$\cr} \right\}\right] \cr}
\eqn\bsgam$$
where $A_W$, $A_{H_1}$ and $A_{H_2}$ are dimensionless functions of
the particle masses arising from the loop graphs involving $W$ and
$\hpm$ exchange.  In particular, $A_{H_1}$ and $A_{H_2}$ approach
finite non-zero constants in the two limiting cases indicated above.
Eq.~\bsgam\ illustrates two of the features encountered in section 3.1.
On one hand, ones sees the non-decoupling of the top-quark in the
large $\mt$ limit.  In this case, the amplitude approaches an
$\mt$-independent constant when $\mt$ is larger than all mass scales
in the problem.  On the other hand, when $\mhpm\gg\mt$, the effect
of the charged Higgs loop decouples quadratically with the Higgs mass.
This behavior
is expected for the same reasons that heavy supersymmetric particles
decouple from oblique radiative corrections in the large $\msusy$ limit.

\FIG\bsgamen{%
The ratio of $BR(B\to K\gamma+X)$ in the two-Higgs-doublet
model relative to its predicted value
in the Standard Model (SM) as a function of the charged
Higgs mass for $\mt=150$ and 200~GeV and various choices for $\tanb$,
assuming an $H^-t\bar b$ coupling given by eq.~\hpmqq.
This graph is based on calculations of ref.~[\buras].}

\REF\pkim{E. Thorndike [CLEO Collaboration], talk at the meeting of the
American Physical Society, Washington, D.C. (1993).}
All that is left to do is to explore the relative sizes
of the coefficients in eq.~\bsgam\ and to incorporate the (leading log)
QCD radiative corrections\refmark\wise\
in order to see how fast the decoupling of a heavy $\hpm$ occurs.
Using the explicit formulae
in ref.~[\buras], I have plotted in fig.~\bsgamen\
the enhancement of the branching ratio
for $b\to s\gamma$ in the two-Higgs-doublet model relative to the
Standard Model rate.  It follows from
eq.~\bsgam\ that the decoupling of the
$\hpm$ contribution is controlled by the factor $m_t^2/\mhpm^2$
[in contrast to $\mz^2/\mhpm^2$ as in eqs.~\shiggsapprox-\thiggsapprox].
Even so, it is somewhat surprising that the significance of the
charged Higgs exchange to $b\to s\gamma$ persists to such large values
of $\mhpm$.
In addition, the $\tan\beta$ dependence of the curves in fig.~\bsgamen\
reflect the contribution in eq.~\bsgam\ proportional to
the square of the Higgs-top quark Yukawa coupling, $\lambda_t^2$.
(The $\tan\beta$-independent part of the charged Higgs contribution
in eq.~\bsgam\ arises from a term proportional to $\lambda_t\lambda_b$.)
The current experimental limit from the CLEO
Collaboration\refmark\pkim\
of $BR(B\to K\gamma+X)<5.4\times 10^{-4}$ (at $95\%$ CL)
already places
interesting limits on the parameters of the charged Higgs sector
(see refs.~[\joanne] and [\joannetwo]
for a recent analysis of these constraints).
Forthcoming improved limits from CLEO (or an observed signal)
could significantly constrain $\mhpm$ and $\tanb$ and may lead to other
important restrictions on supersymmetric particle masses.

\section{Natural Relations of the MSSM Higgs Sector}

In Lecture 1, I constructed the Lagrangian of the MSSM.
This field theory possesses many tree-level natural relations
which reflect the underlying supersymmetry of the model.
In particular, the supersymmetry is broken by soft-supersymmetry-breaking
terms of dimension 2 and 3.  This means that all
dimension-4 interaction terms respect the supersymmetry.  Thus,
the MSSM possesses far fewer independent parameters than the most
general gauge invariant theory with the same particle content.
If one considers a more general (nonsupersymmetric) theory
with the same particle spectrum and gauge symmetries, it is
clear that the imposition of supersymmetry introduces relations among
previously independent parameters.  These relations are called natural
because even if the supersymmetry is softly broken, these
relations are modified by finite (and calculable) radiative
corrections.  In contrast, if we were to add
hard-supersymmetry-breaking
terms, the natural relations would suffer
infinite radiative corrections.  In this case, one must perform
independent renormalizations to remove the infinities, and no vestige of
the natural relation would remain.

\REF\hhgref{For a comprehensive review and a complete set of references,
see Chapter 4 of ref.~[\hhg].}
Examples of natural relations in the MSSM are nicely exhibited in the
Higgs sector.  Consider the scalar potential of the most general
non-supersymmetric two-Higgs-doublet model\refmark\hhgref.
In principle, this model would contain CP-violating as well
as CP-conserving couplings.  For simplicity (and due to lack of knowledge
of the fundamental CP-violating parameters of the underlying
supersymmetric model), I will assume that all CP-violating effects
arising from the Higgs sector are small and can be neglected.\foot{At
tree-level, the Higgs potential of the MSSM is automatically
CP-conserving.}
Let $\Phi_1$ and
$\Phi_2$ denote two complex $Y=1$, SU(2)$\ls{L}$ doublet scalar fields.
The most general gauge invariant scalar potential is given by
\vskip5pt
$$\eqalign{
\calv&=m_{11}^2\Phi_1^\dagger\Phi_1+m_{22}^2\Phi_2^\dagger\Phi_2
-[m_{12}^2\Phi_1^\dagger\Phi_2+{\rm h.c.}]\crrr
&\quad +\half\lambda_1(\Phi_1^\dagger\Phi_1)^2
+\half\lambda_2(\Phi_2^\dagger\Phi_2)^2
+\lambda_3(\Phi_1^\dagger\Phi_1)(\Phi_2^\dagger\Phi_2)
+\lambda_4(\Phi_1^\dagger\Phi_2)(\Phi_2^\dagger\Phi_1)\crrr
&\quad +\left\{\half\lambda_5(\Phi_1^\dagger\Phi_2)^2
+\big[\lambda_6(\Phi_1^\dagger\Phi_1)
+\lambda_7(\Phi_2^\dagger\Phi_2)\big]
\Phi_1^\dagger\Phi_2+{\rm h.c.}\right\}\,.\cr}\eqn\pot$$
\vskip5pt\noindent
In most discussions of two-Higgs-doublet models, the terms proportional
to $\lambda_6$ and $\lambda_7$ are absent.  This can be achieved by
imposing a discrete symmetry $\Phi_1\to -\Phi_1$ on the model.  Such a
symmetry would also require $m_{12}=0$ unless we allow a
soft violation of this discrete symmetry by dimension-two terms.\foot{%
This latter requirement is sufficient to guarantee the absence of
Higgs-mediated tree-level flavor changing neutral currents.}
For the moment, I will refrain from setting any of the coefficients
in eq.~\pot\ to zero.  In principle, $m_{12}^2$, $\lambda_5$,
$\lambda_6$ and $\lambda_7$ can be complex.  However, I shall
ignore the possibility of CP-violating effects in the Higgs sector
by choosing all coefficients in eq.~\pot\ to be real.
The scalar fields will
develop non-zero vacuum expectation values if the mass matrix
$m_{ij}^2$ has at least one negative eigenvalue. Imposing CP invariance
and U(1)$\ls{\rm EM}$ gauge symmetry, the minimum of the potential is
\vskip3pt
$$\langle \Phi_1 \rangle={1\over\sqrt{2}}
\pmatrix{0\cr v_1\cr}, \qquad \langle \Phi_2\rangle=
{1\over\sqrt{2}}\pmatrix{0\cr v_2\cr}\,,\eqn\potmin$$
\vskip3pt\noindent
where the $v_i$ can be chosen to be real and positive.
It is convenient to introduce the following notation:
$$v^2\equiv v_1^2+v_2^2={4\mw^2\over g^2}\,,
\qquad\qquad t_\beta\equiv\tanb\equiv{v_2\over v_1}\,.\eqn\tanbdef$$
Of the original eight scalar degrees of freedom, three Goldstone
bosons ($G^\pm$ and $G^0$)
are absorbed (``eaten'') by the $W^\pm$ and $Z$.  The remaining
five physical Higgs particles are: two CP-even scalars ($\hl$ and
$\hh$, with $\mhl\leq \mhh$), one CP-odd scalar ($\ha$) and a charged
Higgs pair ($\hpm$). The mass parameters $m_{11}$ and $m_{22}$ can be
eliminated by minimizing the scalar potential.  The resulting
squared masses for the CP-odd and charged Higgs states are
\vskip4pt
$$\eqalignno{%
\mha^2 &={m_{12}^2\over \sb\cb}-\half
v^2\big(2\lambda_5+\lambda_6 t_\beta^{-1}+\lambda_7t_\beta\big)\,,
&\eqnalign\massha \cr\crr
m_{H^{\pm}}^2 &=m_{A^0}^2+\half v^2(\lambda_5-\lambda_4)\,,
&\eqnalign\mamthree\cr}$$
\vskip5pt\noindent
where $s_\beta\equiv\sin\beta$ and $c_\beta\equiv\cos\beta$.
The two CP-even Higgs states mix according to the following squared mass
matrix:
$$\eqalign{%
\calm^2 &=m_{A^0}^2 \left(\matrix{\sb^2&-\sb\cb\cr
-\sb\cb&\cb^2}\right)\crrr
&+v^2\left( \matrix{\lambda_1\cb^2+2\lambda_6\sb\cb+\lambda_5\sb^2
 &(\lambda_3+\lambda_4)\sb\cb+\lambda_6
\cb^2+\lambda_7\sb^2\crr
(\lambda_3+\lambda_4)\sb\cb+\lambda_6
\cb^2+\lambda_7\sb^2&
\lambda_2\sb^2+2\lambda_7\sb\cb+\lambda_5\cb^2}\right)\,.\cr
}\eqn\massmhh$$
The physical mass eigenstates are
$$\eqalign{%
\hh &=(\sqrt2\, {\rm Re\,}\Phi_1^0-v_1)\cos\alpha+
(\sqrt2\,{\rm Re\,}\Phi_2^0-v_2)\sin\alpha\,,\crr
\hl &=-(\sqrt2\,{\rm Re\,}\Phi_1^0-v_1)\sin\alpha+
(\sqrt2\,{\rm Re\,}\Phi_2^0-v_2)\cos\alpha\,.\cr}
\eqn\scalareigenstates$$
The corresponding masses are
$$
 m^2_{\hh,\hl}=\half\left[{\cal M}_{11}^2+{\cal M}_{22}^2
\pm \sqrt{({\cal M}_{11}^2-{\cal M}_{22}^2)^2 +4({\cal M}_{12}^2)^2}
\ \right]\,,
\eqn\higgsmasses$$
and the mixing angle $\alpha$ is obtained from
$$\eqalign{%
\sin 2\alpha &={2{\cal M}_{12}^2\over
\sqrt{({\cal M}_{11}^2-{\cal M}_{22}^2)^2 +4({\cal M}_{12}^2)^2}}\ ,\crrr
\cos 2\alpha &={{\cal M}_{11}^2-{\cal M}_{22}^2\over
\sqrt{({\cal M}_{11}^2-{\cal M}_{22}^2)^2 +4({\cal M}_{12}^2)^2}}\ .\cr}
\eqn\alphadefff$$

In the MSSM, the dimension-four terms of the Higgs potential are
constrained by supersymmetry.  Comparing eq.~\mssmhiggspot\
with the nonsupersymmetric Higgs potential given in eq.~\pot,\foot{%
Note that $H_1$ is  a $Y=-1$ field which is related to $\Phi_1$
by $H_1=i\sigma_2\Phi_1^*$.}
one finds the following results for the $\lambda_i$ in the MSSM
$$\eqalign{%
\lambda_1 &=\lambda_2 = \fourth (g^2+g'^2)\,,\cr
\lambda_3 &=\fourth (g^2-g'^2)\,,\cr
\lambda_4 &=-\half g^2\,,\cr
\lambda_5 &=\lambda_6=\lambda_7=0\,.\cr}
\eqn\bndfr$$
Eq.~\bndfr\ is the source of all the natural relations among Higgs
sector observables in the MSSM.  For example, by
inserting eq.~\bndfr\ into eqs.~\massha--\alphadefff, one reproduces
the formulae for the MSSM Higgs masses and mixing angle obtained
in eqs.~\monetwoeq--\cpevenhiggsmass.

\REF\hhprl{H.E. Haber and R. Hempfling, {\sl Phys. Rev. Lett.} {\bf 66}
(1991) 1815.}
\REF\radmssm{Y.
Okada, M. Yamaguchi and T. Yanagida, {\sl Prog. Theor. Phys.} {\bf 85}
(1991) 1; J. Ellis, G. Ridolfi and F. Zwirner, {\sl Phys. Lett.}
{\bf B257} (1991) 83; {\bf B262} (1991) 477.}
\REF\onehiggsrge{Y.
Okada, M. Yamaguchi and T. Yanagida, {\sl Phys. Lett.} {\bf B262}
(1991) 54; R. Barbieri, M. Frigeni,
and F. Caravaglios, {\sl Phys. Lett.} {\bf B258} (1991) 167;
J.R. Espinosa and M. Quiros, {\sl Phys. Lett.} {\bf B267}
(1991) 27.}
\REF\moreradmssm{%
R. Barbieri and M. Frigeni, {\sl Phys. Lett.} {\bf B258} (1991) 395;
A. Yamada, {\sl Phys. Lett.} {\bf B263} (1991) 233.}
\REF\pokorski{
P.H. Chankowski, S. Pokorski and J. Rosiek, {\sl Phys. Lett.}
{\bf B274} (1992) 191; {\bf B281} (1992) 100.}
\REF\llog{H.E. Haber and R. Hempfling, SCIPP 91/33 (1992).}
\REF\berkeley{D.M. Pierce, A. Papadopoulos, and S. Johnson,
{\sl Phys. Rev. Lett.} {\bf 68} (1992) 3678.}
\REF\sasaki{K. Sasaki, M. Carena and C.E.M. Wagner, {\sl Nucl. Phys.}
{\bf B381} (1992) 66.}
\REF\andrea{A. Brignole, {\sl Phys. Lett.} {\bf B281} (1992) 284.}
\REF\diaztwo{M.A. Diaz and H.E. Haber, {\sl Phys.
Rev.} {\bf D46} (1992) 3086.}

Perhaps the most interesting consequence of the Higgs mass
relations of the MSSM is the inequality $\mhl \leq \mz$ [see eq.~%
\cpevenhiggsmass]. If this mass
bound were reliable, it would have significant
implications for future experiments
at LEP-II.  In principle, experiments running at LEP-II operating
at $\sqrt{s}=200$~GeV and design luminosity would either discover the
Higgs boson (via $e^+e^-\to\hl Z$)
or rule out the MSSM.  However, since
$\mhl\leq\mz$ is a consequence of the natural relations of the MSSM
Higgs sector, this inequality need not be respected
when radiative corrections are incorporated.
In section 3.4 I will demonstrate that
in the radiative corrections to the neutral CP-even Higgs squared-mass
matrix, the $22$-element is shifted by a term proportional to
$(g^2 m_t^4/\mw^2)\,\ln(M_{\tilde t}^2/m_t^2)$%
\refmark{\hhprl--\diaztwo}.
Such a term arises
{}from an incomplete cancellation between top-quark and top-squark
loop contributions to the neutral Higgs boson self-energy.
If $m_t$ is large, this term significantly alters the tree-level
natural relations of the MSSM Higgs sector.

\section{Radiative Corrections to MSSM Higgs Masses}

The complete one-loop computation of the MSSM Higgs masses can be
found in the literature\refmark{\pokorski,\andrea}.  However,
the formulae involved are very lengthy and not too transparent.
Instead, I will present here
the results based on a calculation of the radiatively corrected
Higgs masses in which all leading logarithmic terms are
included (see ref.~[\llog] for details).
We take the supersymmetry breaking
scale ($\msusy$) to be somewhat larger than the electroweak scale.
For simplicity, we assume that the masses of all supersymmetric
particles (squarks, sleptons, neutralinos and charginos) are roughly
degenerate and of order $\msusy$.
Although this is a crude approximation, deviations
{}from this assumption will
lead to non-leading logarithmic corrections which tend to be small
if the supersymmetric particles are not widely split in mass.\foot{%
The procedure outlined below can be modified to
incorporate the largest non-leading logarithmic contributions that
arise in the case of multiple supersymmetric particle thresholds and/or
large squark mixing. See ref.~[\llog] for further details.}

The leading logarithmic expressions for the MSSM
Higgs masses are obtained from
eqs.~\mamthree\ and \massmhh\ by treating the $\lambda_i$ as
running parameters evaluated at the electroweak scale, $\mweak$.
In addition, we identify the $W$ and $Z$ masses by
$$\eqalign{\mw^2&=\fourth g^2(v_1^2+v_2^2)\,,\crr
\mz^2&=\fourth (g^2+g'^2)(v_1^2+v_2^2)\,,\cr}\eqn\vmasses$$
where the running gauge couplings are also evaluated at $\mweak$.
Of course, the gauge couplings, $g$ and $g'$ are known from
experimental measurements which are performed at the scale $\mweak$.
The $\lambda_i(\mweak^2)$ are determined from supersymmetry.
Namely, if supersymmetry were unbroken, then the $\lambda_i$ would
be fixed according to eq.~\bndfr.  Since supersymmetry is broken,
we regard eq.~\bndfr\ as boundary conditions for the running
parameters, valid at (and above) the energy scale $\msusy$.  That is,
we take
$$\eqalign{
\lambda_1(\msusy^2)&=\lambda_2(\msusy^2)=\fourth\left[g^2(\msusy^2)
+g'^2(\msusy^2)\right],\crrr
\lambda_3(\msusy^2)&=\fourth\left[g^2(\msusy^2)-g'^2(\msusy^2)\right],\crrr
\lambda_4(\msusy^2)&=-\half g^2(\msusy^2),\crrr
\lambda_5(\msusy^2)&=\lambda_6(\msusy^2)=
\lambda_7(\msusy^2)=0\,,\cr}\eqn\boundary$$
in accordance with the tree-level relations of the MSSM. I shall
assume that $\mha\simeq{\cal O}(\mz)$, so that all Higgs particle
masses are of order $\mweak$ (rather than $\msusy$).  Then, at energy
scales below $\msusy$, the effective low-energy theory is a
two-Higgs-doublet model.  As a result,
the gauge and quartic Higgs couplings evolve
{}from $\msusy$ down to $\mweak$
according to the
renormalization group equations (RGEs) of the non-supersymmetric
two-Higgs-doublet model (with Higgs-fermion couplings given in the
MSSM).\footnote\star%
{If $\mha\sim {\cal O}(\msusy)$, then $H^\pm$, $\hh$ and
$\ha$ would all have masses of order $\msusy$, and the effective
low-energy theory below $\msusy$ would be that of the minimal
Standard Model.  In this case, it would be appropriate to decouple
all but the lightest CP-even Higgs boson, and use the RGEs of the
Standard Model with one Higgs doublet.  For further details,
see refs.~[\onehiggsrge] and [\llog].}
The required one-loop RGEs are of the form given in eq.~\rges,
where the parameter set $p_i$ now includes the Higgs
self-coupling parameters $\lambda_i$.
The corresponding $\beta$-functions are listed below
$$\eqalign{%
\beta_{\lambda_{1}} &={1\over16\pi^2}
   \bigg\{ 6\lambda^2_{1}+2\lambda_3^2+2\lambda_3\lambda_4+
   \lambda_4^2+\lambda_5^2+12\lambda_{6}^2 \bigg.\cr
&\hskip2cm \bigg.        +\threeighth
       \big[2g^4+(g^2+g'^2)^2\big]-2\sum_i N_{ci}h_{d_i}^4\bigg\}
      -2\lambda_{1}\gamma_{1}\crr
\beta_{\lambda_{2}} &={1\over16\pi^2}
   \bigg\{ 6\lambda^2_{2}+2\lambda_3^2+2\lambda_3\lambda_4+
   \lambda_4^2+\lambda_5^2+12\lambda_{7}^2 \bigg.\cr
&\hskip2cm \bigg. +\threeighth
      \big[2g^4+(g^2+g'^2)^2\big]-2\sum_i N_{ci}h_{u_i}^4\bigg\}
      -2\lambda_{2}\gamma_{2}\crr
\beta_{\lambda_3} &={1\over16\pi^2}
   \bigg\{ (\lambda_1+\lambda_2)(3\lambda_3+\lambda_4)+2\lambda_3^2
   +\lambda_4^2+\lambda_5^2+2\lambda_6^2+2\lambda_7^2
   +8\lambda_6\lambda_7 \bigg.\cr
&\hskip2cm +\threeighth   \bigg.
      \big[2g^4+(g^2-g'^2)^2\big]-2\sum_i N_{ci}h_{u_i}^2h_{d_i}^2\bigg\}
      -\lambda_3(\gamma_1+\gamma_2)\crr
\beta_{\lambda_4}  &={1\over16\pi^2}
   \Big[ \lambda_4(\lambda_1+\lambda_2+4\lambda_3+2\lambda_4)+
   4\lambda_5^2+5\lambda_6^2+5\lambda_7^2+2\lambda_6\lambda_7\Big.\crr
&\hskip2cm +\threehalf       \Big.
      g^2g'^2+2\sum_i N_{ci}h_{u_i}^2h_{d_i}^2\Big]
      -\lambda_4(\gamma_1+\gamma_2)\crr
\beta_{\lambda_5} &={1\over16\pi^2}
   \Big[\lambda_5(\lambda_1+\lambda_2+4\lambda_3+6\lambda_4)
   +5\big(\lambda_6^2+\lambda_7^2\big)+2\lambda_6\lambda_7\Big]
   -\lambda_5(\gamma_1+\gamma_2)\crr
\beta_{\lambda_6} &={1\over16\pi^2}
   \Big[\lambda_6(6\lambda_1+3\lambda_3+4\lambda_4+5\lambda_5\big)
   +\lambda_7\big(3\lambda_3+2\lambda_4+\lambda_5\big)\Big]
   -\half\lambda_6(3\gamma_1+\gamma_2)\crr
\beta_{\lambda_7} &={1\over16\pi^2}
   \Big[\lambda_7(6\lambda_2+3\lambda_3+4\lambda_4+5\lambda_5\big)
   +\lambda_6\big(3\lambda_3+2\lambda_4+\lambda_5\big)\Big]
   -\half\lambda_7(\gamma_1+3\gamma_2)\,. \crr}
\eqn\betal$$
In eq.~\betal, the anomalous dimensions of the two
Higgs fields are given by
$$\eqalign{%
\gamma_{1} &={1\over 64\pi^2}
   \Big[9g^2+3g'^2-4\sum_i N_{ci}h_{d_i}^2\Big]\,,\crr
\gamma_{2} &={1\over 64\pi^2}
    \Big[9g^2+3g'^2-4\sum_i N_{ci}h_{u_i}^2\Big]\,,\cr}
\eqn\wavez$$
where the sum over $i$ is taken over three generations of quarks
(with $N_c=3$) and leptons (with $N_c=1$).
The boundary conditions together with the RGEs imply that, at the
leading-log level, $\lambda_5$, $\lambda_6$ and $\lambda_7$ are zero
at all energy scales.  Solving the RGEs with
the supersymmetric boundary conditions at $\msusy$ [eq.~\boundary],
one can determine
the $\lambda_i$ at the weak scale.
The resulting values for $\lambda_i(\mweak)$
are then inserted into eqs.~\mamthree\ and \massmhh\ to obtain the
radiatively corrected Higgs masses.  Having solved the one-loop RGEs,
the Higgs masses thus obtained  include the leading
logarithmic radiative corrections summed to all orders in perturbation
theory.

The RGEs can be solved by numerical analysis on the computer.  But
it is instructive to solve the RGEs iteratively.  In first
approximation, we can take the right hand side of eq.~\rges\ to
be independent of $\mu^2$.  That is, we compute the $\beta_i$ by
evaluating the parameters $p_i$ at the scale $\mu=\msusy$.
Then, integration of the RGEs is
trivial, and we obtain
\vskip5pt
$$p_i(\mweak^2)=p_i(\msusy^2)-\beta_i\,\ln\left({\msusy^2\over\mweak^2}
\right)\,.\eqn\oneloopllog$$
\vskip5pt\noindent
Note that this iterative solution corresponds to computing the
one-loop radiative corrections in which only terms proportional to
$\ln\msusy^2$ are kept.  It is straightforward to work out the
one-loop leading logarithmic expressions for the $\lambda_i$ and
the Higgs masses.  First consider the charged Higgs mass [eq.~%
\mamthree].   Since
$\lambda_5(\mu^2)=0$ at all scales, we need only consider $\lambda_4$.
Evaluating $\beta_{\lambda_4}$ at $\mu=\msusy$, it follows that
\vskip5pt
$$\eqalign{%
\lambda_4(\mw^2)=-\half g^2
-&{1\over{32\pi^2}}\biggl[\bigl({\textstyle{
4\over 3}}N_g+{\textstyle{1\over 6}}N_H-{\textstyle{10\over 3}}
\bigr)g^4+5g^2g'^2\crr
&\qquad\qquad-{{3g^4}\over{2m_W^2}}\left({{m_t^2}\over
{s_{\beta}^2}}+{{m_b^2}\over{c_{\beta}^2}}\right)
+{{3g^2m_t^2m_b^2}\over{s_{\beta}^2c_{\beta}^2m_W^4}}\Biggr]
\ln{{\msusy^2}\over{\mw^2}}\,.\cr}\eqn\lcuaunloop$$
\vskip5pt\noindent
The terms proportional to the number of generations $N_g=3$
and the number of Higgs doublets $N_H=2$ that remain in the
low-energy effective theory at the scale $\mu=\mw$ have their origin in
the running of $g^2$ from $\msusy$ down to $\mw$.
In deriving the above expression, I have taken $\mweak=\mw$.  This is
a somewhat arbitrary decision, since another reasonable choice would
yield a result that differs from eq.~\lcuaunloop\ by a non-leading
logarithmic term.  Comparisons with a more complete
calculation show that
one should choose $\mweak=\mw$ in computations involving the charged
Higgs (and gauge) sector and $\mweak=\mz$ in computations involving the
neutral sector.

\REF\diaz{M.A. Diaz and H.E. Haber, {\sl Phys. Rev.} {\bf D45} (1992)
4246.}
The above analysis also assumes that $m_t\sim {\cal O}(m_W)$.  Since
$\mt>\mw$, one can improve the above result somewhat
by decoupling the $(t,b)$ weak doublet from
the low-energy theory for scales below $m_t$.  The
terms in eq.~\lcuaunloop\ that are proportional to $m_t^2$ and/or
$m_b^2$ arise from self-energy diagrams containing a $tb$ loop.
Thus, such a term should not be present for $\mw\leq \mu\leq m_t$.
In addition, we recognize the term in eq.~\lcuaunloop\ proportional
to the number of generations $N_g$ as arising from the contributions
to the self-energy diagrams containing either quark or lepton loops
(and their supersymmetric partners).
To identify the contribution of the $tb$ loop to this term,
simply
write
\vskip3pt
$$N_g=\fourth N_g(N_c+1)=\fourth N_c+\fourth[N_c(N_g-1)+N_g]\,,
\eqn\ngee$$
\vskip5pt\noindent
where $N_c=3$ colors.  Thus, we identify ${1\over 4}N_c$ as the piece
of the term proportional to $N_g$ that is due to the $tb$ loop.  The
rest of this term is then attributed to the lighter quarks and leptons.
Finally, the remaining terms in eq.~\lcuaunloop\ are due to the
contributions from the gauge and
Higgs boson sector.  The final result is\refmark\diaz\
\vskip5pt
$$\eqalign{\lambda_4(\mw^2) &=-\half g^2
-{N_c g^4\over{32\pi^2}}\left[{1\over 3}
-{1\over{2m_W^2}}\biggl({{m_t^2}\over
{s_{\beta}^2}}+{{m_b^2}\over{c_{\beta}^2}}\biggr)
+{{m_t^2m_b^2}\over{s_{\beta}^2c_{\beta}^2m_W^4}}\right]
\ln{{\msusy^2}\over{m_t^2}}\crr
&\qquad-{1\over 96\pi^2}\left\{\left[N_c(N_g-1)+N_g
+\half N_H-10\right]g^4
+15g^2g'^2 \right\}\ln{\msusy^2\over\mw^2}
\,.\cr}\eqn\lambdaiv$$
\vskip12pt\noindent
Inserting this result (and $\lambda_5=0$) into eq.~\mamthree,
one obtains the one-loop leading-log formula for the charged Higgs mass
\vskip3pt
$$\eqalign{%
m_{H^{\pm}}^2&=m_A^2+m_W^2 +{{N_c g^2}\over{32\pi^2m_W^2}}
\Bigg[{{2m_t^2m_b^2}\over{s_{\beta}^2c_{\beta}^2}}-m_W^2
\bigg({{m_t^2}\over{s_{\beta}^2}}+{{m_b^2}\over{c_{\beta}^2}}\bigg)
+{\textstyle{2\over 3}}m_W^4\Bigg]\ln{{\msusy^2}\over{m_t^2}}  \crr\crr
&\qquad+{{m_W^2}\over{48\pi^2}} \Big\{\left[N_c(N_g-1)+N_g
+\half N_H-10\right]g^2 +15g'^2\Big\}
\ln{{\msusy^2}\over{m_W^2}}\,.
\cr}\eqn\llform$$
\vskip5pt\noindent
\break
Since the low-energy effective theory below \msusy\ is a two-doublet
model, one {\it must} take $N_H=2$ in the
formulae above. In particular, the derivation of eq.~\llform\
is strictly valid only if $\mha$ is not much larger than $\mw$.

Eq.~\llform\ only includes the one-loop
leading log radiative corrections.  This result is
improved by using the full RGE solution to $\lambda_4(\mw^2)$
$$\mhpm^2=\mha^2-\half\lambda_4(\mw^2)(v_1^2+v_2^2)\,,
\eqn\chiggsrge$$
which effectively sums the leading log radiative corrections to all
orders in the perturbation expansion.

\REF\turski{J.F. Gunion and A. Turski, {\sl Phys. Rev.}
{\bf D39} (1989) 2701; {\bf D40} (1989) 2333.}
\REF\berz{A. Brignole, J. Ellis, G. Ridolfi and F. Zwirner,
{\sl Phys. Lett.} {\bf B271} (1991) 123 [E: {\bf B273} (1991) 550].}
\REF\brignole{A. Brignole, {\sl Phys. Lett.} {\bf B277} (1992) 313.}
Although the leading-log formula for $\mhpm$ [eq.~\llform\ or
\chiggsrge] gives
a useful indication as to the size of the radiative corrections,
non-leading logarithmic contributions can also be important
in certain regions of parameter space.  A more complete set of
radiative corrections can be found in the literature%
\refmark{\pokorski,\diaz--\brignole}.
In the numerical results to be
exhibited below, important non-leading corrections to
the charged Higgs mass are also included (as described in ref.~[\diaz]).
However, it should be emphasized that the radiative corrections
to the charged Higgs mass are significant only for $\tanb<1$, a
region of MSSM parameter space not favored in supersymmetric
models.

The computation of the neutral CP-even Higgs masses follows a
similar procedure.  The results are summarized below\refmark\llog.
{}From eq.~\massmhh, we see that we only need results for $\lambda_1$,
$\lambda_2$ and $\widetilde\lambda_3\equiv\lambda_3+\lambda_4+\lambda_5$.
(Recall that in the leading-log analysis,
$\lambda_5=\lambda_6=\lambda_7=0$ at all energy scales.)
By iterating the corresponding RGEs as before, we end up with
$$\eqalign{
\lambda_1(\mz^2)&=~~\fourth[g^2+g'^2](\mz^2)
+{g^4\over384\pi^2\cw^4}\Bigg[ P_t\ln\left({\msusyy\over
m_t^2}\right)\crr
&\qquad+\bigg(12N_c{m_b^4\over\mz^4\cb^4}-6N_c{m_b^2\over\mzz\cb^2}
+P_f+P_g+P_{2H} \bigg)\ln\left({\msusyy\over
m_Z^2}\right)\Bigg]\,,\crr \crr
\lambda_2(\mz^2)&=~~\fourth
[g^2+g'^2](\mz^2)
+{g^4\over384\pi^2\cw^4}\Bigg[\bigg(P_f+P_g+P_{2H}
\bigg)\ln\left({\msusyy\over m_Z^2}\right) \crr
&\qquad+\bigg(12N_c{m_t^4\over\mz^4\sb^4}-6N_c{m_t^2\over\mzz\sb^2}
+P_t\bigg)\ln\left({\msusyy\over m_t^2}\right)\Bigg]\,,\crr\crr
\widetilde\lambda_3(\mz^2)&=-\fourth[g^2+g'^2](\mz^2)
-{g^4\over384\pi^2\cw^4}\Bigg[\bigg(-3N_c{m_t^2\over\mzz\sb^2}
+P_t\bigg)\ln\left({\msusyy\over m_t^2}\right)\crr
&\qquad+\bigg(-3N_c{m_b^2\over\mzz\cb^2}+P_f+P_g'+P_{2H}'
\bigg)\ln\left({\msusyy\over m_Z^2}\right)\Bigg]
\,,\cr}\eqn\dlambda$$
where
$$\eqalign{P_t~&\equiv~~N_c(1-4e_u\sw^2+8e_u^2\sw^4)\,,\crr
P_f~&\equiv~~ N_g\bigg\{N_c\left[2-4\sw^2+8(e_d^2+e_u^2)\sw^4\right]
+[2-4\sw^2+8\sw^4]\bigg\}-P_t\,,\crr
P_g~&\equiv-44+106\sww-62\sw^4\,,\crr
P_g'~&\equiv~~10+34\sww-26\sw^4\,,\crr P_{2H}&\equiv
-10+2\sww-2\sw^4\,,\crr
P_{2H}'&\equiv~~8-22\sww+10\sw^4\,.\cr}\eqn\defpp$$
\vskip5pt\noindent
In the above formulae, the electric charges of the quarks are $e_u
= 2/3$, $e_d = -1/3$, and the subscripts $t, f, g$ and $2H$
indicate that these are the contributions from the
top quark, the fermions (leptons and quarks excluding the top quark),
the gauge bosons and the two Higgs doublets (and corresponding
supersymmetric partners), respectively.
As in the derivation of $\lambda_4(\mw^2)$ above, we have improved
our analysis by removing the effects of top-quark loops below
$\mu=\mt$.  This requires a careful treatment of the evolution of
$g$ and $g'$ at scales below $\mu=\mt$.  The correct procedure is
somewhat subtle, since the full electroweak gauge symmetry is
broken below top-quark threshold; for further details, see
ref.~[\llog].  However, the following pedestrian
technique works: consider the RGE for $g^2+g'^2$ valid for $\mu<\msusy$
$$
{d\over dt}(g^2+g'^2)={1\over 96\pi^2}\Big[\left(8g^4
+\fortythirds g'^4\right)N_g+(g^4+g'^4)N_H-44g^4\Big]\,.
\eqn\grge$$
This equation can be used to run $g^2+g'^2$,
which appears in eq.~\boundary,
{}from $\msusy$ down to $\mz$.  As before, we identify the term
proportional to $N_g$ as corresponding to the fermion loops.
We can explicitly extract the $t$-quark contribution by noting that
$$\eqalign{%
N_g\left(8g^4+\fortythirds g'^4\right)&=
{g^4N_g\over\cw^4}
\Big[\sixtyfourthirds s_W^4-16 s_W^2+8\Big]\crr
&={g^4\over\cw^4} \Big\{N_c\left[1+(N_g-1)\right] (1-4e_u s_W^2+8e_u^2
s_W^4)\crr
&\qquad + N_c N_g(1+4e_d s_W^2+8e_d^2 s_W^4)
+N_g(2-4 s_W^2+8s_W^4)\Big\}\,,\cr}
\eqn\pedest$$
\vskip5pt\noindent
where in the second line of eq.~\pedest, the term proportional
to 1 corresponds to the $t$-quark
contribution while the term proportional
to $N_g-1$ accounts for the $u$ and $c$-quarks; the third line contains
the contributions from the down-type quarks and leptons respectively.
Thus, iterating to one-loop,
$$\eqalign{%
(g^2+g'^2)(\msusy^2)&=
(g^2+g'^2)(\mz^2)+{g^4\over 96\pi^2 c_W^4}
\Bigg\{P_t\ln\left({\msusy^2\over\mt^2}\right)\crr
&\qquad+\left[P_f+(\sw^4+\cw^4)N_H-44\cw^4\right]
\ln\left({\msusy^2\over\mz^2}
\right)\Bigg\}\,.\cr}\eqn\gaugeiter$$
This result and terms that are proportional to $\mt^2$ and $\mt^4$
yield the terms in eq.~\dlambda\ that contain $\ln(\msusy^2/\mt^2)$.

The final step is to insert the expressions obtained in eq.~\dlambda\
into eq.~\massmhh.  The resulting
matrix elements for the mass-squared matrix to one-loop leading
logarithmic accuracy are given by
$$\eqalign{\calm_{11}^2&=\mha^2\sb^2+m_Z^2\cb^2
+{g^2\mzz\cb^2\over96\pi^2\cw^2}\Bigg[
P_t~\ln\left({\msusyy\over m_t^2}\right)\crrr
&\quad+\bigg(12N_c{m_b^4\over\mz^4\cb^4}-6N_c{m_b^2\over\mzz\cb^2}
+P_f+P_g+P_{2H} \bigg)\ln\left({\msusyy\over
m_Z^2}\right)\Bigg] \crr   \crr
\calm_{22}^2&=\mha^2\cb^2+m_Z^2\sb^2
+{g^2\mzz\sb^2\over96\pi^2\cw^2}\Bigg[\bigg(P_f+P_g+P_{2H}
\bigg)\ln\left({\msusyy\over m_Z^2}\right)\crrr
&\quad+\bigg(12N_c{m_t^4\over\mz^4\sb^4}-6N_c{m_t^2\over\mzz\sb^2}
+P_t\bigg)\ln\left({\msusyy\over m_t^2}\right)\Bigg]\crr \crr
\calm_{12}^2&=-\sb\cb\Biggl\{\mha^2+m_Z^2
+{g^2\mzz\over96\pi^2\cw^2}\Bigg[\bigg(P_t-3N_c{m_t^2\over\mzz\sb^2}
\bigg)\ln\left({\msusyy\over m_t^2}\right)\crrr
&\quad+\bigg(-3N_c{m_b^2\over\mzz\cb^2}+P_f+P_g'+P_{2H}'
\bigg)\ln\left({\msusyy\over
m_Z^2}\right)\Bigg]\Biggr\}\,.\cr}\eqn\mtophree$$
\vskip5pt\noindent
Diagonalizing this matrix [eq.~\mtophree] yields
the radiatively corrected CP-even Higgs masses and mixing angle $\alpha$.

The leading-log
formulae presented above are expected to be accurate as long
as: (i) there is one
scale characterizing supersymmetric masses, $\msusy$, which is
large and sufficiently separated from $\mz$
(say, $\msusy\gsim 500$~GeV),
(ii) $m_t$ is somewhat above $\mz$ (say, $\mt\gsim 125$~GeV)
while still being small compared to $\msusy$, and (iii)
the squark mixing parameters are not unduly large.  In particular,
(ii) is an important condition---it is the dominance of the leading
$\mt^4\ln(\msusy^2/\mt^2)$ term in ${\cal M}_{22}^2$ above
that guarantees that the
non-leading logarithmic terms are unimportant.
\foot{In contrast, there
is no leading logarithmic contribution to $\mhpm^2$ that grows with
$\mt^4$.  As a result, the non-leading logarithmic terms tend to be
more important as mentioned earlier.}
Under these conditions, the largest non-leading logarithmic term is
of ${\cal O}(g^2\mt^2)$, which can be identified from a full one-loop
computation as being the subdominant term relative to the leading
${\cal O}(g^2\mt^4\ln\msusy^2)$ term in ${\cal M}_{22}^2$.  Thus,
we can make a minor improvement on our computation of the leading-log
CP-even Higgs squared mass matrix by taking
\vskip5pt
$$
{\cal M}^2 = {\cal M}^2_{\rm LL} +
{N_c g^2\mt^2\over48\pi^2\sb^2c_W^2}\left(\matrix{0&0\cr0&1}\right)
\eqn\nllogtrm
$$
\vskip10pt\noindent
where $\calm^2_{\rm LL}$ is the leading-log CP-even Higgs squared mass
matrix [given to one-loop in eq.~\mtophree].
The shift of the
light Higgs mass due to this non-leading-log correction is of
order 1 GeV.  Finally, the case of multiple and widely separated
supersymmetric particle thresholds and/or
large squark mixing (which is most likely in the top squark sector),
lead to
new non-leading logarithmic contributions to the scalar mass-squared
matrix can become important.  These are discussed further
in ref.~[\llog].

\section{Implications of the Radiatively Corrected MSSM Higgs
Sector}

\FIG\mhtbvma{%
RGE-improved Higgs mass $\mhl$ as a function
of $\tanb$ for (a) $\mt = 150$ GeV and (b) $\mt = 200$ GeV.
Various
curves correspond to $\mha = 0,~20,~50,~100$ and $300$ GeV as labeled in
the figure. All $A$-parameters and $\mu$ are set equal to zero.
The light CP-even Higgs mass varies very weakly with $\mha$
for $\mha>300$ GeV.  Taken from ref.~[\llog].}
In this section I shall briefly survey some of the numerical
results for the radiatively corrected Higgs masses and couplings.
Additional results can be found in ref.~[\llog].
Complementary work can be found in refs.~[%
\radmssm,
\moreradmssm,
\pokorski,
\berkeley--
\brignole].  
In fig.~\mhtbvma (a) and (b) I
plot the light CP-even Higgs mass as a function of
$\tanb$ for $\mt = 150$ and 200 GeV for various choices of $\mha$.
All $A$-parameters and $\mu$ are set equal to zero.
Perhaps the most dramatic consequence of the radiatively corrected
Higgs sector is the large violation of the tree-level bound
$\mhl\leq\mz$.  The new $\mhl$ bound ($\mhl^{\rm max}$)
is saturated when $\mha$ and $\tanb$ are large.  Moreover, when
$\mha\geq\mhl^{\rm max}$, one sees that $\mhl$ is large throughout
the entire $\tanb$ region;
the radiatively corrected $\mhl$ reaches
a maximum (minimum) at $\tanb\simeq\infty$ $(\tanb\simeq1)$.
In particular, there is a substantial
region of Higgs parameter space in which $\mhl$ lies above the
current LEP experimental Higgs mass limits.
Indeed, for $\msusy=1$~TeV, $\mt=200$~GeV,
and $\mha\gsim 200$~GeV, fig.~\mhtbvma(b) indicates that
$\mhl>\mz$ independent of the
value of $\tanb$.  Thus, there is a non-negligible region of parameter
space in which the $\hl$ is kinematically inaccessible to LEP-II
(running at $\sqrt{s}\leq 200$~GeV).  This is quite a departure from
the tree-level expectations of $\mhl\leq\mz$ in which the LEP
discovery of the Higgs boson was assured if the MSSM were correct.

Nevertheless, the LEP Higgs search does rule out some
regions of the Higgs parameter space.  These regions typically
correspond to smaller values of $\mha$ where $\hl$ can be
relatively light.
For fixed $\tanb$, $\mhl$ reaches its minimum
value, $\mhl^{\rm min}$, when $\mha\to 0$.  In contrast to the
tree-level behavior (where $\mhl\leq \mha$),
the Higgs mass does not vanish
as $\mha\to 0$. Moreover, $\mhl^{\rm min}$
increases as $\tanb$ decreases but
exhibits only a moderate dependence on $\mt$ and $\msusy$.
One interesting consequence is that
there exists a range of parameters for which
the tree-level bound, $\mhl\leq\mha$ is violated.  In fact,
the results of fig.~\mhtbvma\ indicate that in the region of small
$\tanb$ and small $\mha$, it is possible to have $\mhl>2\mha$, thereby
allowing a new decay-mode $\hl\to\ha\ha$ which is kinematically
forbidden at tree-level.

\REF\colemanweinberg{S. Coleman and E. Weinberg, {\sl Phys.~Rev.}
{\bf D7} (1973) 1888.}
\REF\hhgreftwo{See section 2.5 of ref.~[\hhg].}
\REF\lepsearch{See, \eg, D.~Decamp \etal\ [ALEPH Collaboration],
{\sl Phys. Rep.} {\bf 216} (1992) 253.}
\REF\franco{E. Franco and A. Morelli, {\sl Nuovo Cim.} {\bf 96A} (1986)
257.}
One other difference between the tree-level prediction for $\mhl$ and the
results of fig.~\mhtbvma\ is noteworthy.   From eq.~\cpevenhiggsmass,
we see that for $\tanb=1$, $\mhl=0$ at tree-level.  The results of
fig.~\mhtbvma\ indicate that the radiative corrections to $\mhl$ are
substantial for $\tanb=1$, particularly when $\mt$ is large.  This
is again a consequence of the $g^2 m_t^4\ln(\msusy^2/\mz^2)$ enhancement
of ${\cal M}_{22}^2$.
The $\tanb=1$ limit is analogous to the
Coleman-Weinberg limit\refmark\colemanweinberg\ of
the Standard Model, in which the mass of the Higgs boson arises
entirely from radiative corrections.  However, in the Standard Model,
the Coleman-Weinberg mechanism cannot be operative
if $\mt\gsim\mw$\refmark\hhgreftwo\ (and
in any case, the Higgs mass that arises from this mechanism cannot be
larger than about 10 GeV, which is ruled
out by the LEP Higgs search\refmark\lepsearch).
Clearly, no such restriction exists in the
MSSM\refmark{\franco,\diaztwo}.  The difference lies
in the large positive contribution to the Higgs squared mass from
a loop of top squarks.  From fig.~\mhtbvma(b), we see that for
$\mt=200$~GeV and $\tanb=1$, a value of $\mhl$ as large as 100~GeV is
possible.  Thus, LEP cannot yet rule out the possibility that the mass
of the lightest CP-even Higgs boson arises entirely from radiative
corrections\refmark\diaztwo.

\FIG\regime{%
The range of allowed Higgs masses for
large $\mha$ (in these plots, $\mha=300$ GeV).
The lower limit corresponds to $\tanb=1$.
The upper limit corresponds to the limit of large
$\tanb$ (we take $\tanb=20$). In (a) and (b)
$\mt$ is varied for $\msusy= 1$ and $0.5$
TeV, respectively. In (c) and (d)
$\msusy$ is varied and $\mt= 150$ and $200$~GeV,
respectively. The solid (dashed) curves in (c) and (d)
correspond to the computation in which the RGEs are solved numerically
(iteratively to one-loop order).  Taken from ref.~[\llog].}

In the limit $\mha\to\infty$, the couplings of $\hl$ to gauge
bosons and matter fields are identical to the Higgs couplings of
the Standard Model so that the Higgs sector of the two models cannot be
phenomenologically distinguished.
However, supersymmetry does impose constraints on the quartic Higgs
self-coupling at the scale $\msusy$, and this  influences
the possible values of $\mhl$. To illustrate this point, I have plotted
in fig.~\regime\
the range of allowed $\mhl$ in the case of large $\mha$ (taken here to
be $\mha=300$ GeV).
As noted above, the lower limit for $\mhl$ is attained
if $\tanb\simeq1$ and the upper limit is attained in the limit of large
$\tanb$ (taken to be $\tanb=20$ in fig.~\regime).\foot{A second
maximum for $\mhl$ would arise for very
small $\tanb$; however, this lies outside the permitted region indicated
in figs.~\tanlima\ and \tanlimb.}
Suppose the
top quark mass is known
and that $\hl$ is discovered with Standard Model couplings. If
$\mhl$ does not lie in the allowed mass region displayed in fig.~\regime,
one could conclude that the MSSM is ruled out.
Fig.~\regime\ also exhibits the sensitivity to the choice of $\msusy$.
The larger the value of $\msusy$, the more significant the corrections
to the Higgs mass due to full renormalization group improvement.
In fig.~\regime(c) and (d), the dashed lines (labeled 1LL for one-loop
leading-log) correspond to computing $\mhl$ by exactly diagonalizing the
squared mass matrix given in eq.~\mtophree.  The solid lines (denoted
by RGE) are obtained by solving numerically the RGEs for the $\lambda_i$,
inserting the results into eq.~\massmhh, and computing the eigenvalue of
the lighter CP-even Higgs scalar.  For $\msusy=1$~TeV, the largest
discrepancy between the RGE and 1LL results occurs for
large $\mt$ and $\mha$.  For example, for $\tanb=1$, $\mha=300$~GeV
and $\mt = 200$~GeV, we find $(\mhl)_{\rm RGE} = 96.8$~GeV
while $(\mhl)_{\rm 1LL} = 104.4$~GeV.   Values of $\msusy$ much larger
than 1~TeV would be in conflict with the philosophy of low-energy
supersymmetry.

\FIG\massesra{%
The masses of $\hl$, $\hh$ and $\hpm$ in the MSSM
for $\mha=50$ and 200~GeV.
The neutral CP-even Higgs masses are obtained from a calculation
that includes the leading-log one-loop radiative corrections
[based on eq.~\mtophree].
The charged Higgs mass is obtained from a similar calculation, but
important non-leading logarithmic effects have also been included%
\refmark\diaz.
All supersymmetric masses are assumed to be roughly degenerate
of order $\msusy=1$~TeV.
The two curves for each Higgs mass shown
correspond to $\mt=150$ and 200 GeV. The larger neutral Higgs mass
corresponds to the larger $\mt$ choice.  In the case of $\hpm$,
$\mhpm$ increases [decreases] with $\mt$ for large [small] $\tanb$.}

Consider next the predictions of the one-loop radiatively
corrected Higgs sector for the other physical Higgs bosons of the MSSM.
In fig.~\massesra, I plot the radiatively corrected MSSM Higgs masses
as a function of $\tanb$ for $\msusy=1$~TeV and for
two choices of $\mt$ and $\mha$.  (As above, all $A$ and $\mu$ parameters
are set to zero.)  The neutral Higgs masses have been obtained by
diagonalizing eq.~\mtophree.  Full RGE-improvement, which is not
included in fig.~\massesra, would change these results by no more than
about $5\%$.  In the case of the charged Higgs mass, important
non-leading logarithmic contributions have also been included, as
described in ref.~[\diaz].  Note that
the tree-level bound $\mhpm\geq\mw$ can
be violated, but only if $\tanb\lsim 0.5$ and $\mha$ is small.
The small $\tanb$ region corresponds to an enhanced
Higgs-top quark Yukawa coupling.  This also explains the increase of
$\mhh$ in this region, which is being controlled by the $\mt^4/\sb^2$
factor in ${\cal M}_{22}^2$ [eq.~\mtophree].
Of course, this same
factor is responsible for the violation of the bound $\mhl\leq\mz$.

\vskip30pt
\leftline{\fourteenbf Concluding Remarks}
\vskip\headskip
Is there supersymmetry in our future?  The definitive answer to
this question will not be known
until future colliders present us with compelling evidence
for supersymmetric particles in their data. Nevertheless, the theoretical
motivation for supersymmetry is quite strong.  The fact that gravity
exists almost certainly means that supersymmetry must exist at some
energy scale, since supersymmetry is intimately involved in
string theory---the only known consistent quantum mechanical
framework for the union of gravitational theory and the theory of
elementary particle physics.  However, supersymmetry is clearly not
an exact symmetry of nature.  The existence of gravity implies
nothing about the energy scale of supersymmetry breaking.  If this
scale turns out to be the Planck scale, then supersymmetry will
never be relevant for theories of particle physics that can be
directly tested by experiment.  It is the hierarchy and naturalness
problems that provide the hint for the
connection between supersymmetry and the
electroweak scale.  As a result, much effort has been spent
during the past decade in the
construction of low-energy supersymmetric models and the detailed
exploration of its phenomenology.

The MSSM has been particularly
successful for providing a specific framework for the study of
physics beyond the Standard Model.  However, our present understanding of
supersymmetry is far from complete.  The origin of supersymmetry
breaking at the fundamental level remains poorly
understood.  Future theoretical breakthroughs will almost
certainly require the assistance of
our experimental colleagues.  Perhaps with a detailed low-energy
supersymmetric particle spectrum in hand, one could begin to unravel
some of the secrets of the fundamental underlying theory.  The
successful unification of the gauge coupling constants within the grand
unified supersymmetric framework provides some hope that it may be
possible to probe physics all the way up to the Planck scale based on
our limited information at the TeV scale and below.  The discovery
of low-energy supersymmetry is essential if this hope is to be
realized.  If so, it would truly be a remarkable achievement of
theoretical physics.

\vskip30pt
\centerline{\fourteenbf Acknowledgments}
\vskip\headskip
I would like to thank Joe Polchinski, Jeff Harvey and
K.T. Mahanthappa for a well-organized and stimulating TASI-92.
Their hospitality and support during my stay in Boulder is greatly
appreciated.  I am also grateful to Michael Dine for his critical
comments on a preliminary version of this manuscript.
This work is partially supported by the U.S.
Department of Energy.
\vskip15pt
\refout
\endpage
\figout
\bye